\documentclass{aa}  

\usepackage{graphicx}
\usepackage{txfonts}
\usepackage[utf8]{inputenc}
\usepackage{hyperref}
\usepackage{tabularx} 
\usepackage{caption}
\usepackage{subcaption}
\usepackage[table]{xcolor}
\usepackage[flushleft]{threeparttable}
\usepackage{multirow}
\usepackage[rightcaption]{sidecap}
\usepackage{placeins}
\usepackage{orcidlink}

\hypersetup{
  colorlinks=true,   
  urlcolor=blue,     
  linkcolor=blue,     
  citecolor=blue
}

\begin{document}

   \title{Investigating the metallicity dependence of the mass-loss rate relation of red supergiants}

   \author{K. Antoniadis
          \inst{\ref{noa}, \ref{nkua}},
        E. Zapartas\inst{\ref{crete}}, A. Z. Bonanos \inst{\ref{noa}}, G. Maravelias\inst{\ref{noa},\ref{crete}}, S. Vlassis\inst{\ref{crete},\ref{unicrete}}, G. Mu\~noz-Sanchez\inst{\ref{noa}, \ref{nkua}}, \\ C. Nally\inst{\ref{edinburg}}, M. Meixner\inst{\ref{jpl}}, O. C. Jones\inst{\ref{uk}}, L. Lenki\'c\inst{\ref{jpl},\ref{ipac}}, P. J. Kavanagh\inst{\ref{maynooth}} 
          }

   \institute{IAASARS, National Observatory of Athens, 15236 Penteli, Greece \\
              \email{k.antoniadis@noa.gr} \label{noa}
         \and
             National and Kapodistrian University of Athens, 15784 Athens, Greece \label{nkua}
        \and
            Institute of Astrophysics FORTH, 71110 Heraklion, Greece \label{crete}
        \and
            Physics Department, University of Crete, 71003 Heraklion, Greece \label{unicrete}
        \and
            Institute for Astronomy, University of Edinburgh, Blackford Hill, Edinburgh EH9 3HJ, UK \label{edinburg}
        \and
            Jet Propulsion Laboratory, California Institute of Technology, 4800 Oak Grove Drive, Pasadena, CA 91109, USA \label{jpl}
        \and
            UK Astronomy Technology Centre, Royal Observatory, Blackford Hill, Edinburgh EH9 3HJ, UK \label{uk}
        \and
            IPAC, California Institute of Technology, 1200 E. California Blvd., Pasadena, CA 91125, USA
            \label{ipac}
        \and
             Department of Physics, Maynooth University, Maynooth, Co. Kildare, Ireland \label{maynooth}
            }

  \abstract
   {Red supergiants (RSGs) are cool and evolved massive stars exhibiting enhanced mass loss compared to their main sequence phase, affecting their evolution and fate. However, despite recent advances, the theory of the wind-driving mechanism is not well-established and the metallicity dependence has not been determined.}
   {We aim to uniformly measure the mass-loss rates of large samples of RSGs in different galaxies with $-0.7\lesssim[Z]\lesssim0$ to investigate whether there is a potential correlation with metallicity.}
   {We collected photometry from the ultraviolet to the mid-infrared for all our RSG candidates to construct their spectral energy distribution (SED). Our final sample includes 893 RSG candidates in the Small Magellanic Cloud (SMC), 396 in NGC 6822, 527 in the Milky Way, 1425 in M31, and 1854 in M33. Each SED was modelled using the radiative transfer code \texttt{DUSTY} under the same assumptions to derive the mass-loss rate.}
   {The mass-loss rates range from approximately $10^{-9} \ M_{\odot}$ yr$^{-1}$ to $10^{-5} \ M_{\odot}$ yr$^{-1}$ with an average value of $1.5\times10^{-7} \ M_{\odot}$ yr$^{-1}$. We provided a new mass-loss rate relation as a function of luminosity and effective temperature for both the SMC and Milky Way and compared our mass-loss rates with those derived in the Large Magellanic Cloud (LMC). The turning point in the mass-loss rate vs. luminosity relation differs by around 0.2 dex between the LMC and SMC. The mass-loss rates of the Galactic RSGs at $\log(L/L_\odot)<4.5$ were systematically lower than those determined in the other galaxies, possibly due to uncertainties in the interstellar extinction. We found 60--70\% of the RSGs to be dusty, while 14\% of the LMC and 2\% of the SMC RSGs were significantly dusty. The results for M31 and M33 are inconclusive because of significant blending of sources at distances above 0.5 Mpc, given the resolution of \textit{Spitzer}, which compromises the mid-IR photometry.}
   {Overall, we found similar mass-loss rates among the galaxies, indicating no strong correlation with metallicity other than the location of the turning point. More accurate mid-IR photometry is needed to determine the metallicity dependence.}


   \titlerunning{Investigating the metallicity dependence of the mass-loss rate relation of red supergiants}
   \authorrunning{Antoniadis et al.}

   \maketitle
%
 
\section{Introduction}

Red supergiants (RSGs) are cool evolved massive stars with initial masses $8\lesssim M_\mathrm{init}\lesssim30 \ M_{\odot}$ \citep{Meynet_2003, Heger_2003}. They exhibit enhanced mass loss \citep{vLoon_2025} compared to the main sequence phase. Their winds affect the stellar evolution, their fate, the resulting type of supernovae \citep{Eldridge_2013, Yoon_2017, Beasor_2021, Zapartas_2025}, and the resulting mass of the compact object. Additionally, the winds contribute to the chemical enrichment of a galaxy, enriching it with gas and dust and contributing to star formation. 

There are several hypotheses and explanations about the driving mechanism of the RSG winds, but no concrete theory yet. Radial pulsations and the highly convective envelope of RSGs are expected to contribute to this mechanism \citep{Yoon_2010}. However, \citet{Arroyo-Torres_2015} showed that these factors alone cannot lift enough material for dust condensation. Recent studies suggest that atmospheric turbulence dominantly drives the RSG mass loss \citep{Josselin_2007, Ohnaka_2017, Kee_2021}. Previous empirical studies have shown that the mass-loss rates are between $10^{-9}\lesssim\dot{M}/(M_\odot \ \mathrm{yr}^{-1})\lesssim10^{-4}$ for a range of luminosities \citep[e.g.][]{deJager_1988,vLoon_2005, Goldman_2017, Beasor_2020, Humphreys_2020, Beasor_2022, Yang_2023, Antoniadis_2024}. These studies rely on the emission of the dust shell around the RSG in the infrared, which appears as an excess in the observed spectral energy distribution (SED). Tracing mass loss from atomic hydrogen, through the detection of the 21 cm H\textsc{i} line \citep{Gerard_2024}, would be a more direct method. Given the weakness of the line, such a measurement has not been achieved yet.

In \citet{Antoniadis_2024}, we showed that the discrepancies in the mass-loss rate by two to three orders of magnitude for the same luminosity between some of these works arise from the different assumptions on the RSG mass loss mechanism and dust shell properties. We also studied one of the largest samples of RSGs to provide a more accurate mass-loss rate prescription than those that are commonly used in stellar evolution models, which are dated and based on only a few dust-enshrouded RSGs \citep[e.g.][]{deJager_1988, vLoon_2005}. Gas-dependent methods, either using CO molecular emission \citep{Decin_2024} or theoretical models \citep{Kee_2021, Fuller_2024}, agree with our results in \citet{Antoniadis_2024}, providing stronger support for the assumptions we used. \citet{Zapartas_2025} showed that implementing prescriptions, which predict high mass-loss rates ($\sim10^{-4}\ {M_\odot \ \mathrm{yr}^{-1}}$), strips off the envelope of luminous RSGs, driving them to hotter temperatures before their collapse. This can explain the low number of luminous observed RSGs but is inconsistent with the lack of luminous, yellow supernova progenitors (see Fig.\ 8 in \citealt{Zapartas_2025}). 
Thus, quiescent steady-state RSG winds cannot strip the star of the hydrogen-rich envelope (also suggested in \citealt{Beasor_2020}). Instead, a superwind phase or outbursts occurring throughout the RSG lifetime \citep[e.g.][]{Decin_2006, Montarges_2021, Dupree_2022, Munoz-Sanchez_2024} or binary interaction \citep[e.g.][]{Ercolino_2024} are necessary for stripping the RSG envelope.

In contrast to the winds of hot stars that are line-driven and are well-established to depend on metallicity \citep[e.g.][]{Vink_2001, Vink_2005, Puls_2008, Vink_2022}, the effect of metallicity on the mass loss of RSGs is not yet clear. Previous studies found little to no effect of metallicity on the mass-loss rates, comparing results between sources from the Milky Way (MW) and the Magellanic Clouds \citep{vLoon_2005, Groenewegen_2009, Mauron_2011, Jones_2012, Goldman_2017}. However, these studies were based on dust-enshrouded stars, including AGB stars, and either had small samples or did not analyse the different samples uniformly and comprehensively. Metallicity can affect the formation of dust around the star, hence the radiation pressure on the dust grains, and the terminal outflow velocity \citep[e.g.][]{Goldman_2017}, but how dominant this factor is in the RSG wind is yet unknown. Finally, recent theoretical models of the RSG wind \citep{Kee_2021, Fuller_2024} did not find a strong direct dependence on metallicity. However, these studies did not investigate any indirect effects of metallicity, such as its impact on the dust shell and the RSG radius, which could contribute to the wind.

This paper aims to uniformly derive the mass-loss rates of RSGs in different metallicity environments, in the range $-0.7\lesssim[Z]\lesssim0$, to determine the metallicity dependence of the RSG winds. We present an analysis of large RSG samples from the Small Magellanic Cloud (SMC), NGC 6822, the Milky Way, M31, and M33. We then compare the results with our result of the Large Magellanic Cloud (LMC) from \citet{Antoniadis_2024}. In Sect.~\ref{sec:sample}, we describe the sample selection for each galaxy and in Sect.~\ref{sec:models}, we present the models and fitting methodology. We demonstrate the results in Sect.~\ref{sec:results} and discuss and compare them in Sect.~\ref{sec:discussion}. Finally, we present a summary in Sect.~\ref{sec:conclusion}.

\section{Sample} \label{sec:sample}

To investigate the effect of metallicity on RSG mass loss, we select Local Group galaxies with a large number of RSGs that span a range of metallicities with photometric surveys over a broad wavelength range. These include the SMC ($[Z]\simeq-0.75\pm0.3$; \citealt{Davies_2015, Choudhury_2018}), NGC 6822 ($[Z]=-0.5\pm0.2$; \citealt{Muschielok_1999, Venn_2001, Patrick_2015}), the Milky Way ($[Z]\simeq0\pm0.2$; \citealt{Anders_2017}), M31 ($[Z]\simeq0.13$; \citealt{Zurita_2012}) and M33 ($[Z]\simeq-0.18\pm0.2$; \citealt{U_2009}). The results from these galaxies were then compared to the results from the LMC ($[Z]=-0.37\pm0.14$; \citealt{Davies_2015}) from our previous study \citep{Antoniadis_2024}.

\subsection{SMC}
We used the RSG sample from \citet{Yang_2023}, which is a combination of the \citet{Yang_2020} and \cite{Ren_2021} RSG catalogues and consists of 2121 RSG candidates, selected through photometric criteria. We additionally removed the binary RSGs from \cite{Patrick_2022}. Binarity could affect the assumed spherical symmetry of the dust shell, and the hot companion could destroy part of the dust, leading to an inaccurate estimation of the mass-loss rate. However, the binary candidates did not significantly affect the results in the LMC \citep{Antoniadis_2024}. In this study, we removed them to have more secure mass-loss rate estimates. We used the photometry provided in the catalogue, replacing the \textit{Gaia} EDR3 astrometry and photometry with \textit{Gaia} DR3 \citep{gaia1, gaia2}. We also excluded photometry brighter than the saturation limit of the VISTA Magellanic Cloud Survey (VMC), $J\leq12$ mag and $K_s\leq11.5$ mag. Then, we required that each source has at least one observation at a wavelength $\lambda>8 \ \mu$m to better constrain the infrared emission of the dust shell with our models. Furthermore, we excluded 5 sources with extreme values of \textit{Gaia} DR3 proper motion or parallax, which are not consistent with the distribution of the sample, indicating that they are foreground stars. The final number of RSG candidates is 893. Finally, to correct for the foreground Galactic extinction, we assumed $R_V=3.1$ and $E(B-V)=0.04$~mag \citep{Massey_2007} and the extinction law and the $A_\lambda/A_V$ values from \cite{Wang_2019} for each band.

\subsection{Milky Way}
We used the RSG catalogue from \citet{Healy_2024}, a compilation of literature spectroscopic RSGs and a $Gaia$-based method (i.e. based on the colour and magnitude of the stars in the \textit{Gaia} bands). We added optical photometry from SkyMapper DR2 \citep{Onken_2019}, mid-IR photometry from AllWISE \citep{allwise} with signal-to-noise ratio $S/N>3$ in all bands, and from the \textit{Spitzer} Enhanced Imaging Products (SEIP)\footnote{\hyperlink{https://www.ipac.caltech.edu/doi/irsa/10.26131/IRSA3}{https://www.ipac.caltech.edu/doi/irsa/10.26131/IRSA3}} with \texttt{SExtractor} flags $\leq4$ in the IRAC bands. We required that there is available photometry in at least one band in the optical, the near-IR and the mid-IR ($\geq8 \ \mu$m) for each source. In addition, sources that have binary flags in \cite{Healy_2024} were removed. The final sample is reduced to 527 (from 651) RSGs after applying these constraints.

In the case of the Galactic RSGs, we used the provided values of $A_V$ from \cite{Healy_2024} to correct the foreground interstellar extinction and the values of $A_\lambda/A_V$ from \cite{Wang_2019} for the corresponding bands.

\subsection{NGC 6822}

For NGC 6822, we compiled the sample of RSG candidates from \cite{Yang_2021} and \cite{Ren_2021}, including 234 and 465 photometric RSG candidates, respectively. We cross-matched their coordinates within $1''$ to remove duplicates. \citet{Nally_2024} identified RSG candidates using James Webb Space Telescope (JWST) photometric observations, covering the central part of the galaxy. After applying a cut at 17.4 mag on the F200W filter to exclude low luminosity sources, we found 82 new RSG candidates that did not exist in the two previously mentioned catalogues. 

We used the photometry from different surveys in \cite{Yang_2021} and \cite{Ren_2021}, but when data were missing, we added or updated photometric data from the following surveys: \textit{Gaia} DR3; Pan-STARRS1 \citep{Chambers_2016}, AllWISE with ccf = 0, ex = 0 or 1 and $S/N>3$ in all bands; \textit{Spitzer} \citep{Khan_2015}; and JWST \citep{Nally_2024}. We did not use the \textit{WISE} $W3$, $W4$ and \textit{Spitzer} [24] bands because the angular resolution of these filters is $6.5''$, $12''$, and $6''$, respectively, where $6''$ corresponds to a radius of around 13.1 pc at the distance of NGC 6822. This led to unusually high fluxes in these bands, possibly due to blending or contamination from nearby sources. \citet{Lenckic_2024} found that a large number of \textit{Spitzer}-identified young stellar objects (YSOs) in this galaxy are not recovered in the JWST data due to blending, even though the case of RSGs may not be the same since they are brighter sources.

The Tip of the Red Giant Branch (TRGB) of NGC~6822 was estimated to be at 17.36 mag \citep{Hirschauer_2020} in the $K$ band. We applied a cut at $K=16.9$ mag to the sources from \cite{Ren_2021}, following \cite{Yang_2021} to be consistent between the two samples. We additionally used the boundary at \textit{Spitzer} $[3.6]=17.16$ mag from \cite{Hirschauer_2020}, which also defines the TRGB. Only one of the sources had [3.6] mag above that limit. Then, we applied the $J-K$ vs.\ $K$ cuts from \cite{Yang_2021} to the photometry of all instruments available and removed the sources that did not pass the photometric constraints of at least one instrument. For the sources that did not have JWST mid-infrared photometry, we required that they at least have \textit{Spitzer} [8.0]. This may not be ideal to constrain the SED during the model fitting, but as we show in the \autoref{app:ML_6822}, it can give sufficient results for the LMC (although with larger uncertainties).

We removed foreground sources using the \textit{Gaia} DR3 astrometry and the method described in \citet{Maravelias_2025} \citep[similar to][]{Maravelias_2022}. We additionally removed one flagged source as a quasar candidate from $Gaia$. Our final sample for NGC 6822 consists of 396 RSG candidates. Finally, we implemented the extinction law and values from \cite{Wang_2019} for each band with $R_V=3.1$ and $E(B-V)=0.22$~mag \citep{Massey_2007} to correct for foreground extinction. \autoref{tab:filters} presents all the photometric surveys and filters used, whenever available, in each galaxy to construct the SEDs.

\subsection{M31 and M33}
\citet{Wang_2021} derived the mass-loss rates of more than 1000 RSG candidates near solar metallicity in M31 and M33. We used their catalogues with near-IR photometry, adding observations from \textit{Gaia} DR3, AllWISE, and \textit{Spitzer} \citep{Khan_2015, Khan_2017}. We corrected for foreground extinction as in the other galaxies, using $E(B-V)=0.06$ and 0.05 mag for M31 and M33, respectively \citep{Massey_2007}. Initially, we attempted to model and recalculate the mass-loss rates with our assumptions to compare with the other galaxies. However, we found that the \textit{Spitzer} and \textit{WISE} photometry is not reliable at these distances ($0.7\lesssim d\lesssim0.9$ Mpc), providing abnormally high fluxes in the mid-IR, especially in the [24] band. We show some example SEDs in \autoref{app:m31}. {\citet{Javadi_2015} also found that the photometry of the \textit{Spitzer} MIPS instrument was unreliable in M33, possibly affected by nearby sources.} \citet{Javadi_2013} derived mass-loss rates for RSGs in M33 using \textit{Spitzer} photometry up to [8.0], which has better resolution ($\sim2''$) than the MIPS [24]. We tried this approach, but still found similar mass-loss rates to those we present in \autoref{app:m31}. Since we are not able to evaluate the quality of the photometry, e.g.\ by comparing with data from a higher resolution instrument, and without being able to constrain the SED at longer wavelengths, we do not present these results here.

\begin{table}[h]
    \tiny
    \centering
    \caption{Photometric surveys and filters used for the spectral energy distributions in each galaxy.}
    \renewcommand{\arraystretch}{1.3}
    \begin{tabular}{l l c c c}
        \hline\hline
       Survey  & Filters & SMC & NGC 6822 & MW \\
         \hline
       \textit{GALEX}  & FUV, NUV                           & \checkmark & - & - \\
       SkyMapper & $u, v, g, r, i, z$              & \checkmark & - & \checkmark \\
       \citet{Massey_2002}         & $U, B, V, R$  & \checkmark & - & - \\
       LGGS                 & $U, B, V, R, I$      & - & \checkmark & - \\
       NSC DR2  & $u, g, r, i, z, Y$               & \checkmark & - & - \\
       \textit{Gaia} DR3  & $G_{BP}, \ G, \ G_{RP}$& \checkmark & \checkmark & \checkmark \\
       Pan-STARRS & $g, r, i, z, y$                & - & \checkmark & - \\
       OGLE-III & $V, I$                           & \checkmark & - & - \\
       DENIS & $I, J, K_s$                & \checkmark & - & - \\
       VMC or VHS & $Y, J, K_s$           & \checkmark & \checkmark & - \\ 
       2MASS  & $J, H, K_s$               & \checkmark & - & \checkmark \\
       IRSF  & $J, H, K_s$                & \checkmark & \checkmark & - \\
       UKIRT  & $J, H, K$                          & - & \checkmark & - \\
       HAWK-I  & $J, K$                            & - & \checkmark & - \\
       AKARI  & N3, N4, S7, S11, L15, L24          & \checkmark & - & - \\
       AllWISE & [3.4], [4.6], [12], [22]          & \checkmark & \checkmark & \checkmark \\
       \multirow{2}{*}{\textit{Spitzer}}   & [3.6], [4.5], [5.8], & \multirow{2}{*}{\checkmark} & \multirow{2}{*}{\checkmark} & \multirow{2}{*}{\checkmark} \\
                   &  [8.0], [24] &   &    & \\
       \multirow{3}{*}{JWST}   & F115W, F200W, F356W, & \multirow{3}{*}{-} & \multirow{3}{*}{\checkmark} & \multirow{3}{*}{-} \\
              &  F444W, F770W, F1000W, &  &  & \\
               &  F1500W, F2100W &  &  & \\
       \hline \\     
    \end{tabular}
    \
    \label{tab:filters}
\end{table}

\subsection{Luminosity and effective temperature}

We calculated the luminosity, $L$, of each source by integrating its observed spectral energy distribution using a distance of $d=62.44 \pm 1.28$ kpc for the SMC \citep{Graczyk_2020}, $d\simeq 0.45\pm0.01$ Mpc for NGC 6822 \citep{Gorski_2011,Zgirski_2021}, and the distances provided in \cite{Healy_2024} using the estimations from \cite{Bailer-Jones_2021} for the Galactic RSGs, based on the \textit{Gaia} DR3 astrometry. Considering the photometric and distance errors, the $\log{(L/L_\odot)}$ uncertainty for the RSGs in the SMC and NGC 6822 is $\sim0.03$ dex.

It should be mentioned that we found a luminosity of $\log{(L/L_\odot)}\sim 6$ for RW Cep using a distance estimate of $6.7^{+1.6}_{-1.0}$ kpc from \cite{Bailer-Jones_2021}. However, RW Cep is considered to be a member of Cep OB1 association \citep{Melnik_2020} at a distance of 3.4 kpc \citep{Rate_2020} or the Berkeley 94 star cluster at a distance of 3.9 kpc \citep{Delgado_2013}. Thus, we adopted a distance value of 3.9 kpc for this RSG with 3.4 and 6.7 the lower and upper limits, respectively. This demonstrates the uncertainties in the distance estimations from \citet{Bailer-Jones_2021}.

\begin{figure}[h]
    \centering
    \includegraphics[width=\columnwidth]{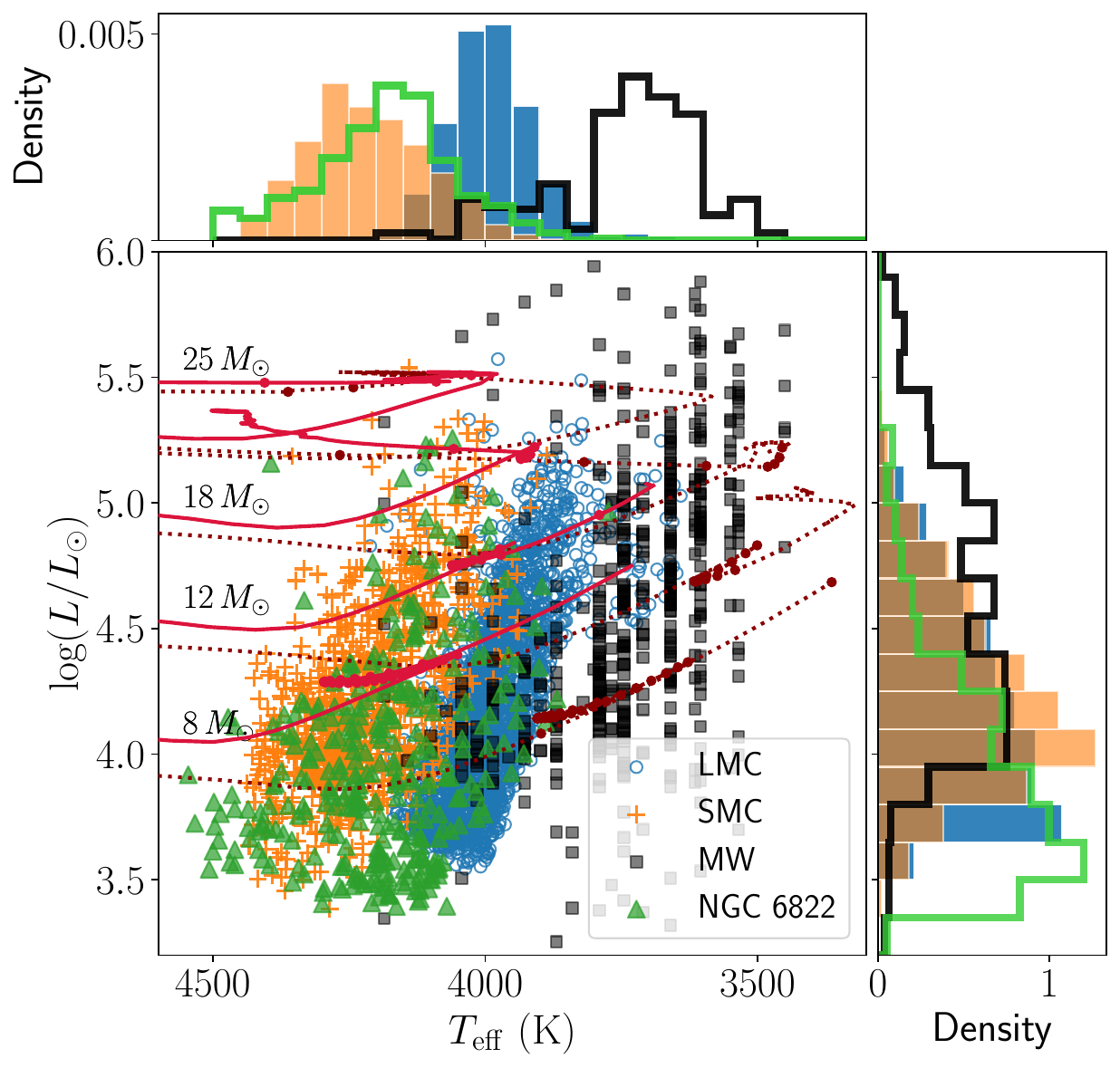}
    \caption{HR diagram of the RSG samples in the LMC (blue circles), SMC (orange crosses), Milky Way (black squares), and the NGC 6822 (green triangles). Four POSYDON evolutionary tracks for $[Z]=-0.7$ and $[Z]=0$ are shown in red and dotted dark-red lines, respectively, and the points indicate intervals of 10,000 yr. The upper and right panels show the density histograms (area equals to 1) of the effective temperature and luminosity, respectively.}
    \label{fig:HRD}
\end{figure}

We used empirical relations to calculate the effective temperatures, $T_\mathrm{eff}$, for the SMC \citep{Britavskiy_2019smc}:
\begin{equation}
    T_\mathrm{eff} = 5449 - 1432 \times (J-K_s)_0, \label{eq:Teff_smc}
\end{equation}
with an error of 140 K, and for NGC 6822 the $Z$-dependent relation calibrated from synthetic models from \citet{deWit_2024} (see their Table 5). In the case of Galactic RSGs, we use the $T_\mathrm{eff}$ provided by \citet{Healy_2024} which is an estimation from the spectral type, with an average error of 160 K. We present the luminosity and effective temperature distributions in a Hertzsprung-Russell (HR) diagram for all galaxies including the LMC from \citet{Antoniadis_2024} along with density histograms in Fig.~\ref{fig:HRD}. The lines represent indicative evolutionary tracks for four different initial masses at metallicity $Z=0.2$ Z$_\odot$ or $[Z]=-0.7$ using POSYDON \citep{Fragos_2023,Andrews_2024}, which is a grid of MESA models \citep{Paxton2011,Paxton2013,Paxton2015,Paxton2018,Paxton2019}.

At lower metallicities we find higher $T_\mathrm{eff}$, which is expected from theory \citep[e.g.][]{Maeder_2001} and has been verified from observations \citep[e.g.][]{Tabernero_2018, Gonzalez-Tora_2021}. However, the $T_\mathrm{eff}$ of the Galactic RSGs can be quite uncertain since it is an estimation from the spectral type. In addition, if the spectral classifications originate from optical wavelengths, which have the TiO bands as a diagnostic, $T_\mathrm{eff}$ could appear systematically lower \citep[see][]{Davies_2013}. The differences in luminosity between the galaxies, especially at the lower limit, originate from the different cuts and selection criteria in each galaxy. We expect most of the sources with $\log(L/L_\odot)<4$ to be lower-mass stars; therefore, we do not consider them further.

\section{Dust shell models} \label{sec:models}

We used the 1D radiative transfer code \texttt{DUSTY} (V4)\footnote{{\hyperlink{https://github.com/ivezic/dusty}{https://github.com/ivezic/dusty}}} \citep{Ivezic_1997} to model the SED of each RSG. \texttt{DUSTY} solves the radiative transfer equations for a central source in a spherically symmetric dust shell and we assumed that the shell extends to $10^4$ times the inner radius, $R_\mathrm{in}$. We describe the parameters and assumptions used in more detail in \citet{Antoniadis_2024}. We assumed a constant terminal outflow velocity, $v_\infty$, scaled with the luminosity as defined in \citet{Antoniadis_2024} and density distribution of the dust shell proportional to the radius as $\rho\propto r^{-2}$.

The mass-loss rate is calculated as 
\begin{equation}
    \dot{M} = \frac{16\pi}{3}\frac{a}{Q_V} R_{\mathrm{in}}\tau_V\rho_d v_\infty r_{\mathrm{gd}} , \label{eq:dotM}
\end{equation}
where $a/Q_V$ is the ratio of the dust grain radius over the extinction efficiency in the $V$ band, $\rho_d$ is the bulk density, $R_\mathrm{in}$ is the inner shell radius, and $r_\mathrm{gd}$ is the gas-to-dust ratio \citep{Beasor_2016}. We assumed an average $r_\mathrm{gd}$ for each galaxy; $r_\mathrm{gd}=1500^{+1000}_{-1000}$ \citep{Roman-Duval_2014, Clark_2023} for the SMC and the typical value of $r_\mathrm{gd}=200^{+200}_{-30}$ for the Milky Way, while in \citet{Antoniadis_2024} we used $r_\mathrm{gd}\simeq300$ \citep{Clark_2023}. Since there is no measurement of $r_\mathrm{gd}$ in NGC 6822 and it scales with metallicity \citep[e.g.][]{vLoon_2000, Li_2019, Clark_2023}, we assumed $r_\mathrm{gd}=800^{+500}_{-500}$ for this galaxy. We also used some nominal values for the uncertainties on $r_\mathrm{gd}$ because they vary between studies.

We assumed a modified Mathis-Rumpl-Nordsieck (MRN) grain size distribution $n(a)\propto a^{-q}$ for $a_{min}\leq a \leq a_{max}$ \citep{MRN} with $a_{min}=0.1 \ \mu\mathrm{m}$, $a_{max}=1 \ \mu\mathrm{m}$ and $q=3$. Finally, we used the \textsc{MARCS} model atmospheres \citep{Gustafsson_2008} as input SEDs for the central source with typical parameters for RSGs and metallicity $[Z]=-0.75$ for the SMC, $[Z]=-0.5$ for NGC 6822, and $[Z]=0$ for the Milky Way.

We fitted the models with the observations calculating the minimum modified $\chi^2$, $\chi^{2}_{mod}$ \citep{Yang_2023},
\begin{equation}
    \chi^2_{mod}=\frac{1}{N-p-1} \sum{\frac{[1-F(\mathrm{Model}, \lambda)/F(\mathrm{Obs}, \lambda)]^2}{F(\mathrm{Model}, \lambda)/F(\mathrm{Obs}, \lambda)}},
\end{equation}
where $F(\lambda)$ is the flux at a specific wavelength, $N$ is the number of photometric data points for each source, and $p$ is the number of free parameters. The error models were defined as those within the minimum $\chi_{mod}^2+0.002$, a crude estimation of the errors from the fitting procedure, following the reasoning from \citet{Beasor_2020} as we described in \citet{Antoniadis_2024}. 

\begin{figure*}[h]
     \centering
    \begin{subfigure}[t]{0.47\textwidth}
        \raisebox{-\height}{\includegraphics[width=\textwidth]{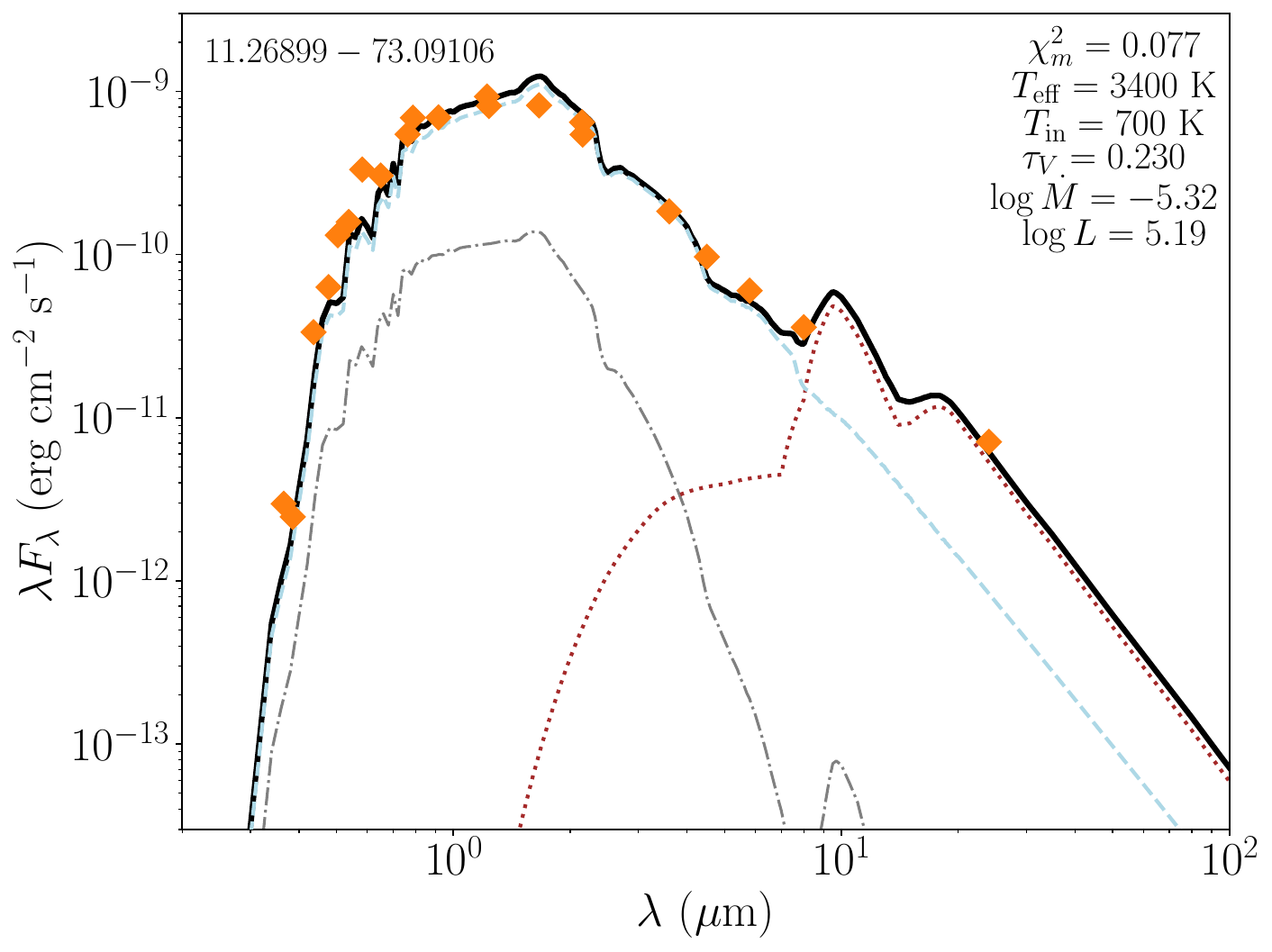}}
    \end{subfigure}
    \hfill
    \begin{subfigure}[t]{0.47\textwidth}
        \raisebox{-\height}{\includegraphics[width=\textwidth]{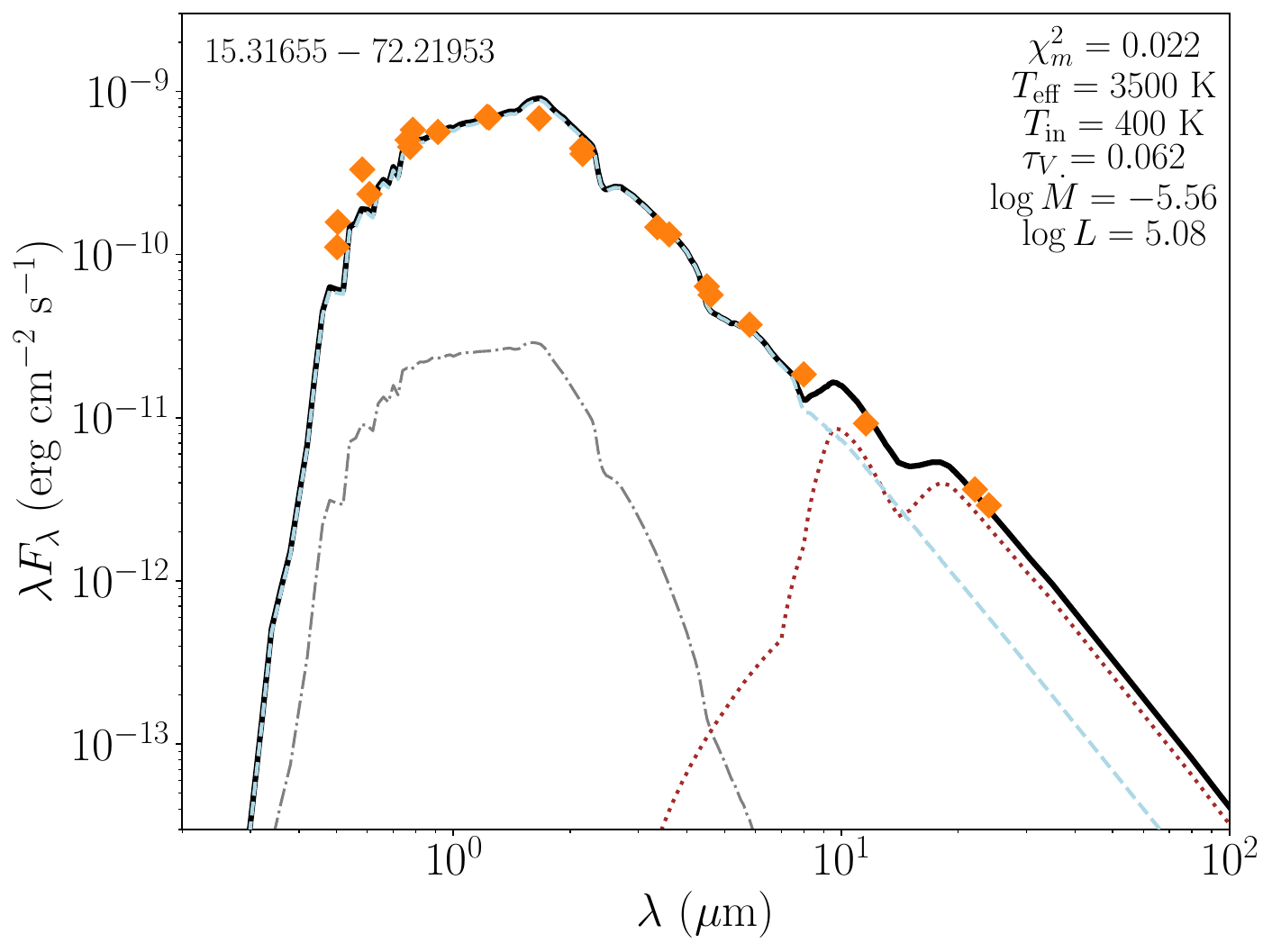}}
    \end{subfigure}
    \begin{subfigure}[t]{0.47\textwidth}
        \raisebox{-\height}{\includegraphics[width=\textwidth]{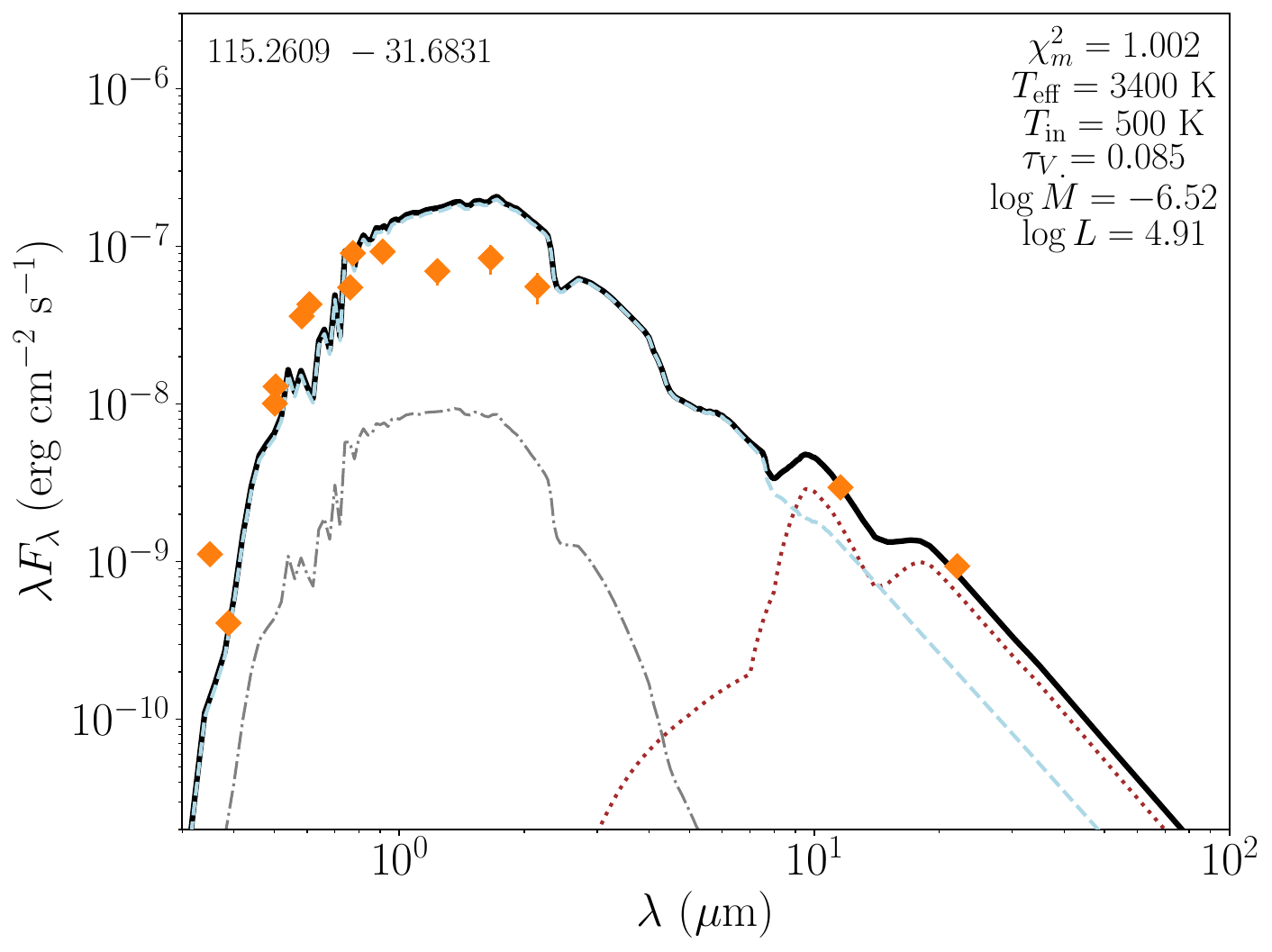}}
    \end{subfigure}
    \hfill
    \begin{subfigure}[t]{0.47\textwidth}
        \raisebox{-\height}{\includegraphics[width=\textwidth]{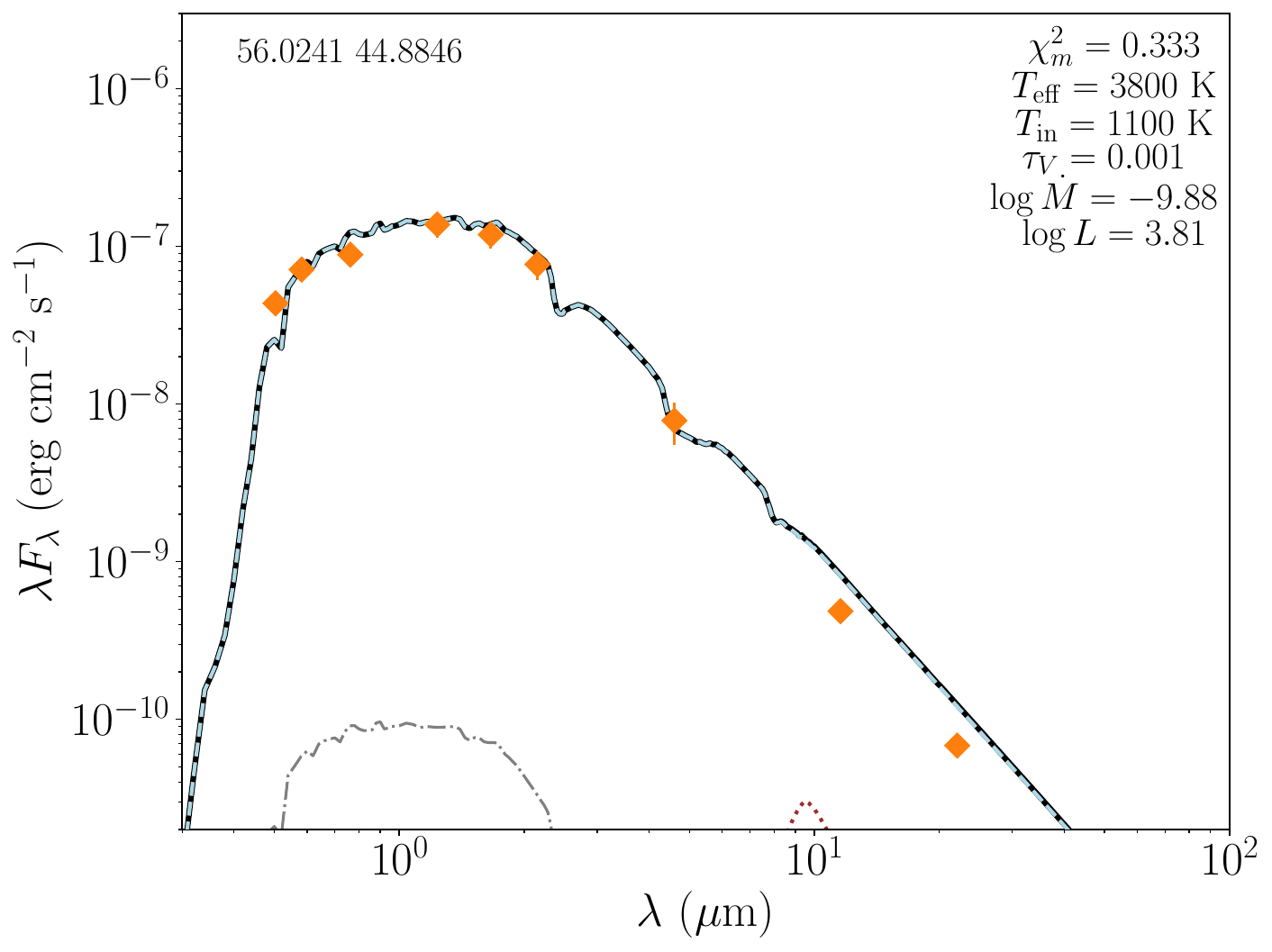}}
    \end{subfigure}
    \begin{subfigure}[t]{0.47\textwidth}
        \raisebox{-\height}{\includegraphics[width=\textwidth]{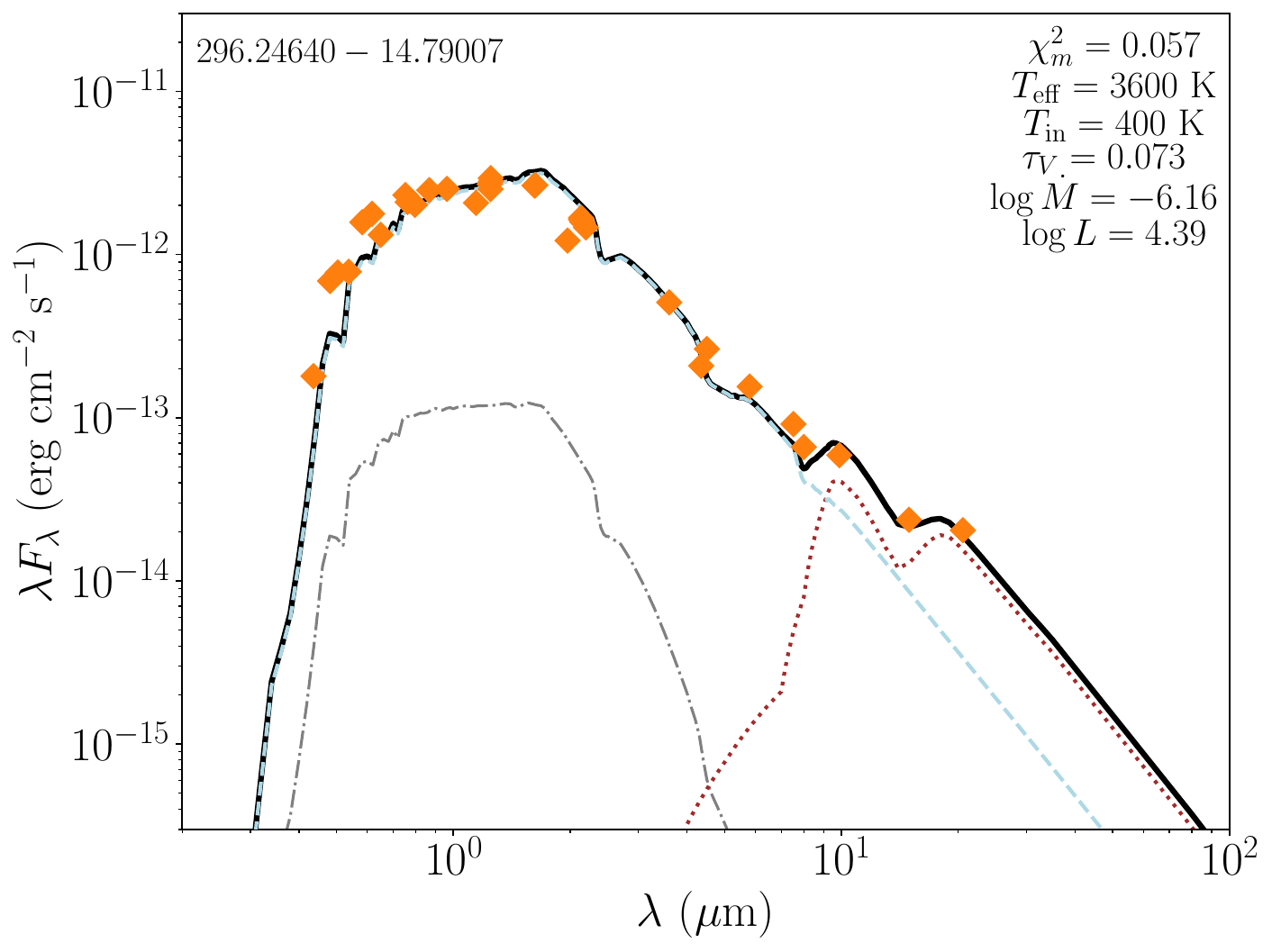}}
    \end{subfigure}
    \hfill
    \begin{subfigure}[t]{0.47\textwidth}
        \raisebox{-\height}{\includegraphics[width=\textwidth]{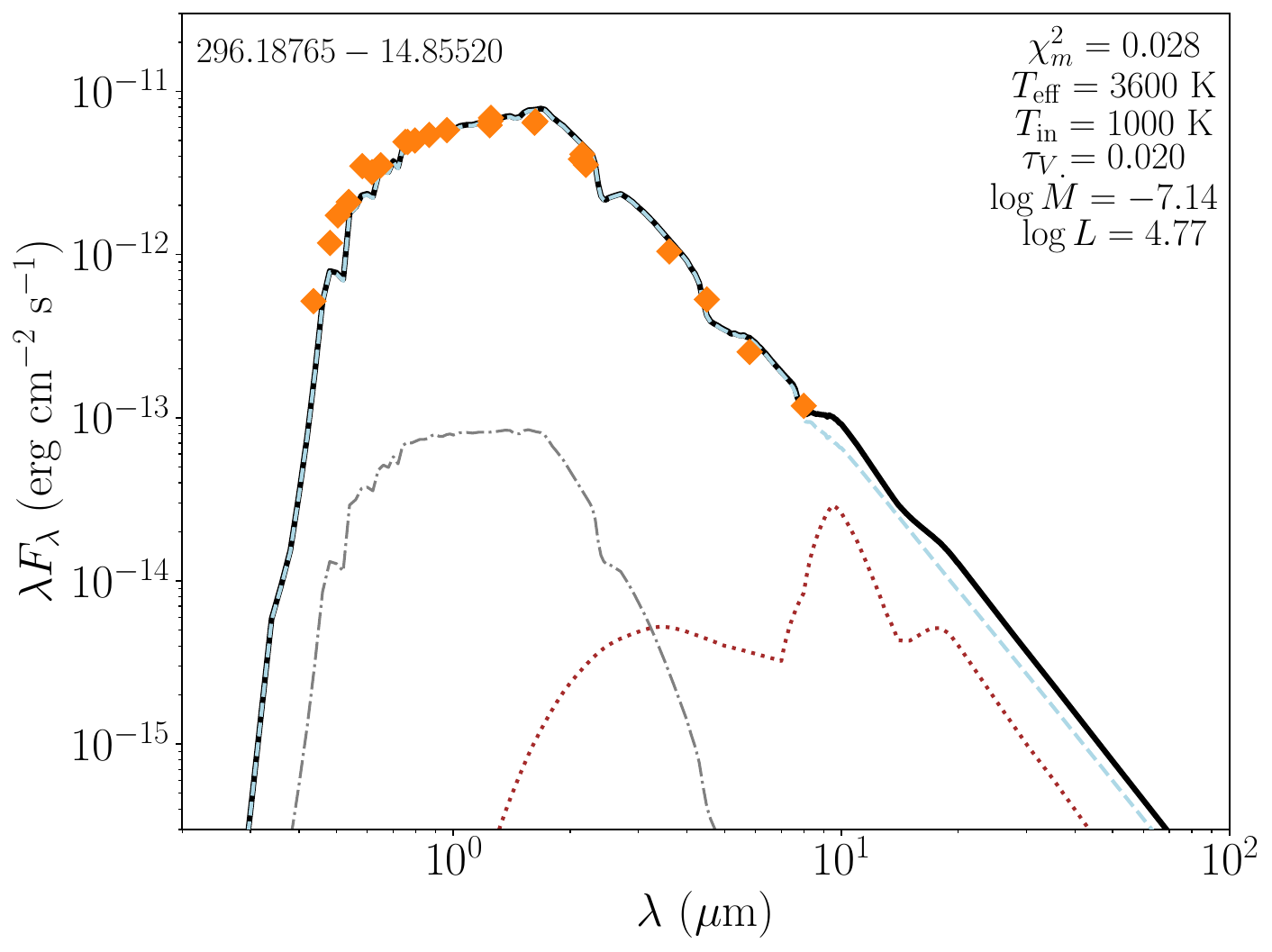}}
    \end{subfigure}

    \caption{Examples of SED fits of SMC (top row), Galactic (middle), and NGC 6822 (bottom row) RSGs. The orange diamonds show the observations and the black line is the best-fit model consisting of the attenuated flux (dashed light blue), the scattered flux (dot-dashed grey) and dust emission (dotted brown). The top left corner includes the coordinates of each source in degrees. \textit{Notes:} All SED fit figures are available in an electronic form via \url{https://zenodo.org/records/16736627}.} \label{fig:sed}
\end{figure*}

\autoref{fig:sed} shows example SEDs of RSGs from each of the three galaxies of our sample. The excess at around 10 $\mu$m indicates the presence of silicate dust. The observed near-IR photometry was below the model in many model fittings of the Galactic RSGs, attributing these to difficulties in estimating the interstellar extinction within the MW or the distance of the RSG.

\section{Results} \label{sec:results}

\begin{figure}[h]
    \centering
    \includegraphics[width=0.9\columnwidth]{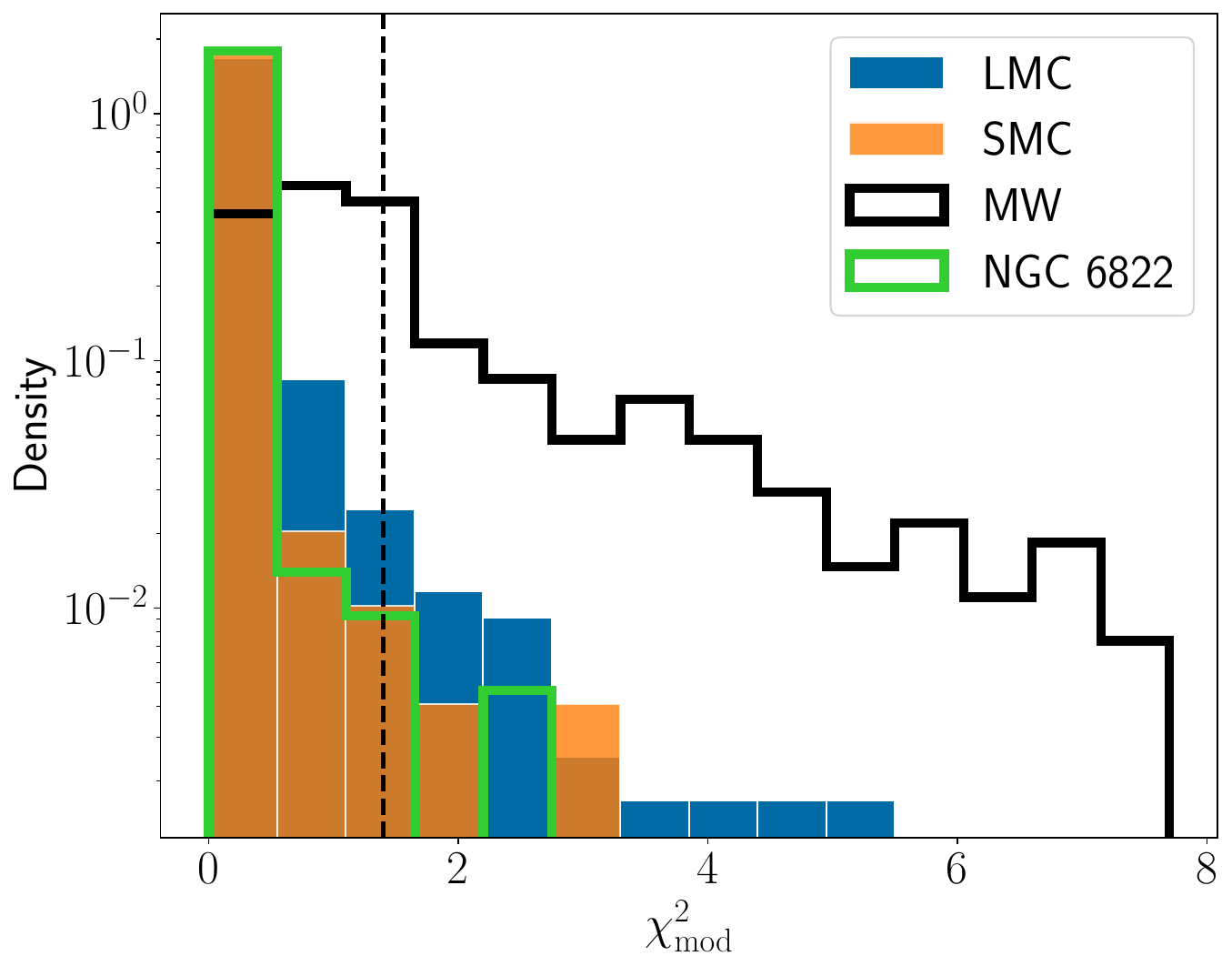}
    \caption{Distribution of the minimum $\chi^2_\mathrm{mod}$ for the LMC (blue bar), SMC (orange bar), MW (black line), and NGC 6822 (green line). The dashed vertical line indicates the considered limit for a good fit at $\chi_\mathrm{mod}^2=1.4$.
 }
    \label{fig:chi2}
\end{figure}

We obtained the dust shell properties from the best-fit \texttt{DUSTY} models to the observed SEDs. One of the most significant properties is the optical depth, $\tau_V$, which indicates how dusty a RSG is and is needed to calculate the mass-loss rate (see Eq.~\ref{eq:dotM}). \autoref{fig:chi2} shows the histogram of the minimum $\chi_\mathrm{mod}^2$ for all the galaxies of our study (and $\log(L/L_\odot)>4$). Using the $\chi_\mathrm{mod}^2$ distribution of the LMC \citep{Antoniadis_2024}, which is the largest sample, we consider a good fit one with $\chi_\mathrm{mod}^2<1.4$, the limit at roughly $3\sigma_{\chi^2}$, where the standard deviation in the LMC is $\sigma_{\chi^2}=0.41$ \citep{Antoniadis_2024}. We limit our results to the best-fit models with this constraint. The Galactic RSGs had a significant fraction of bad fits (39\%) mainly due to uncertainties in the distance and interstellar extinction estimations. We present the dereddened photometry, the $Gaia$ astrometry, the derived stellar and dust shell properties, and the mass-loss rates of RSGs in the SMC, MW, and NGC 6822 in the corresponding tables in \autoref{app:catalogues}.

\begin{figure}[h]
    \includegraphics[width=\columnwidth]{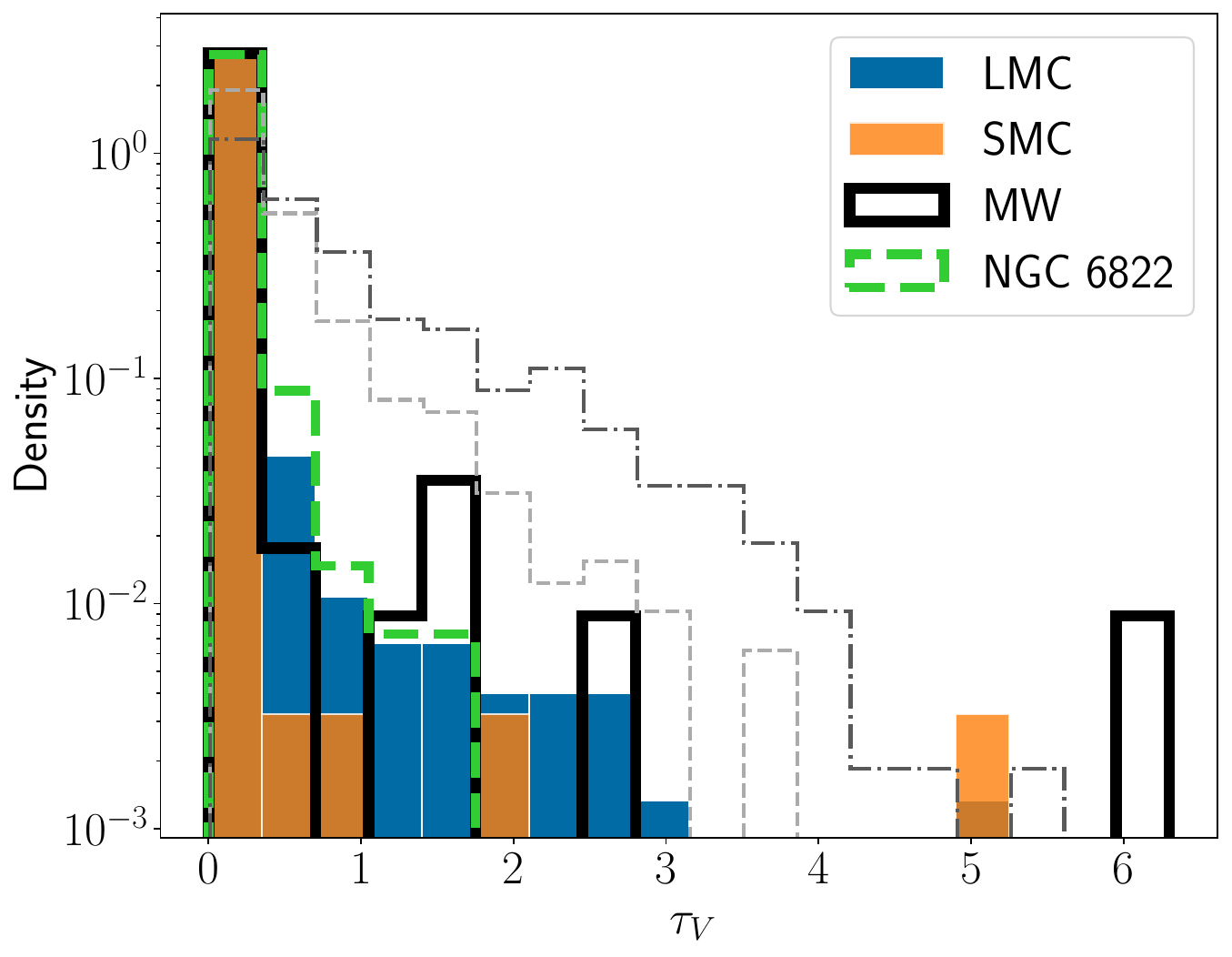}
    \caption{Distribution of the minimum $\tau_V$ for the RSGs in the LMC (blue bar), SMC (orange bar), MW (black line), and NGC 6822 (green dashed line). The dashed and dot-dashed histograms correspond to the RSGs in M31 and M33, respectively.
 }
    \label{fig:tau}
\end{figure}

\autoref{fig:tau} shows the distribution of $\tau_V$ for each galaxy. The LMC has more RSGs with higher values of $\tau_V$ than NGC 6822 and the SMC. This implies thicker dust shells and more dust production, which is expected in higher metallicity environments. We interpret the unusually high $\tau_V$ values for M31 and M33 to result from the contamination from nearby sources due to the insufficient angular resolution of \textit{Spitzer} at that distance ($6''$ corresponds to a radius of $\sim20$ pc). Thus, the derived mass-loss rates in \citet{Wang_2021} were significantly affected by this factor and are not accurate. We demonstrate our results for these two galaxies in \autoref{app:m31} and do not consider them further.

\subsection{SMC}

\begin{figure*}[h!]
    \centering
    \includegraphics[width=0.47\textwidth]{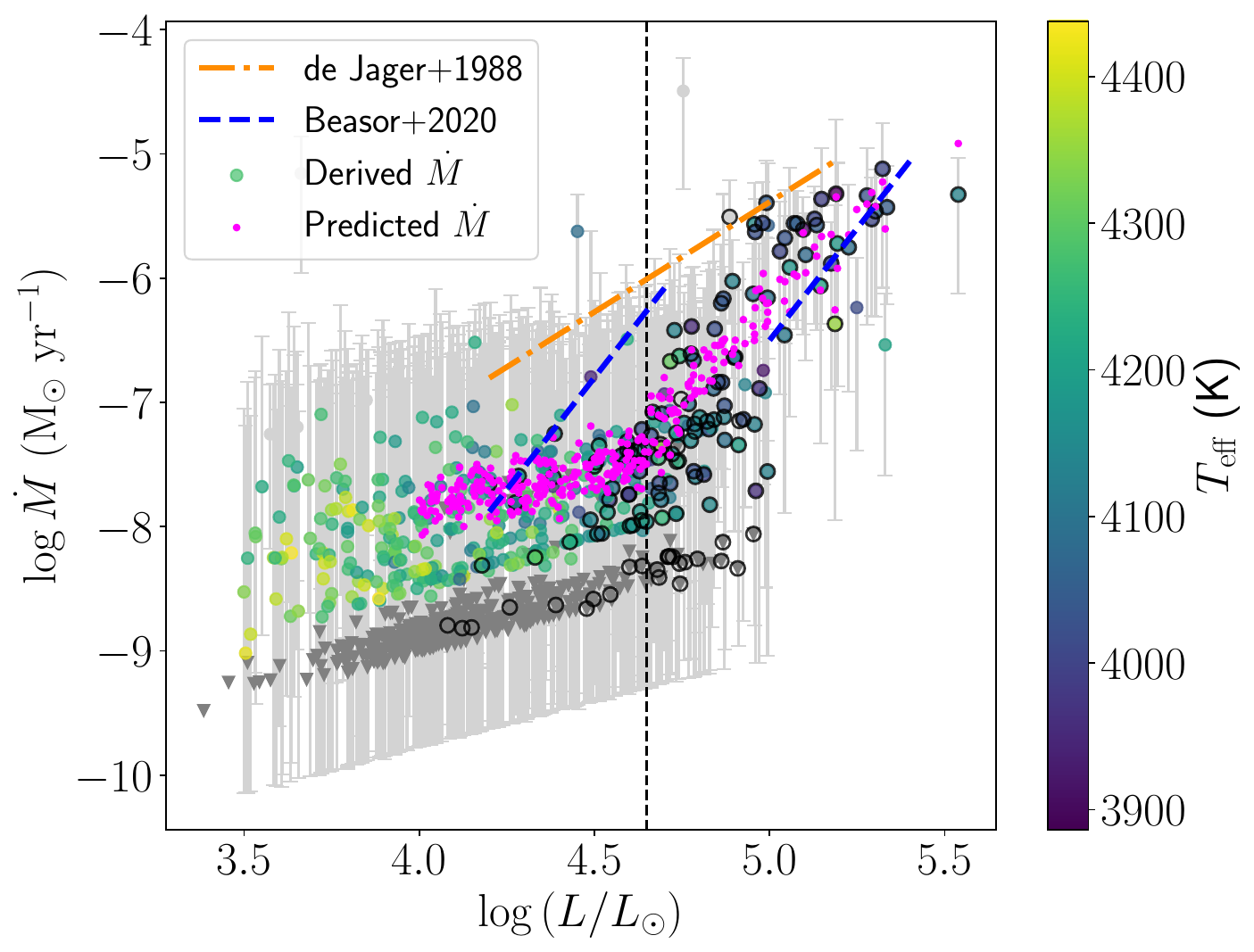}
    \includegraphics[width=0.47\textwidth]{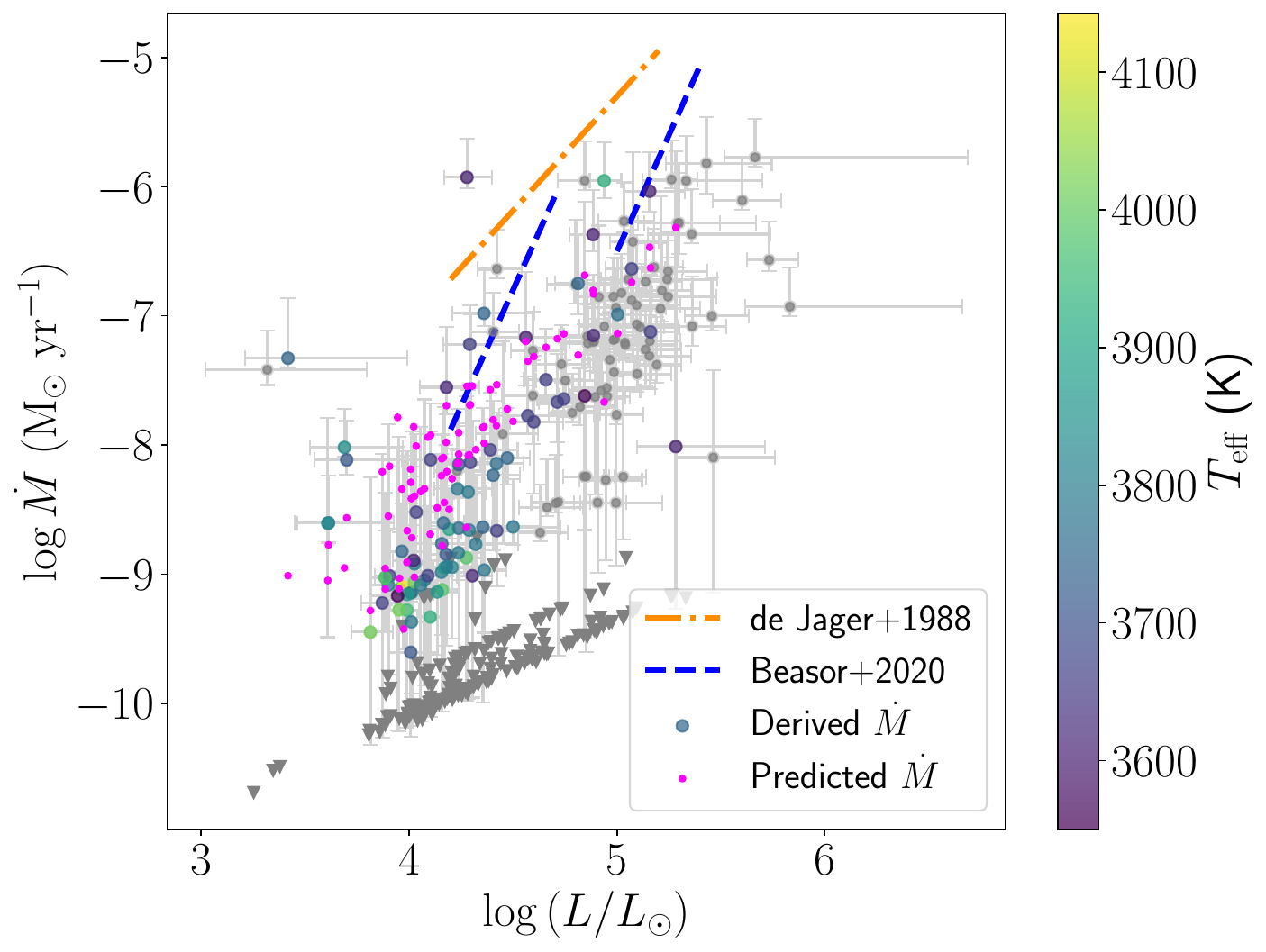}
    \includegraphics[width=0.47\textwidth]{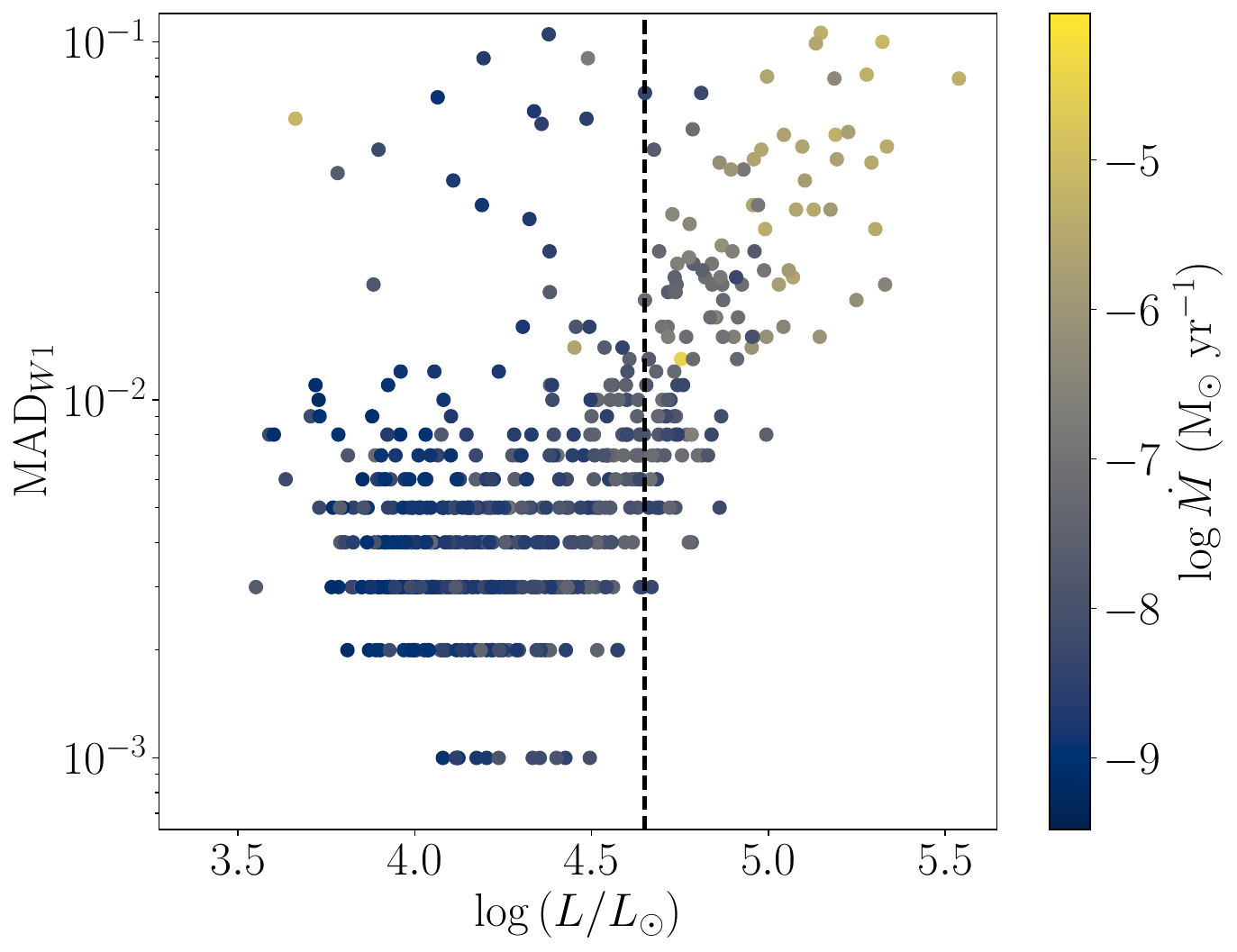} 
    \includegraphics[width=0.47\textwidth]{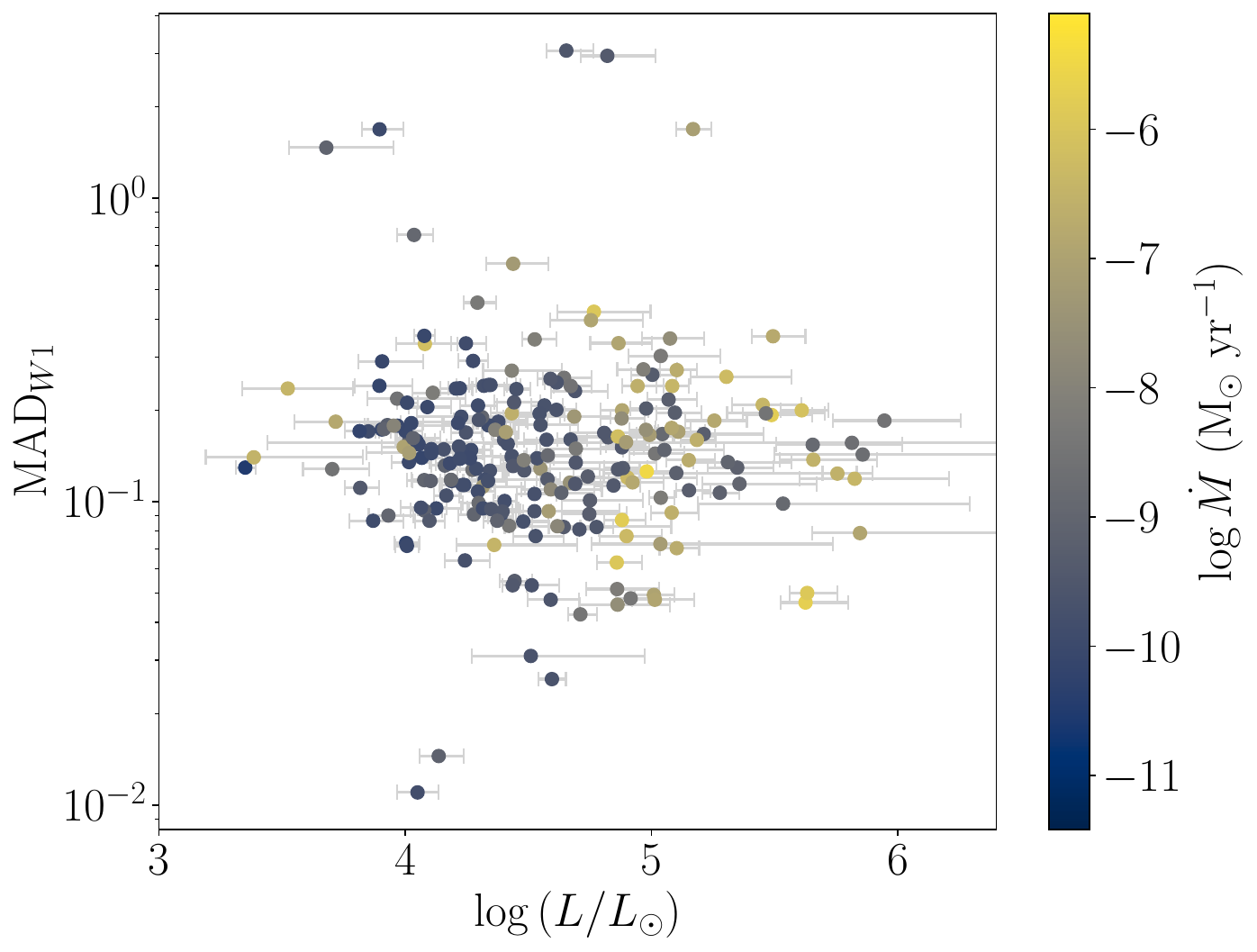} 
    \includegraphics[width=0.47\textwidth]{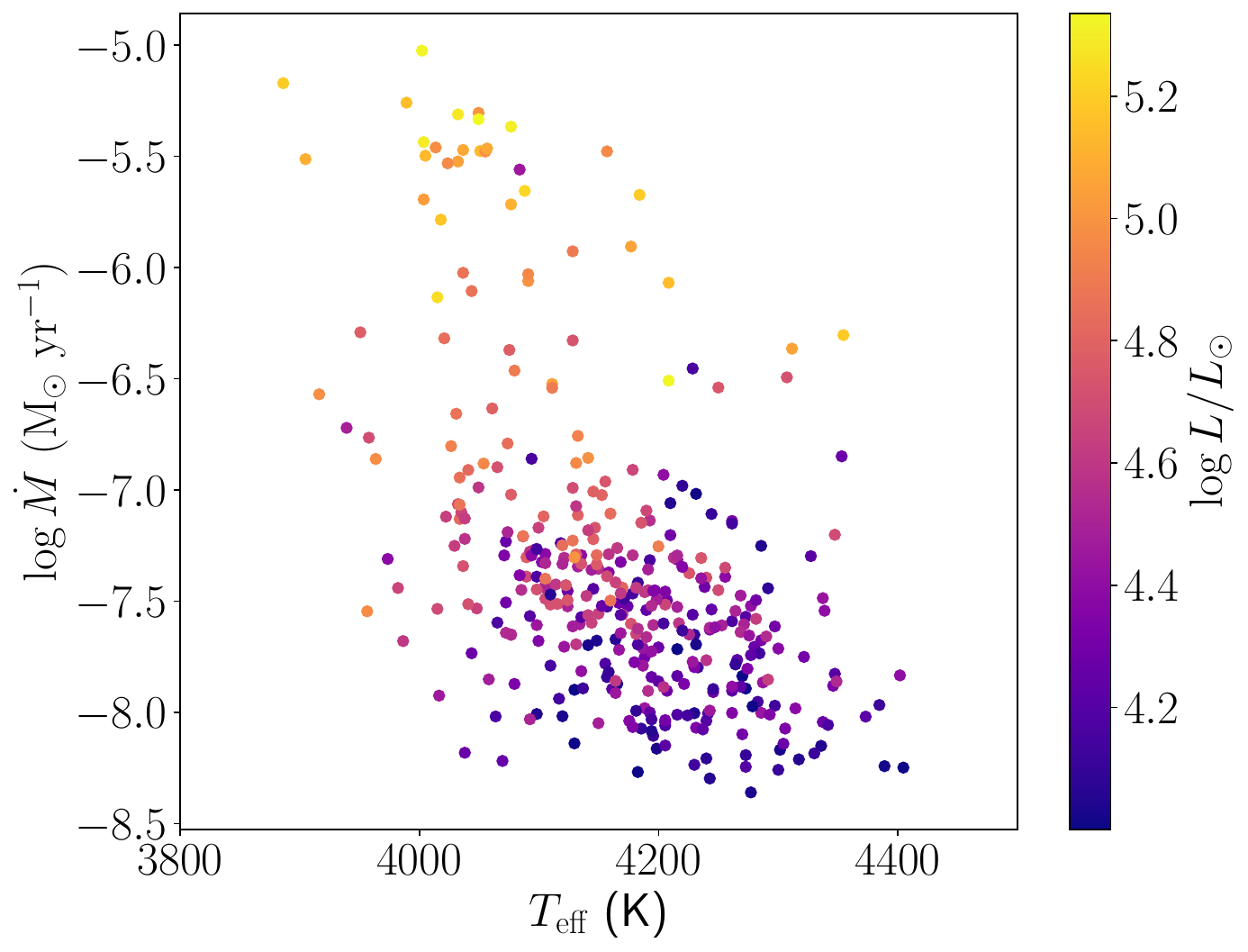}    
    \includegraphics[width=0.47\textwidth]{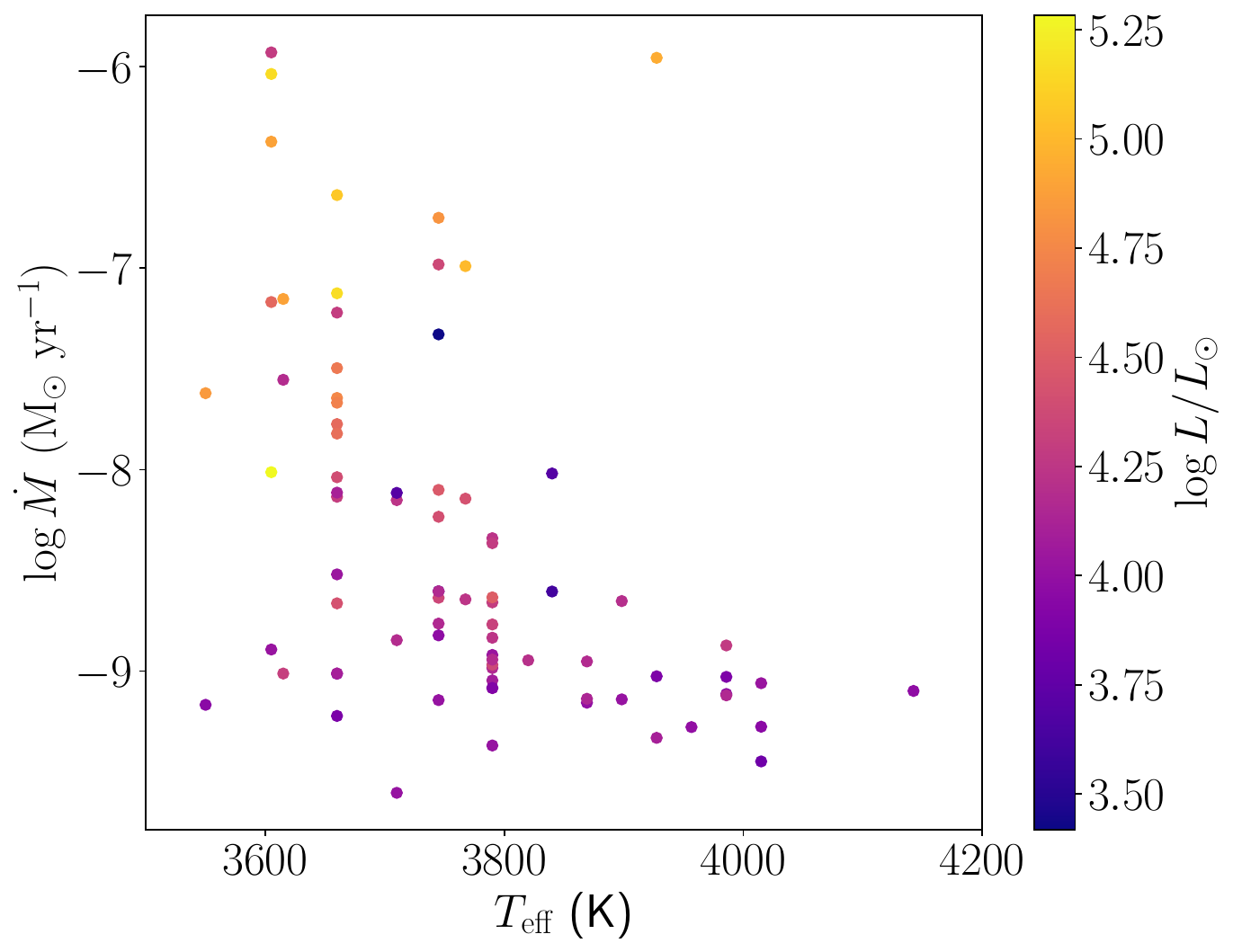}  
    \caption{\textit{Top:} Mass-loss rates vs. luminosity for SMC (left) and MW (right). The colour bar shows the $T_\mathrm{eff}$. The grey triangles represent the upper limits, and the open circles indicate the RSGs with spectral classifications. The magenta points correspond to the prediction of the derived $\dot{M}(L, T_\mathrm{eff})$ relation. The dashed blue lines show the prescription from \citet{Beasor_2020} for initial masses 10 and 20 $M_\odot$ and the orange dot-dashed line is the one from \citet{deJager_1988}. \textit{Middle:} MAD of $W1$ band vs. luminosity diagram for the SMC (left) and the MW (right). The colour bar shows the mass-loss rate. \textit{Bottom:} Mass-loss rates vs. the effective temperature for the SMC (left) and the MW (right) indicating the luminosity with colour.
 }
    \label{fig:Mdot_L_smc_mw}
\end{figure*}

We recalculated the $\dot{M}$ of the RSGs in the SMC from \citet{Yang_2023} using similar assumptions among samples to compare with the rest of the galaxies. In Fig. \ref{fig:Mdot_L_smc_mw}, we demonstrate the $\dot{M}$ vs.\ $L$ and the median absolute deviation (MAD) of $W1$ band from NEOWISE epoch photometry versus $L$ for SMC. We observed a turning point or kink at around $\log(L/L_\odot)=4.65$ and a correlation of $\dot{M}$ vs $L$ as in \citet{Yang_2023} (see their Fig. 15).

We derived a broken $\dot{M}$ relation as a function of $L$ and $T_\mathrm{eff}$, similar to \citet{Antoniadis_2024},
\begin{equation}
    \log{\dot{M}} = c_1 \log{L} + c_2\log{\left(\frac{T_\mathrm{eff}}{4000}\right)} + c_3 ,\label{eq:Mdot}
\end{equation}
where $\dot{M}$ is the mass-loss rate in $M_{\odot} \ \mathrm{yr}^{-1}$, $L$ is the luminosity in $L_\odot$, and $T_\mathrm{eff}$ is the effective temperature in $\mathrm{K}$. Table \ref{tab:coef} includes the best-fit parameters. Figure \ref{fig:Mdot_L_smc_mw} shows the prediction of the relation and the observed anti-correlation of $\dot{M}$ with $T_\mathrm{eff}$. We derived the relation without considering the non-dusty RSGs (upper limits with $\tau_V<0.002$, grey triangles). We also applied a cut at $\log{(L/L_\odot)}=4$, approximately above the least luminous RSG with a spectroscopic classification and around the corresponding $L$ for an 8 $M_\odot$ star, where RSGs dominate over the lower mass stars (see also \citealt{Massey_2021} and our discussion in Sect.~\ref{sec:discuss_kink}). This limit can be slightly higher (around 4.2), but regardless, the slope does not change up to the position of the kink. 

The steep dependence on the $T_\mathrm{eff}$ is mainly because the $T_\mathrm{eff}$ was calculated using the relation from \citet{Britavskiy_2019smc}, based on \textit{i}-band measurements from \citet{Tabernero_2018}, which shows a steep correlation with the $(J-K_s)_0$ colour. In addition, since the dispersion is large, the error on $c_2$ is high. Taking into account the error on the measured $T_\mathrm{eff}$ for each RSG, the exact value of $c_2$ can have high uncertainties, but one can also use an average value for $T_\mathrm{eff}$ for an $L$-only dependent $\dot{M}$ relation. However, our $T_\mathrm{eff}$ dependence reproduces well the results and partly accounts for the dispersion. Mass-loss rate prescriptions are usually extrapolated to the yellow phase up to 10,000 K, but we do not suggest using our $T_\mathrm{eff}$ dependence above around 4,500 K as it would lead to negligible mass-loss rates. Using a nominal $T_\mathrm{eff}$ to create an $L$-only dependent relation is more appropriate for that regime.

Furthermore, we compare our results with the prescriptions from \citet{Beasor_2020} (corrected by \citealt{Beasor_2023}) and \citet{deJager_1988}. Finally, the fraction of RSGs without dust is 40\% (this is an upper limit because they can reach $\tau_V>0.002$ considering the errors) and the fraction of optically thick or dusty RSGs ($\tau_V>0.1$) is 2\%.

\begin{table}[h]
    \small
    \centering
    \caption{Best-fit parameters of Eq. (\ref{eq:Mdot}) for the SMC.}
    \renewcommand{\arraystretch}{1.2}
    \begin{tabular}{c | c c c}
        \hline\hline
        $\log{L/L_\odot}$ & $c_1$   & $c_2$  & $c_3$ \\
         \hline 
        $<4.65$       & $0.18\pm0.37$ & $-17.96\pm8.21$  & $-7.88\pm1.69$ \\
        $\gtrsim4.65$ & $2.36\pm0.31$  & $-15.32\pm6.81$  & $-17.78\pm1.56$ \\
       \hline
    \end{tabular}
    \
    \label{tab:coef}
\end{table}

\subsection{Milky Way}

We show the corresponding $\dot{M}$ vs.\ $L$ diagram for  Galactic RSGs from \citet{Healy_2024} in the right panel of Fig. \ref{fig:Mdot_L_smc_mw}. The $\dot{M}$ is systematically lower than the other galaxies, mainly at $\log(L/L_\odot)<4.5$. We denote with grey points those with $JHK_s$ photometry significantly lower than the best-fit model SED, as in the middle left in Fig.~\ref{fig:sed}, which could result in an underestimated $\dot{M}$, and we did not consider those in our analysis. The three outliers with high $\dot{M}$ and $\log(L/L_\odot)<4.5$ and the three dust-free sources near $\log(L/L_\odot)=3$ lie in `Region E', denoted as less likely RSGs in \citet{Healy_2024}. The fraction of RSGs without dust is 32\%. The middle panel of Fig.~\ref{fig:Mdot_L_smc_mw} also shows the MAD diagram of $W1$ versus $L$. Contrary to the LMC and SMC corresponding results, we do not observe any correlation with luminosity in this case. 

\citet{Healy_2024} estimated the effective temperatures from the average spectral classification. This may not be an accurate measurement, but we can still see the anti-correlation with $\dot{M}$ in the bottom part of Fig.~\ref{fig:Mdot_L_smc_mw}. This figure also presents our derived $\dot{M}(L,T_\mathrm{eff})$ from Eq. (\ref{eq:Mdot}) with magenta points (using only those with a good fit of the $JHK_s$), with best-fit parameters $c_1=1.22\pm0.1$, $c_2=-24.98\pm3.25$, and $c_3=-13.58\pm0.4$. The derived $\dot{M}$ relation is more uncertain than those in the LMC or SMC. The data are scarcer at high $L$ in the Milky Way, and the exact value of $c_2$ can be less accurate since the $T_\mathrm{eff}$ is estimated from the spectral type. An average value for $T_\mathrm{eff}$ can also be used to remove its dependence.

\subsection{NGC 6822}
\begin{figure}[h]
    \includegraphics[width=\columnwidth]{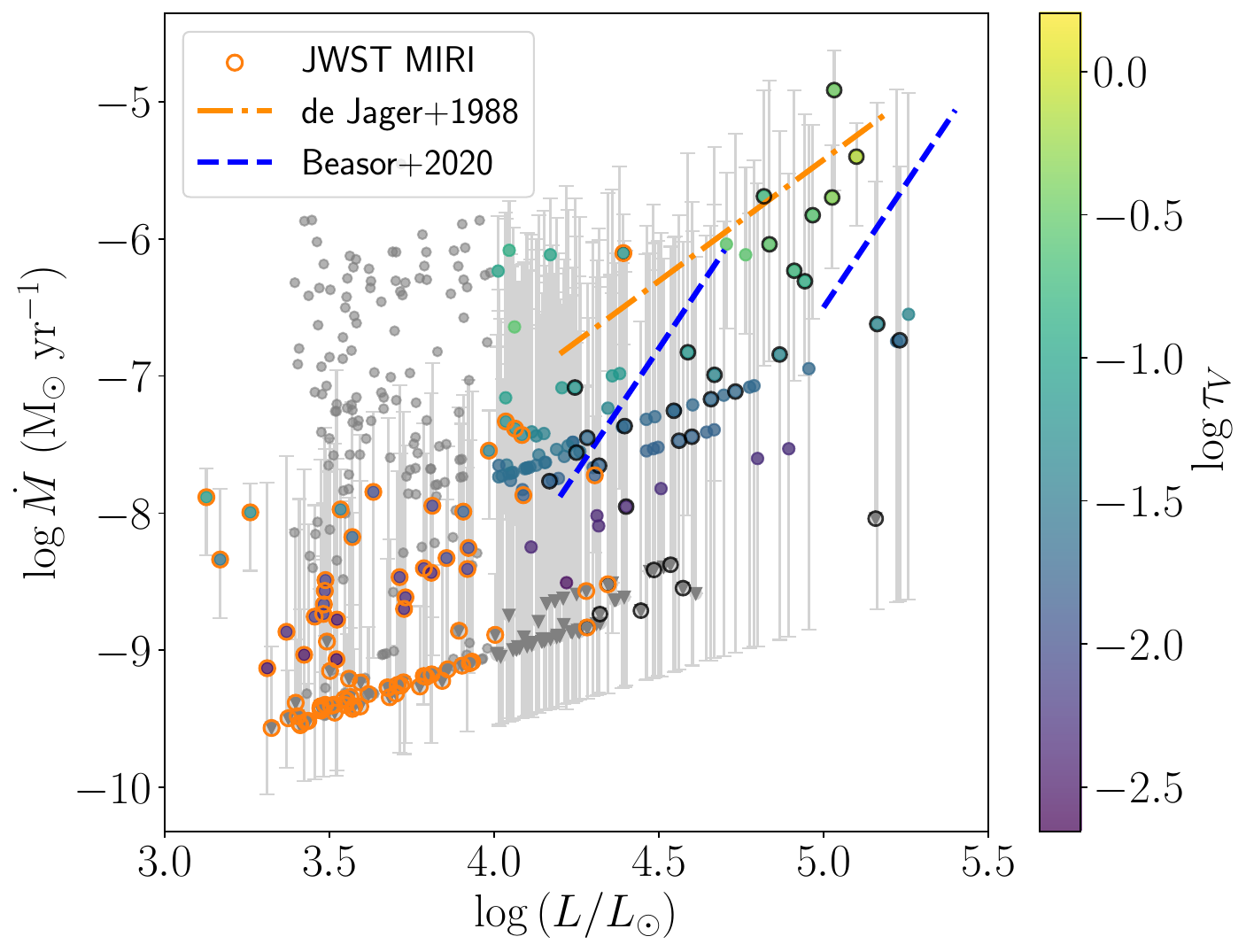}
    \caption{Derived mass-loss rate as a function of the luminosity of each RSG candidate in NGC 6822. The colour bar shows the best-fit $\tau_V$. RSGs with spectral classifications are outlined in black. Orange circles denote the sources that have JWST MIRI photometry. Blue dashed lines show the prescription from \citet{Beasor_2020} for initial masses of 10 and 20 $M_\odot$; the orange dot-dashed line is from \citet{deJager_1988}.} \label{fig:Mdot_ngc}
\end{figure}

Finally, we present the $\dot{M}$ versus $L$ of the RSG candidates in NGC 6822 in Fig. \ref{fig:Mdot_ngc}. We denote with orange circles the sources that have JWST Mid-Infrared Instrument (MIRI) photometry, providing more accurate results for the estimation of $\dot{M}$. However, most sources with JWST MIRI have $\log(L/L_\odot)<4$, and are likely lower-mass stars and not RSGs. The uncertainties are large for those without MIRI observations because their SED is constrained up to 8 $\mu$m, above which wavelength the dust emission becomes significant. The grey points indicate those without MIRI observations and $\log(L/L_\odot)<4$, at which point the results become even more uncertain. We also show the RSGs with spectral classifications from \citet{Massey_1998}, \citet{Levesque_2012}, \citet{Patrick_2015}, and \citet{deWit_2025} and compare with the prescriptions from \citet{deJager_1988} and \citet{Beasor_2020}. The fraction of the RSGs without dust is 37\% and the fraction of dusty RSGs is 8\%. 

We also used a machine-learning (ML) algorithm to predict the \textit{Spitzer} [24] photometry using the other \textit{Spitzer} bands. In this way, we were able to constrain the model SED better. The $\dot{M}$ for luminous RSGs agrees with our result without the ML-predicted [24] band. This agreement supports our findings using observations up to [8.0]. The less luminous RSGs resulted in lower $\dot{M}$ but agreed within the uncertainties with our main result. We present the method and result in \autoref{app:ML_6822}. 

We did not derive a $\dot{M}(L, T_\mathrm{eff})$ relation for this galaxy because of the large uncertainties and there is no clear correlation with $T_\mathrm{eff}$. Since the metallicity is similar to the SMC, one can use our derived relation for that case. In addition, we did not find a significant correlation of the MAD of $W1$ with $L$ or $\dot{M}$ because the errors of the NEOWISE $W1$ photometry for this galaxy are high compared to the MAD value.

\section{Discussion} \label{sec:discussion}

This work is a first step to extend the study of \citet{Yang_2023} and \citet{Antoniadis_2024} and uniformly study the mass-loss rates of large samples of RSGs in different metallicity environments, i.e. different galaxies. Overall, the values per luminosity bin do not vary significantly. In \citet{Antoniadis_2024}, we examined the effect of varying assumptions on the mass-loss mechanism and dust shell properties to resolve the large discrepancy between different studies. We chose the optimal parameters based on theory and observational constraints. However, many uncertainties can still arise, as we briefly mention in Sect.~\ref{sec:caveats} and in more detail in \citet{Antoniadis_2024}.

\begin{SCfigure*}[0.55]
    \includegraphics[width=1.3\linewidth]{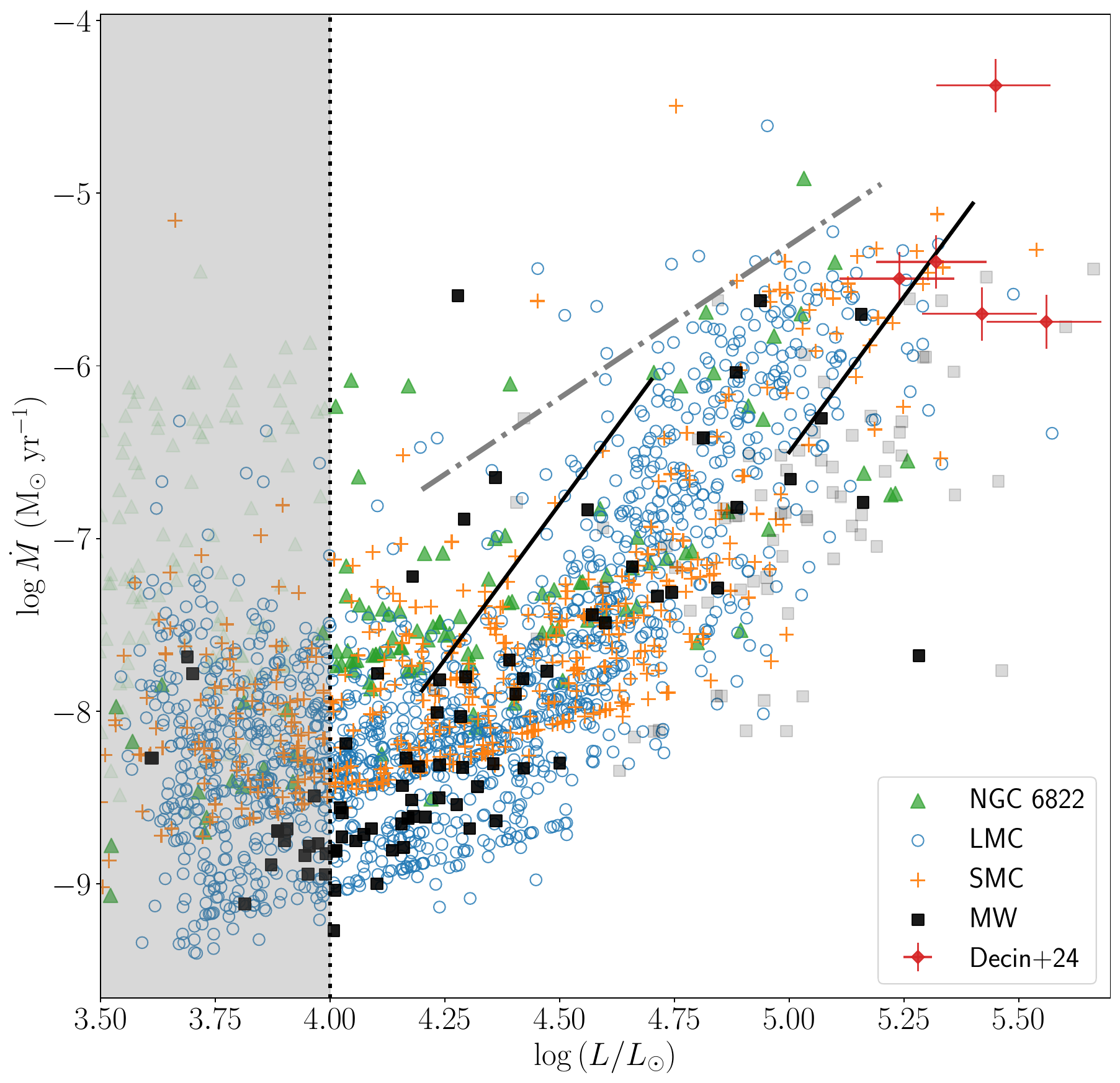}
    \caption{Comparison of the mass-loss rates from all the galaxies studied, the LMC (blue circles), SMC (orange crosses), MW (black squares), and NGC 6822 (green triangles). The red points denote mass-loss rates derived from gas diagnostics. The solid lines show the prescription from \citet{Beasor_2020} for initial masses of 10 and 20 $M_\odot$, and the dot-dashed line is the relation from \citet{deJager_1988}. The transparent symbols for MW and NGC 6822 indicate uncertain results. The shaded region includes stars with $\log(L/L_\odot)<4$, which are less likely RSGs.}\label{fig:Mdot_all}
\end{SCfigure*}

Figure~\ref{fig:Mdot_all} compares the results from all galaxies\footnote{The foreground extinction correction for AKARI S11 in \citet{Antoniadis_2024} was incorrect, leading to a slight overestimation of the $\dot{M}$ for the sources that had this photometric data point. Those sources were around 30\% of the total catalogue of the LMC, and we found that it did not affect the overall trend of the $\dot{M}(L)$ relation. Figure~\ref{fig:Mdot_all} shows the results for the corrected value.}. The red points indicate the results from \citet{Decin_2024} for five Galactic RSGs using gas diagnostics (CO rotational line emission). Their method is independent and does not rely on the properties of the dust and the SED fitting, which can lead to several uncertainties. Their results agree with ours, providing stronger support for the assumptions we used (e.g. steady-state wind, grain sizes). We found on average lower $\dot{M}$ in all galaxies than the most commonly used prescription from \citet{deJager_1988}, but within the uncertainties and with a steeper slope at $\log(L/L_\odot)\gtrsim4.5$. Our results seem to agree with \citet{Beasor_2023}, who followed similar assumptions. Furthermore, \citet{Fuller_2024} developed a theoretical model for RSG mass loss, and their results agree with ours at $\log(L/L_\odot)\gtrsim 4.5$. For a detailed discussion on the comparison between our results and other past works, see \citet{Antoniadis_2024}.

We found systematically lower rates for our Galactic RSGs, mainly at $\log(L/L_\odot)\lesssim 4.5$. This may be due to uncertainties in the distances, which result in uncertain $L$. Furthermore, estimating the interstellar extinction within the MW is also difficult and additionally contributes to these uncertainties. In case the extinction is underestimated in the near-IR or overestimated in the optical, the models will not accurately reproduce the observed SED and result in low $\tau_V$, as in the SED shown in the middle left panel of Fig.\ \ref{fig:sed}. To assess this issue, we obtained the extinction values using the 3D dust extinction maps from \cite{Lallement_2022} and \citet{Vergely_2022}\footnote{Available through the G-Tomo tool \url{https://explore-platform.eu/sda/g-tomo}.} and we found no significant differences from those presented in \citet{Healy_2024}.

\sidecaptionvpos{figure}{c}

\subsection{Enhancement of mass loss - kink} \label{sec:discuss_kink}

We found the position of the turning point or kink for the SMC at approximately $\log(L/L_\odot)= 4.65$\footnote{We refer to the position of the kink as the point where the higher end of $\dot{M}$ creates a "turn" as a function of $L$. Due to the dispersion of $\dot{M}$ at a specific $L$, this point occupies the region around $4.6<\log(L/L_\odot)<5$.} as in \citet{Yang_2023}, in which they used radiatively-driven winds (RDW) as the mass-loss mechanism assumption. This turning point is around 0.2 dex higher than the LMC. One factor that could contribute to this difference is that a lower-$Z$ star results in a higher $L$ for the same initial mass, as we also discussed in \citet{Antoniadis_2024}. Although not clear, the turning point seems to lie at the same $L$ for NGC 6822, which is expected if it is $Z$-dependent. Notably, there is no clear turning point for the results in the MW, possibly due to the small number of RSGs and their higher uncertainties. There seems to be an enhancement of mass loss at roughly $\log(L/L_\odot)= 5$, as also noticed in the Galactic sample by \citet{Humphreys_2020}. Furthermore, the variability of the Galactic RSGs remains roughly constant with $L$ (see Fig.~\ref{fig:Mdot_L_smc_mw}). This might indicate contamination of the NEOWISE photometry for several sources, making the variability not evident. 

A parameter that could be responsible for the observed kink is the gas-to-dust ratio, $r_{\rm gd}$. \citet{vLoon_2025} suggested that $r_{\rm gd}$ {could be systematically higher for low-luminosity RSGs}, meaning that the kink appears due to the assumed constant value for the whole range of $L$. More specifically, {hotter and earlier-type RSGs} could form {less} dust, leading to a {higher} $r_{\rm gd}$ and vice versa. We would expect that this effect would also cause a noticeable shift in the $\dot{M}$ vs.\ $T_{\rm eff}$ relation, but we did not observe such a shift. In addition, according to the measurements presented in \citet{Goldman_2017}, $r_{\rm gd}$ appears to be proportional to $L$, {although those authors did not explicitly claim this}. Other parameters could affect the $r_{\rm gd}$, so its correlation with $L$ is not clear. The kink may also arise from a combination of different factors, such as the $L/M$ ratio as suggested by \citet{Vink_2023}. Thus, it is difficult to determine the relevance of $r_{\rm gd}$ to the kink without additional measurements of this parameter with respect to $L$. 

\begin{figure}
    \centering
    \includegraphics[width=0.99\columnwidth]{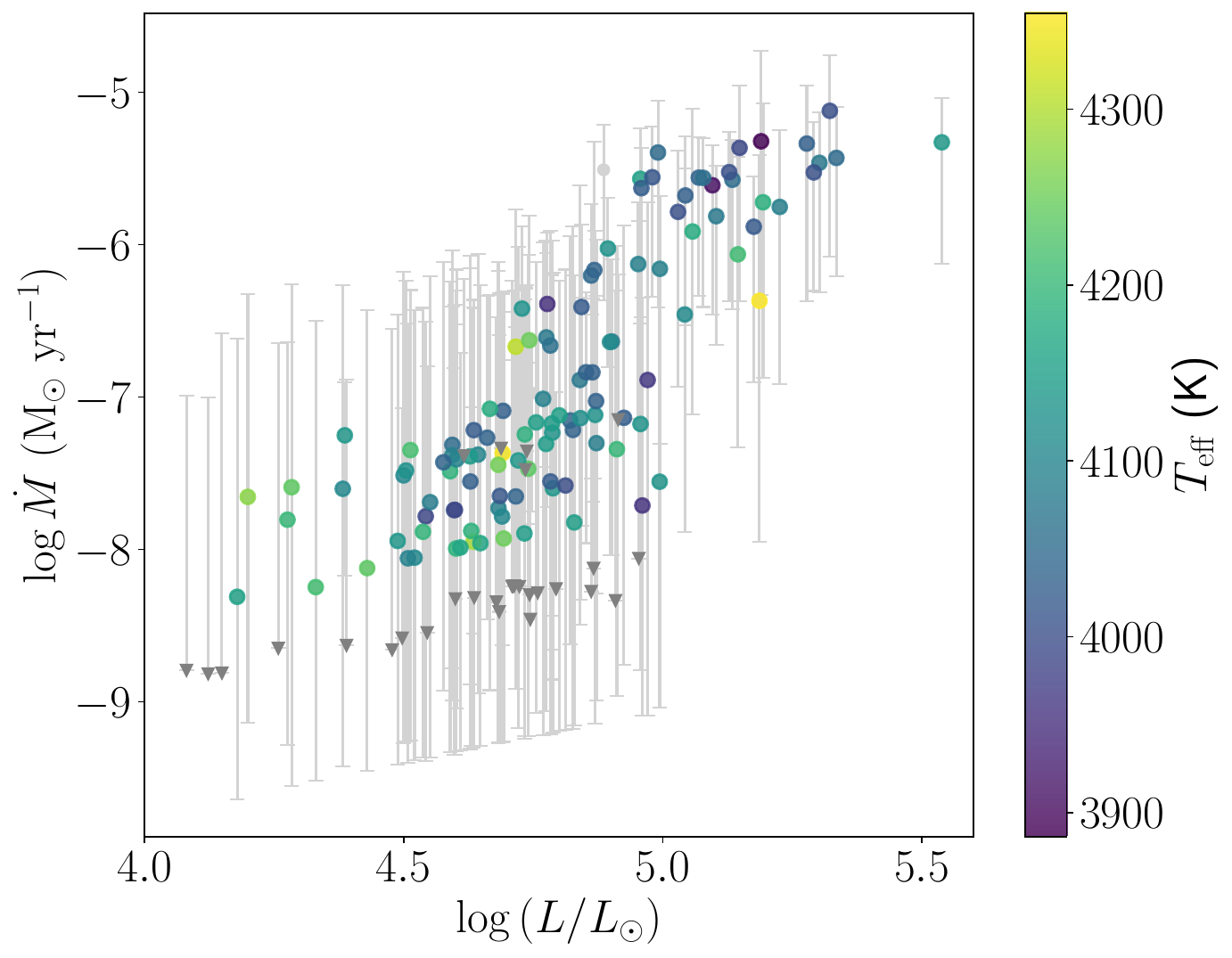}
    \caption{Mass-loss rates of RSGs in the SMC with spectroscopic classifications from the literature. Triangles indicate upper limits and the colour indicates the effective temperature using Eq. (\ref{eq:Teff_smc}).}
    \label{fig:mdot_specSMC}
\end{figure}

\begin{figure}
    \centering
    \includegraphics[width=1\columnwidth]{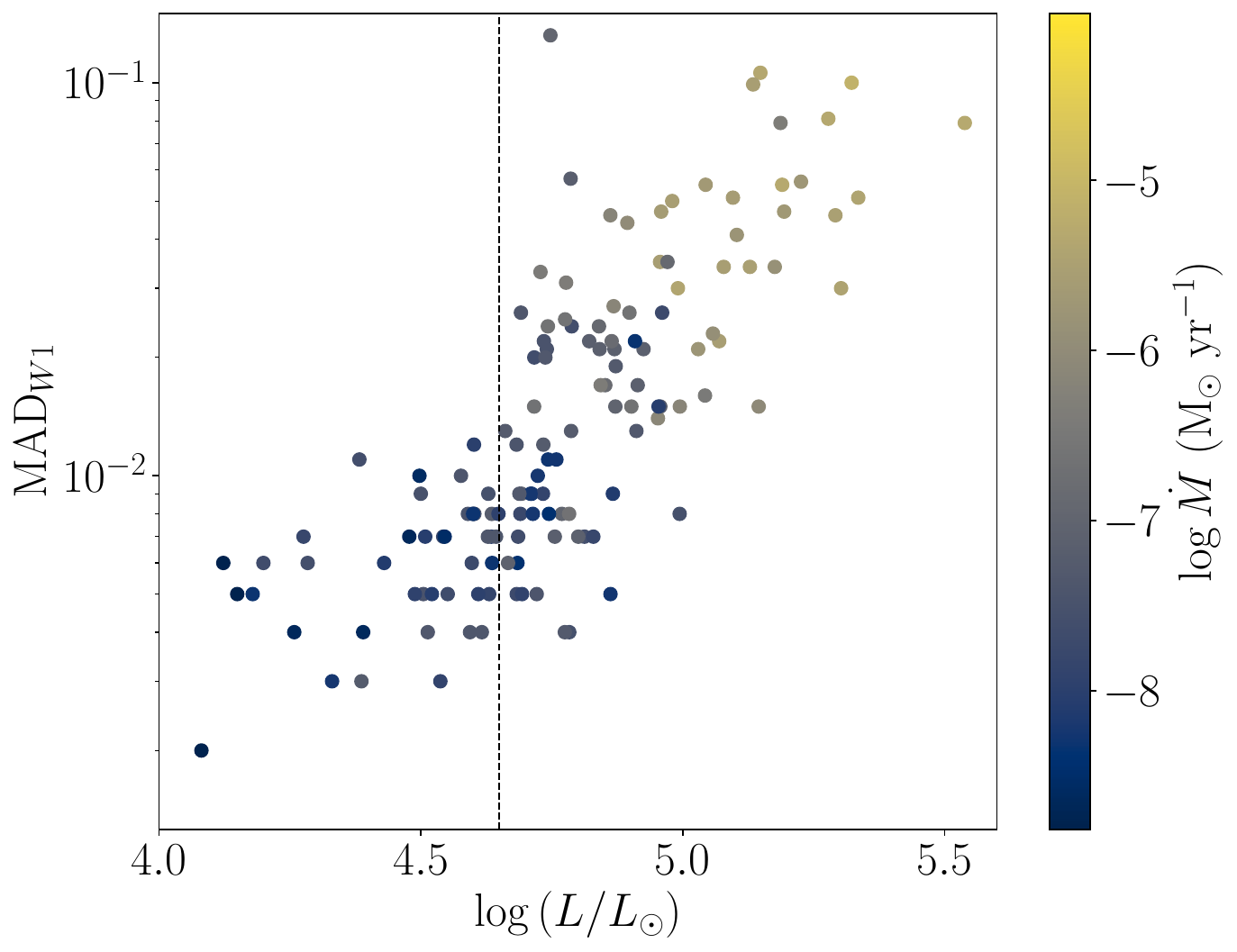}
    \caption{Median-absolute-deviation (MAD) diagram of $W1$ band of RSGs in the SMC with spectroscopic classifications from the literature. The colour indicates the corresponding mass-loss rates. The dashed vertical line shows the position of the kink from Fig.~\ref{fig:Mdot_L_smc_mw} at $\log(L/L_\odot)=4.65$.}
    \label{fig:mad_specSMC}
\end{figure}

We investigated whether contamination from giant stars could cause the kink, since the samples in the LMC, SMC, and NGC 6822 are photometrically classified. Asymptotic-giant-branch (AGB) stars have high mass-loss rates and could appear at the higher end of $\dot{M}$ vs.~$L$ at $\log(L/L_\odot)\lesssim4.5$, creating an artificial turning point or kink (although AGB stars also span a wide range of $\dot{M}$ similar to RSGs). We followed two ways to assess this. First, we examined the mass-loss rate relation of confirmed RSGs (i.e. with spectral classifications) in the SMC and LMC to verify the turning point. In the LMC, there are not many spectroscopic observations at $\log(L/L_\odot)\lesssim4.4$, so, in this case, we cannot distinguish a kink. In the SMC, there are more observations at $\log(L/L_\odot)\lesssim4.6$ to confirm the presence of a kink. We show $\dot{M}$ vs. $L$ and the MAD diagram for the SMC in Fig.~\ref{fig:mdot_specSMC} and \ref{fig:mad_specSMC}. In both figures, especially in the MAD diagram of $W1$, we observe a turning point at $\log(L/L_\odot)\simeq 4.65$, although in the $\dot{M}(L)$ relation the uncertainties of the $\dot{M}$ are significant. The presence of the kink for the spectrally classified RSGs implies that AGB contamination is not causing it. We demonstrate the fitted relation on the spectroscopic data assuming a broken law or a linear relation for the SMC in \autoref{app:spec}.

\begin{figure}
    \centering
    \includegraphics[width=0.9\columnwidth]{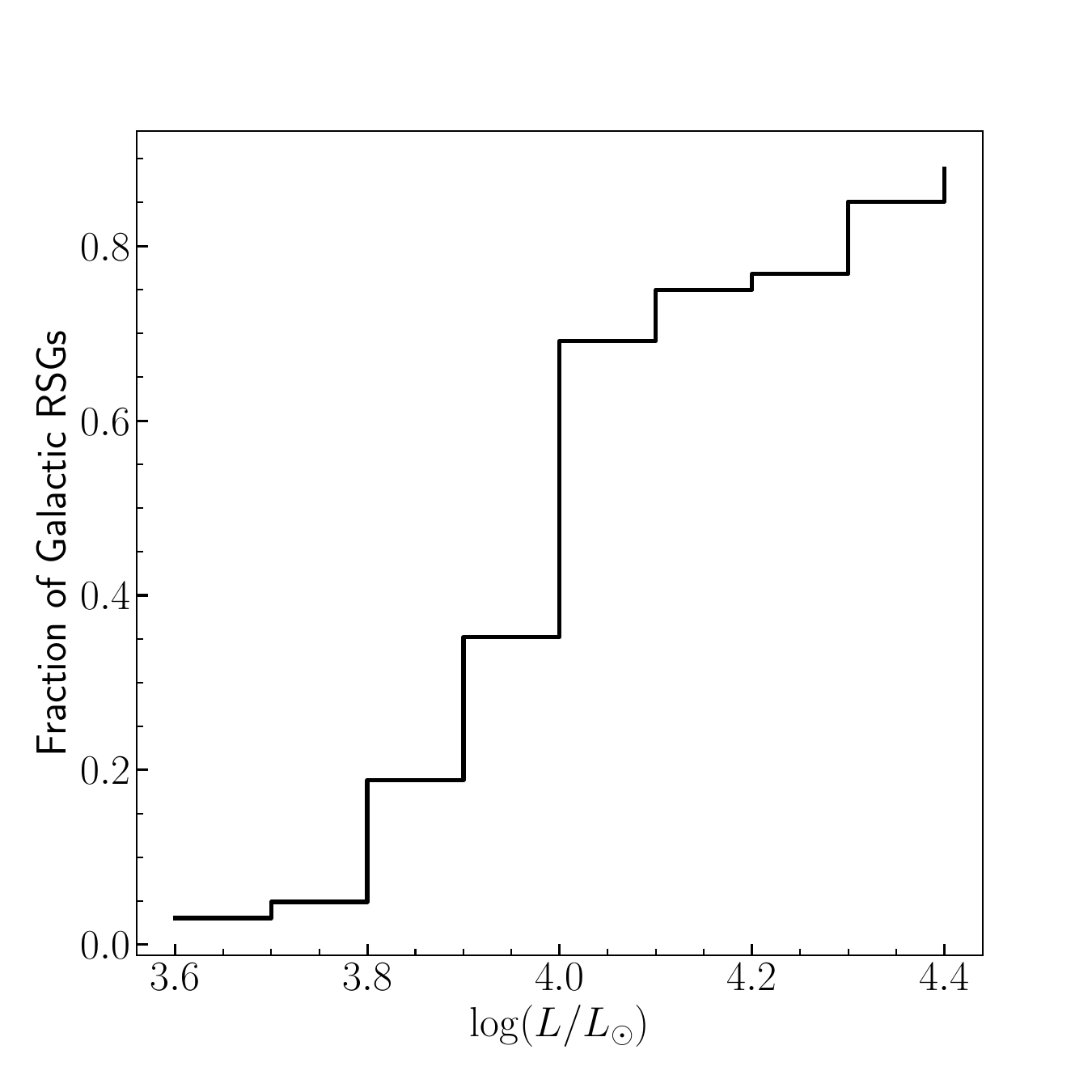}
    \caption{Fraction of the Galactic RSGs from \citet{Healy_2024} over the total number of K or M-type bright stars from their initial sample.}
    \label{fig:rsg_frac}
\end{figure}

Second, we quantified the contamination from giant stars with the same colours as RSGs in the MW to use it as a probe for the other galaxies. \citet{Healy_2024} compiled both a catalogue of all spectrally classified K and M-type stars in the Milky Way and a final catalogue consisting of only RSGs. This means that the initial sample included AGB stars with similar $T_\mathrm{eff}$ (and colour) as RSGs. Since our RSG samples in the LMC and SMC are defined from colour-magnitude diagrams (CMDs), i.e. the colour of the sources, we used their full (\textit{all late-type bright stars}) catalogue to calculate the fraction of RSGs over the total number of all K or M-type stars in the MW in the range $3.6\leq \log{(L/L_\odot)}<4.5$. Figure~\ref{fig:rsg_frac} presents this fraction as a function of $L$. We found that RSGs consist of around 70\% of the total sample at $\log{(L/L_\odot)}=4$ and this fraction increases with increasing $L$. At $3.8\leq \log{(L/L_\odot)}<4$, the fraction drops to around 20 to 40\%. We should note that the same initial mass corresponds to a higher $L$ at lower metallicity. This means that the corresponding fraction of 70\% RSGs at $\log{(L/L_\odot)}\sim 4$ in the MW could be around $\log{(L/L_\odot)}\sim 4.3$ for the LMC or 4.4 for the SMC. 

According to \citet{Healy_2024}, their sources at $3.36< \log{(L/L_\odot)}<3.92$ at what they defined as `Region C' in the HR diagram (see their Fig. 4 and 5) consist of intermediate-mass AGB stars, which are 4.9\% of the total AGB sample, and lower mass ($\sim9$ $M_\odot$) RSGs, which consist of almost 0\% of the total RSG sample\footnote{\citet{Healy_2024} reports these values but in their Fig.\ 4 there are 5 out of 156 RSGs, which would correspond to around 3\%. This applies similarly to the AGB stars and their fractions, which are probably typos. In any case, for our discussion, these regions consist of small fractions of RSGs and intermediate-mass AGB stars.}. However, they mention that this region is biased against observations of RSGs. Hence, it is difficult to robustly quantify the AGB contamination in this range of $L$ and probe it to the corresponding region in the LMC and SMC. Furthermore, we estimated the mass-loss rates for the whole catalogue of \textit{all bright late-type stars} from \citet{Healy_2024} to examine if an artificial turning point is created from AGB stars and we found no such indication. Similarly, \citet{Goldman_2017} included lower-$L$ Galactic AGB stars and higher-$L$ LMC RSGs in their study but did not observe a turning point to justify such an argument (see their Fig.\ 20).

Finally, \citet{Neugent_2020} mention that the strong AGB contamination is up to $\log{(L/L_\odot)}=4$ in M31, above which point there is a distinct stellar population in the $J-K_s$ CMD (see their Fig.~8). This limit could be around 4.2 - 4.3 if we want to be more conservative. Our analysis shows that the AGB contamination cannot be that significant to create an artificial kink. Further spectroscopic observations at $4<\log{(L/L_\odot)}<4.5$ of our studied LMC and SMC photometric RSG samples would resolve this issue.

\subsection{On the metallicity dependence}

\begin{figure}
    \centering
    \includegraphics[width=1\columnwidth]{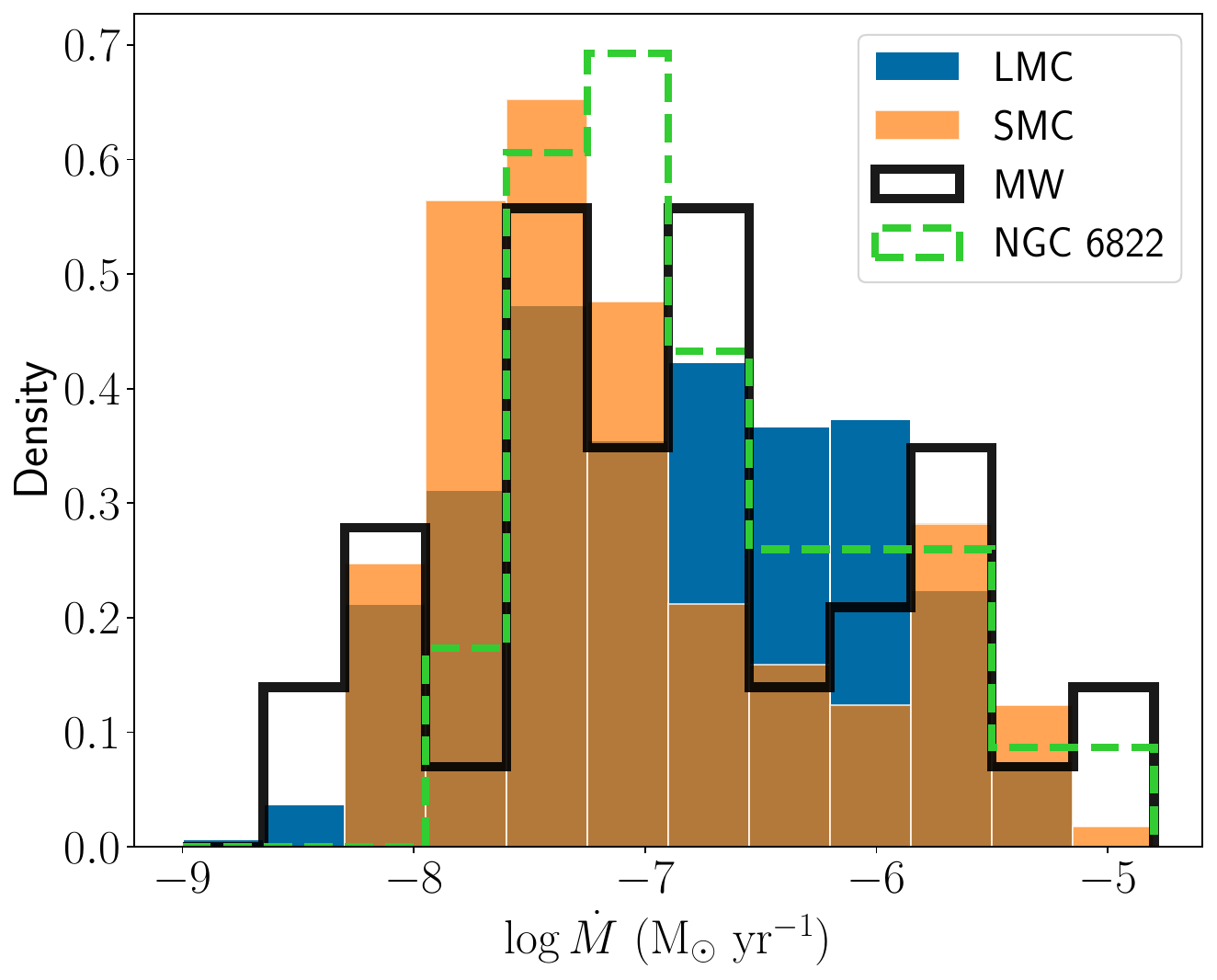}
    \caption{Distribution of the mass-loss rate of RSGs with $\log{(L/L_\odot)}>4.5$ for the LMC (blue bar), SMC (orange bar), MW (black line), and NGC 6822 (green dashed line).}
    \label{fig:mdot_all_hist}
\end{figure}

\begin{figure}
    \centering
    \includegraphics[width=1\columnwidth]{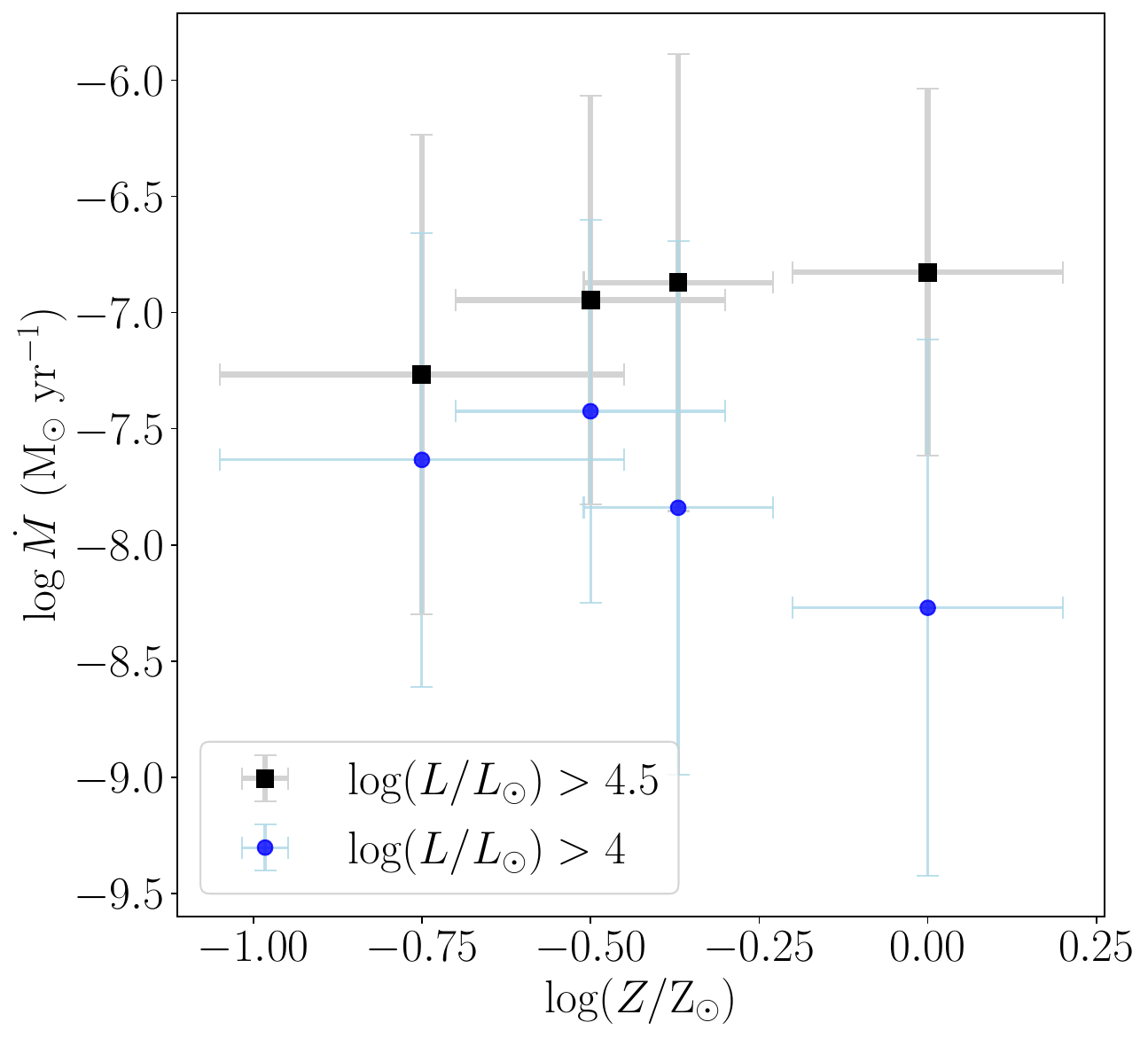}
    \caption{The median mass-loss rate vs. metallicity for each galaxy of our study and two luminosity ranges.}
    \label{fig:mdot_Z}
\end{figure}

Overall, we found no strong correlation between the mass-loss rate and metallicity. Considering the uncertainties of the results from the MW, it is difficult to determine the metallicity dependence of RSG mass loss accurately. The shift of the kink by around 0.2 dex in luminosity between the LMC and SMC might be due to metallicity, as also mentioned by \citet{Wen_2024}. However, we could attribute part of it to the different resulting luminosity for the same initial mass at different metallicity. We compare the distribution of $\dot{M}$ for each galaxy for $\log{(L/L_\odot)}>4.5$, so that we remove all the uncertainties or biases regarding the kink, in Fig.~\ref{fig:mdot_all_hist}. The distributions seem to agree without clear differences. However, the LMC has several sources with higher $\dot{M}$. Table \ref{tab:median} presents the median and mean values of the $\dot{M}$ and other parameters for each galaxy. Correspondingly, Fig.~\ref{fig:mdot_Z} shows the median $\dot{M}$ as a function of $Z$ for all RSGs ($\log(L/L_\odot)>4$) and for high $L$ RSGs ($\log(L/L_\odot)>4.5$). The error bars on the $y$ axis correspond to the deviation on the median. The median $\dot{M}$ of the LMC appears to be slightly higher than the SMC, likely due to the position of the turning point at lower $L$ creating a steeper slope after that point. However, everything lies within the uncertainties of the gas-to-dust ratio, which is significant for the SMC, and the dispersion of the $\dot{M}$. 

Finally, we performed a Kolmogorov–Smirnov (KS) test on the $\dot{M}$ distributions with $\log{(L/L_\odot)}>4.5$ to assess if two samples originate from the same distribution. We found a $p$-value of $5\times10^{-5}$ for the LMC and SMC indicating a significant difference, implying a $Z$-dependence. However, this may not be completely valid for two reasons. First, the KS test is more sensitive close to the centres of the distributions, rather than the whole shapes. Second, the different positions of the kink and the assumption of an average gas-to-dust ratio could contribute to that difference. The combinations of the other galaxies result in $p>0.1$, so we cannot reject the null hypothesis.

\begin{table*}[h]
    \centering
    \small
    \caption{Properties of RSGs with $\log(L/L_\odot)>4$ in each galaxy}
    \begin{tabular}{c c c c c c c}
        \hline\hline
         Galaxy & [Z] & Median $T_\mathrm{eff}$ (K) & Mean $\tau_V$ & Mean $\log\dot{M}$ & Mean $\log\dot{M_d}$ & Median $\log\dot{M}$ at $\log (L/L_\odot)>4.5$ \\ \hline \\ [-0.3cm] 
         SMC & $-0.75\pm0.3$ $^1$ & $4190\pm130$ & 0.04 & $-6.68$ & $-10.63$ & $-7.27\pm1.03$ \\ [+0.01cm]
         NGC 6822 & $-0.5\pm0.2$ $^2$ & $4160\pm380$ & 0.06 & $-6.66$ & $-10.11$ & $-6.95\pm0.88$ \\ [+0.01cm]
         LMC & $-0.37\pm0.14$ $^3$ & $3960\pm100$ & 0.11 & $-6.69$ & $-10.06$ & $-6.87\pm0.98$ \\ [+0.01cm]
         Milky Way & $0\pm0.2$ $^4$ & $3750\pm190$ & 0.15 & $-7.88$ & $-10.19$ & $-6.83\pm0.79$ \\ \hline

    \end{tabular}
    \tablebib{(1) \citet{Davies_2015, Choudhury_2018}, (2) \citet{Muschielok_1999, Venn_2001, Patrick_2015}, (3) \citet{Davies_2015}, (4) \citet{Anders_2017}.}
    \label{tab:median}
\end{table*}

Previous studies \citep{vLoon_2005, Groenewegen_2009, Goldman_2017} concluded that metallicity had little to no effect on the $\dot{M}$ studying samples consisting of both RSGs and AGB stars.  However, these studies had significantly smaller samples than ours. In addition, \citet{Mauron_2011} suggested a scaling of $\dot{M} \propto(Z/Z_\odot)^{0.7}$ to the \citet{deJager_1988} relation derived from Galactic RSGs. However, they considered small samples of RSGs in the SMC and LMC from \citet{Bonanos_2010}, who estimated the mass-loss rates using a different method from \citet{deJager_1988}. 

Metallicity could affect the dust-driven wind. \citet{Kee_2021} developed a theoretical model for the RSG mass loss assuming turbulent pressure as the dominant mechanism behind it, which is not strongly dependent on metallicity. However, metallicity affects dust production and if radiation pressure on the dust grains becomes important, metallicity could play a role. Similarly, the theoretical model of \citet{Fuller_2024} does not directly depend on metallicity, but they did not investigate its role on the dust shell. Finally, \citet{Goldman_2017} found a metallicity correlation with the outflow velocity but the effect was mostly cancelled out with the anti-correlation of the gas-to-dust ratio.

\begin{figure}
    \centering
    \includegraphics[width=1\columnwidth]{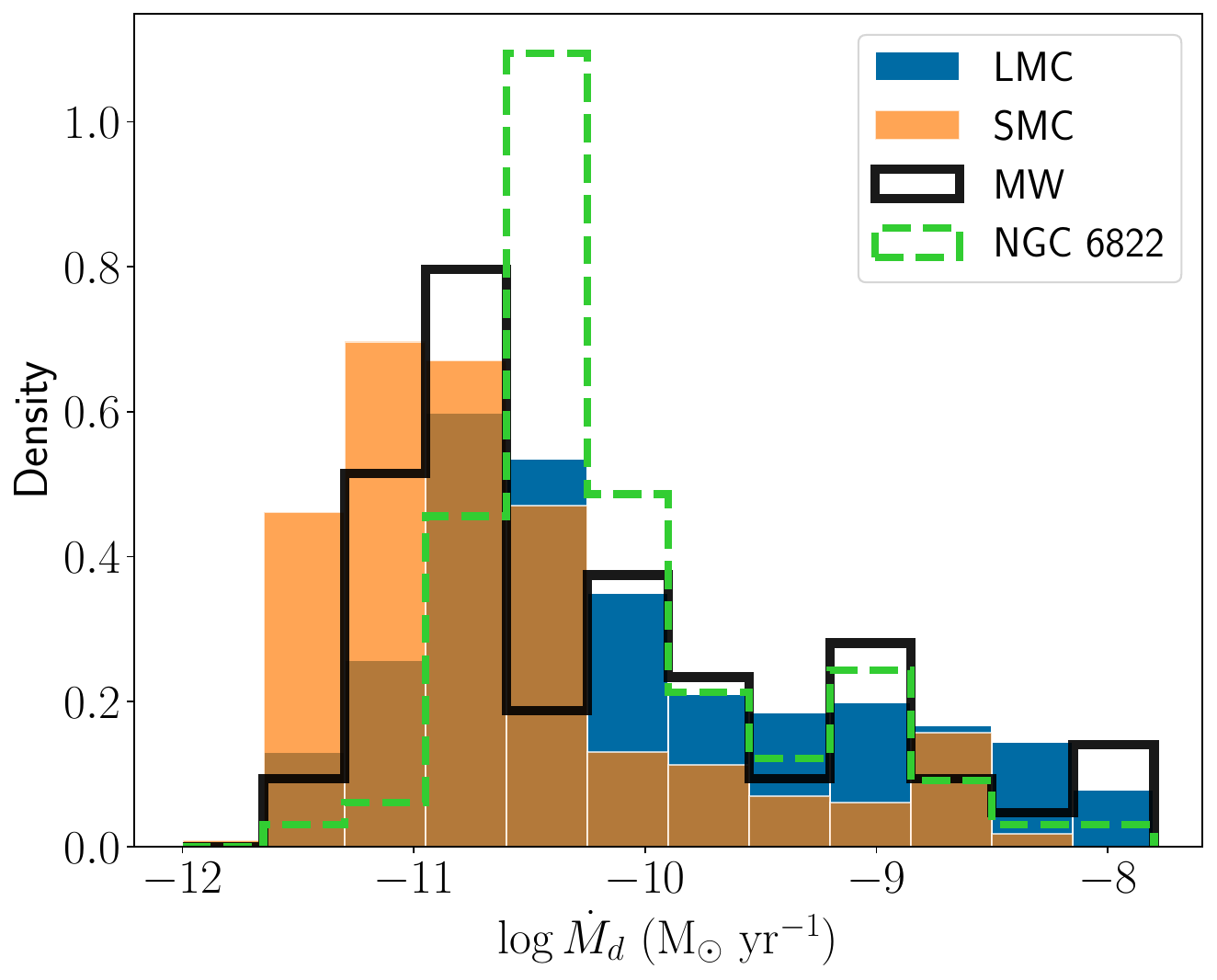}
    \caption{Distribution of the dust-production rate of RSGs with $\log{(L/L_\odot)}>4$ for the LMC (blue bar), SMC (orange bar), MW (black line), and NGC 6822 (green dashed line).}
    \label{fig:dpr_all_hist}
\end{figure}

More dust is expected to form in higher metallicity environments. In Fig.~\ref{fig:tau}, we observe that $\tau_V$ increases with increasing $Z$, also depicted in the mean $\tau_V$ in \autoref{tab:median}. \autoref{fig:dpr_all_hist} demonstrates the distribution of the dust-production rate, $\dot{M}_d$, for RSGs with $\log{(L/L_\odot)}>4$, which we calculated from Eq.~(\ref{eq:dotM}) removing the factor of $r_{\mathrm{gd}}$. As in the distribution of $\tau_V$, the dust-production rate is higher with increasing $Z$. However, this is not clear for the Galactic RSGs, which have a wide range of $\dot{M}_d$, and it is quite low, especially at $\log{(L/L_\odot)}<4.5$ compared to the other galaxies. This could arise from the uncertainties of the Galactic RSG results we mentioned and the reason that the data are more scarce at high $L$ compared to the SMC and LMC. The trend of $\dot{M_d}$ with $Z$ becomes clearer at $\log{(L/L_\odot)}>4.5$, as we examined. Furthermore, considering as dusty RSGs those with $\tau_V>0.1$, we found 14\% of the RSGs in the LMC and 2\% of the RSGs in the SMC to be dusty. This agrees with the value of 12\% reported by \citet{deWit_2024} for RSGs at subsolar metallicity (defining the dusty ones as $[3.6]-[4.5]>0.1$~mag). Finally, we show mid-IR colour (related to dust emission) diagrams with respect to the $\dot{M}$ in \autoref{app:colour}.

\subsection{Caveats} \label{sec:caveats}
The caveats regarding the estimation of mass-loss rates from the properties of the dust shell, the SED fitting, and the different assumptions (grain size, gas-to-dust ratio, wind mechanism) have been described in \citet{Antoniadis_2024}. We assumed a uniform average value of $r_\mathrm{gd}$, which can vary within a galaxy and for RSGs with different $L$, especially in the SMC where it can reach values of up to around 10,000 \citep{Clark_2023}. We also assumed an average $Z$, which can vary between RSGs in the same host galaxy \citep[e.g.][]{Davies_2015}. Nevertheless, we could distinguish the overall trend by studying large populations of RSGs. There seems to be a negligible upward trend in Fig.\ \ref{fig:mdot_Z}. However, if the $Z$ dependence is minor compared to the mentioned uncertainties in $\dot{M}$, it becomes more difficult to accurately compare RSGs from different galaxies and draw a concrete conclusion.

\section{Summary} \label{sec:conclusion}

We compared the mass-loss rates of the largest sample of RSGs in four galaxies with different metallicity environments, the SMC, Milky Way, NGC 6822, and the LMC; the latter from \citet{Antoniadis_2024}. We measured the dust shell properties and mass-loss rates by fitting the observed SEDs with models from the radiative transfer code \texttt{DUSTY}. The mass-loss rates range from approximately $10^{-9} \ M_{\odot}$ yr$^{-1}$ to $10^{-5} \ M_{\odot}$ yr$^{-1}$ with an average value of around $1.5\times10^{-7} \ M_\odot \  \mathrm{yr}^{-1}$. There was no evidence of a significant correlation with metallicity, apart from the shift of the kink by 0.2 dex in the $\dot{M}(L)$ relation between the LMC and SMC. Concerning the sample of Galactic RSGs, there was no clear indication of a kink. However, the results from the Galactic RSGs are inconclusive due to uncertainties in the distance and interstellar extinction estimations, while the results of NGC 6822 have uncertainties due to worse mid-IR photometric quality from \textit{Spitzer} at that distance. 

We derived $\dot{M}$ relations as a function of $L$ and $T_\mathrm{eff}$ for the SMC and Milky Way, and relations using only the spectroscopic RSGs in the LMC and SMC. We excluded the upper limits in the derivation of these relations; thus, they are valid for RSGs with dust ($\tau_V\geq0.002$). The $\dot{M}(L)$ relation of the spectroscopic RSGs in the SMC indicates the presence of a kink, but more observations are needed to confirm it. Using the sample from \citet{Healy_2024}, we showed that RSGs dominate the stellar population at $\log(L/L_\odot)>4$ in the Milky Way, implying that the kink does not arise from AGB contamination. However, this limit may be at a higher $L$ for lower metallicity. Moreover, we confirmed that metallicity affects the formation of dust by showing that the optical depth of the dust shell of RSGs in higher-$Z$ galaxies is larger on average. This could mean that radiation pressure on the dust shell becomes more significant at higher $Z$ and would result in higher mass-loss rates if this force was dominant in the driving of the wind.

Considering the dust properties of the RSG sample, we found that 30-40\% of the RSGs in all the galaxies do not have any dust, most of which have $\log(L/L_\odot)\lesssim4.7$. Also, 14\% of the RSGs in the LMC and 2\% of the RSGs in the SMC were found to be significantly dusty ($\tau_V>0.1$). Finally, we studied RSG samples in M31 and M33, but the \textit{Spitzer} photometry, especially at [24], becomes unreliable at these distances ($d>0.5$ Mpc). JWST data will allow the investigation of RSGs in different metallicity galaxies with better quality mid-IR photometry and resolve the metallicity dependence of the RSG mass-loss rates.

\section{Data availability}

All SED fit figures are available in an electronic form via \url{https://zenodo.org/records/16736627}. Tables C.1, C.2, and C.3 are only available in electronic form at the CDS via anonymous ftp to \url{cdsarc.u-strasbg.fr} (130.79.128.5) or via \url{http://cdsweb.u-strasbg.fr/cgi-bin/qcat?J/A+A/} .

\begin{acknowledgements}
KA, AZB, GMS, and GM acknowledge funding support from the European Research Council (ERC) under the European Union’s Horizon 2020 research and innovation program ("ASSESS", Grant agreement No. 772086). EZ acknowledges support from the Hellenic Foundation for Research and Innovation (H.F.R.I.) under the ``3rd Call for H.F.R.I. Research Projects to Support Post-Doctoral Researchers'' (Project No: 7933). This work is based in part on observations made with the NASA/ESA/CSA James Webb Space Telescope. The data were obtained from the Mikulski Archive for Space Telescopes at the Space Telescope Science Institute, which is operated by the Association of Universities for Research in Astronomy, Inc., under NASA contract NAS 5-03127 for JWST. These observations are associated with program \#1234. We thank Stephan de Wit for his presence and feedback throughout a significant part of this work, Emma Beasor and Nathan Smith for valuable discussions regarding the analysis of the results, Sarah Healy for answering questions about their Galactic RSG catalogue, and Despina Hatzidimitriou for discussions on the results. Finally, we thank the referee, Dr.\ Jacco van Loon, for his valuable comments and suggestions.
\end{acknowledgements}

\bibliographystyle{aa}
\bibliography{ref}

\begin{appendix}

\section{Using Machine-Learning for \textit{Spitzer} [24] in NGC 6822} \label{app:ML_6822}

Contamination is a key issue for the MIPS [24] band for \textit{Spitzer}. Especially at larger distances ($d>0.5$ Mpc) the angular resolution for this band results in unreliable photometric measurements, due to the blending of sources. Therefore, the number of sources with valuable and trustworthy measurements decreases dramatically. Hence, we applied a machine-learning method to predict the [24] photometry using the \textit{Spitzer} IRAC-band values. We first built and explored the efficiency of a machine-learning regressor in the LMC and the SMC independently, before applying the final trained model on NGC 6822. 

We used 2218 and 892 sources for the LMC and SMC from \citet{Antoniadis_2024} and this work, respectively. First, we removed all sources with negative and zero values in any of the \textit{Spitzer} bands. This resulted in 1291 sources in the LMC and 320 in the SMC. Although this decreased our samples, we secured informative measurements across all bands. That allowed us to create the mock samples with missing [24] band values. We optimised our models with these samples since we calculated their performance by comparing our predictions with the true values. We examined different metrics, but we chose the Mean Relative Error (MRE), which provides an estimate of how well our model is performing, and the R2 score, which estimates how well the model generalizes to unseen data (i.e. to avoid overfitting). 

We opted to use directly the magnitudes of \textit{Spitzer} bands for the SMC case as features. For technical reasons, it was better to work with positive numbers. To consider the distance effects on magnitudes for the LMC and NGC 6822, we converted their magnitudes to the distance of the SMC according to their distance moduli (i.e. 18.88 mag, 18.47 mag, and 23.50 mag for the SMC, LMC, and NGC 6822, respectively\footnote{These values correspond to the mean values from all available resources as found in the NED - NASA/IPAC Extragalactic Database (accessed, July 24, 2024), for each galaxy. They might slightly differ from the distances used in the main part of the paper, but do not affect the result.}).

As baseline models, we examined several classical approaches to missing data imputation, such as replacing the missing values with the mean or the median value of [24] band, derived from all sources without missing values (achieving an MRE of $\sim19.5\%$). Then we tested various machine-learning methods, such as linear regression, random forest, extra trees, and XGBoost, as well as a stacking approach in which the predictions of a method, e.g. random forest, are used as an input to a linear regressor. We kept 30\% of each original sample separate to use as a test sample. The rest (70\%) of the samples were used for the training and the optimisation through a randomised 5-fold cross-validation technique to accelerate implementation and enhance accuracy compared to the classic approach of a full-grid cross-validation. All these methods easily achieved MRE~$\sim3\%$ and R2~$\sim0.9$, outperforming the baseline models. 

From the methods mentioned above, the best results were obtained from a stacking approach (random forest and linear regression) when working with the SMC and LMC independently, and the extra trees method when combining the two samples, achieving MREs of 3.3\%, 3.5\%, and 3.2\%, respectively. For the final application, we opted to use the model trained on the SMC, since this model achieved the best MRE and R2 score on the (independent) test sample. Furthermore, its metallicity is similar to that of NGC 6822. There were 396 NGC 6822 sources, but only 88 had [24] band values on which we could test our model. We estimated our performance at MRE $\sim5\%$. However, the R2 score ($\sim0.3$) was low for the corresponding number of sources, indicating a rather poor performance. This compelling different result alerted us to potential issues with the photometry. 

\begin{figure}[h]
    \centering
    \includegraphics[width=\columnwidth]{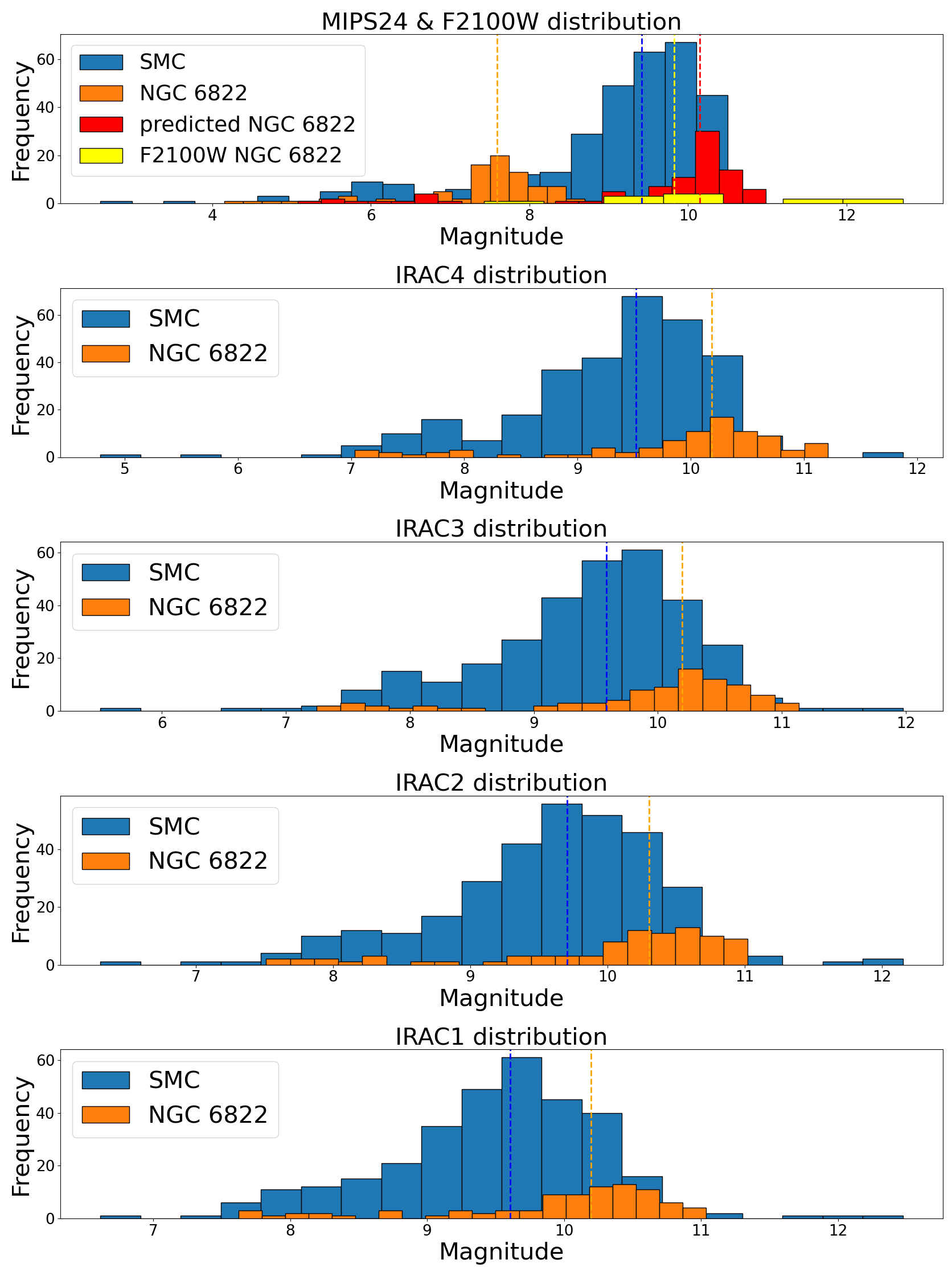}
    \caption{Distribution of different \textit{Spitzer} bands for the SMC (blue) and NGC 6822 (orange). The red histogram shows the predicted MIPS [24] band and the yellow demonstrates the JWST [21] band. The values for NGC 6822 were corrected for the distance difference between the two galaxies.}
    \label{fig:spitzer_distributions}
\end{figure}

Indeed, when examining the \textit{Spitzer} magnitude distributions for NGC 6822 we noticed that although the distributions of the IRAC bands were similar to the ones of the LMC and SMC, the distribution of [24] was significantly off. We compare the SMC and NGC 6822 distributions in Fig. \ref{fig:spitzer_distributions}. Due to the increased distance of NGC 6822 and the large spatial resolution of MIPS, there is significant blending of the sources, leading to brighter photometry. To verify that our hypothesis is correct, we took advantage of the JWST [21] band (F2100W filter) in this galaxy \citep{Nally_2024}. Although data was available for 12 sources (yellow histogram in Fig. \ref{fig:spitzer_distributions}), their values lie where we would expect them compared to the SMC \textit{Spitzer} [24]. From these sources, only four had \textit{Spitzer} data. Consequently, using our model we predicted values of MIPS [24] band with a relative error of 7.5\%, 11.2\%, 3\%, and 23.5\% compared to the JWST [21] band (considered as the true value). Collectively, the final estimate of our performance is an MRE~$\sim11\%$ (with an improved value for R2~$\sim0.4$ for these four sources)\footnote{The mean values for the SMC and NGC 6822 are 9.60 mag and 10.19 mag for IRAC1, 9.59 mag and 10.03 mag for IRAC2, 9.45 mag and 9.90 mag for IRAC3, 9.34 mag and 9.86 mag for IRAC4, 9.02 mag and 7.36 mag for MIPS [24] band, for each galaxy, respectively. The mean values of the JWST [21] band (yellow) and our predictions for [24] (red) are found at 10.29 mag and 9.58 mag, respectively.}.

\begin{figure}[h]
    \centering
    \includegraphics[width=1\columnwidth]{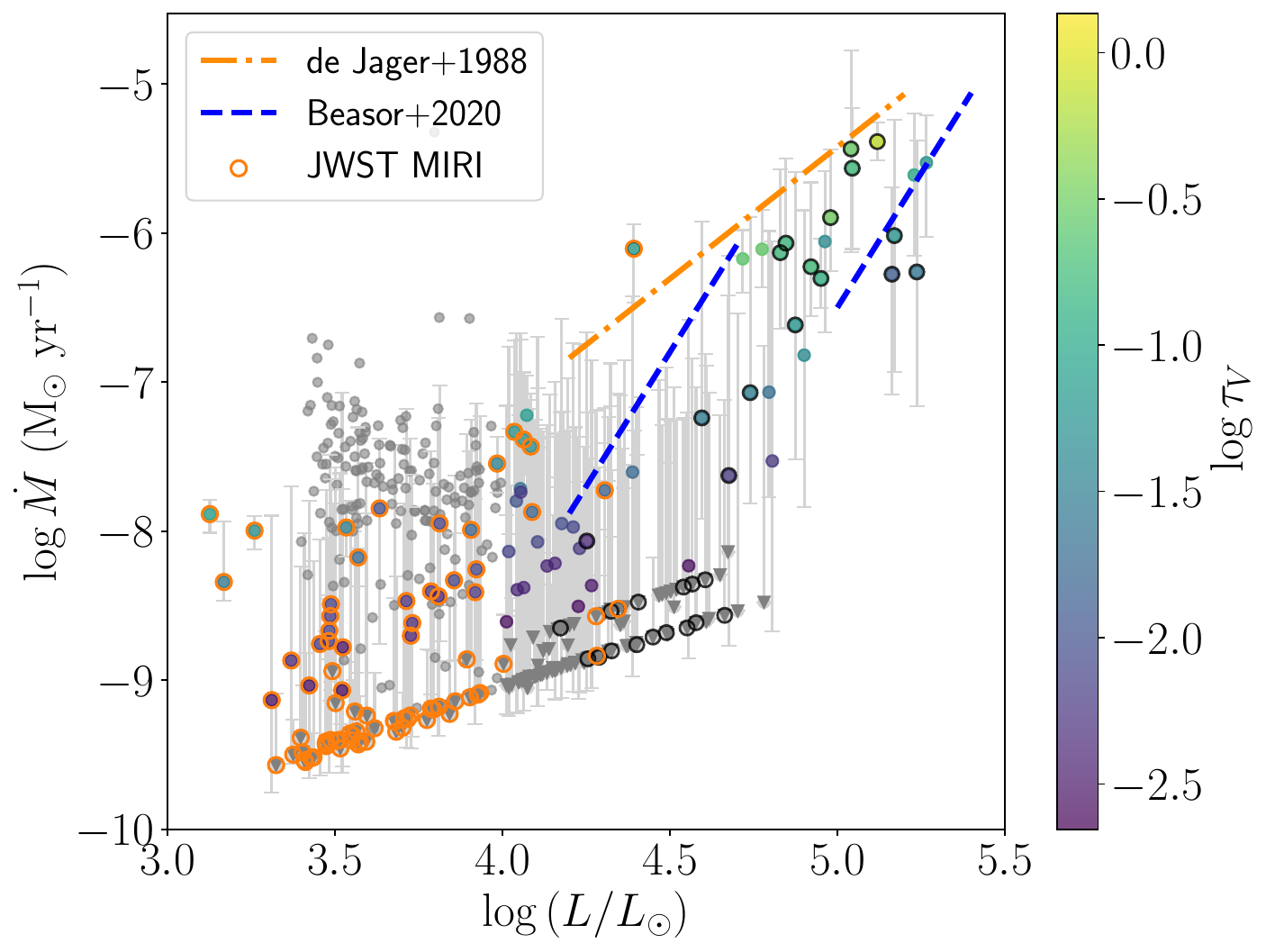}
    \caption{Same as Fig.~\ref{fig:Mdot_ngc} but using the ML predicted photometry for \textit{Spitzer} [24] in the SED fitting procedure to derive the mass-loss rate.}
    \label{fig:mdot_6822_ML}
\end{figure}

We derived the mass-loss rates and we present the results with respect to luminosity and the best-fit optical depth in Fig.~\ref{fig:mdot_6822_ML}. In the cases where JWST mid-IR photometry existed, we did not use the ML [24] predicted values. Comparing Fig.~\ref{fig:mdot_6822_ML} with Fig.~\ref{fig:Mdot_ngc}, we notice that a significant number of RSGs at $4<\log{(L/L_\odot)}\lesssim4.7$ have now lower $\dot{M}$ appearing as dust-free, but they agree within the errors. The more luminous RSGs almost agree and 4 RSGs that appeared as low $\dot{M}$ outliers in Fig.~\ref{fig:Mdot_ngc}, now follow the rest of the distribution.

\begin{figure}[h]
    \centering
    \includegraphics[width=0.93\columnwidth]{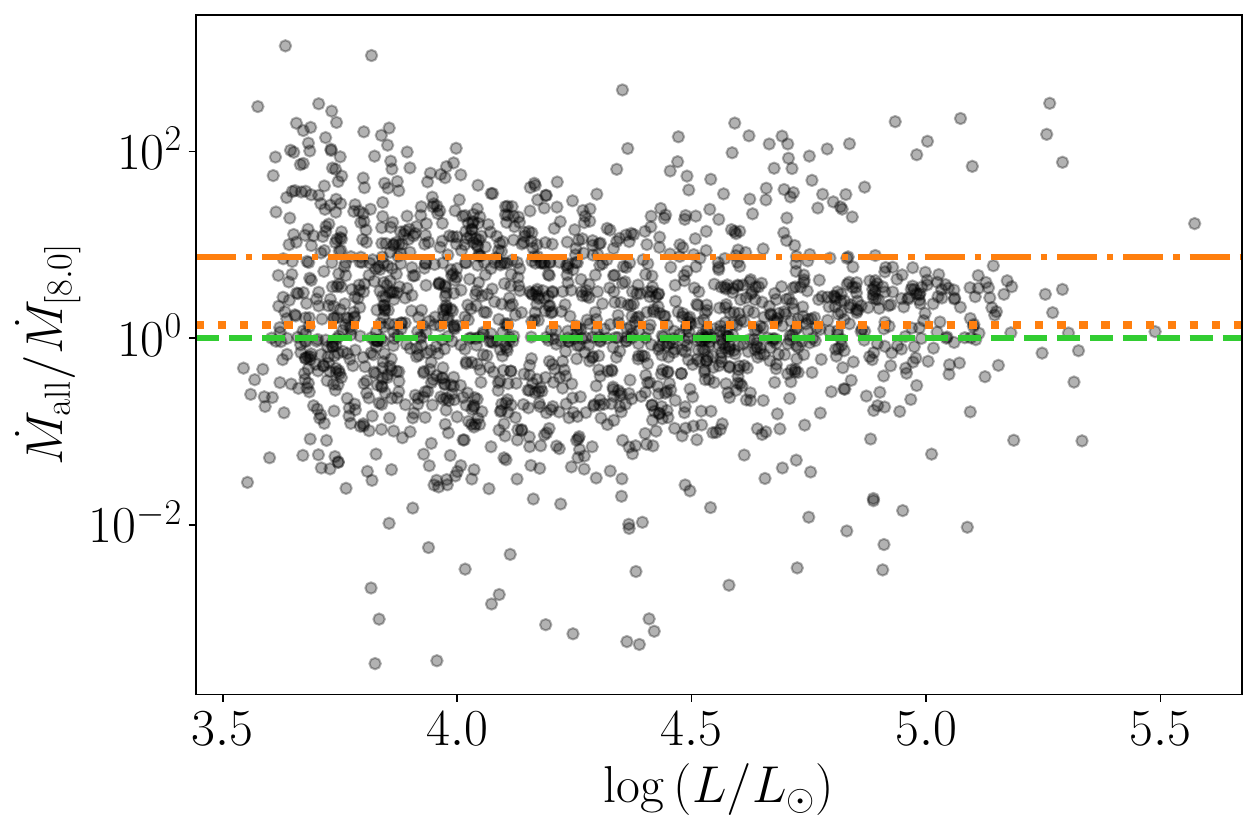}
    \caption{Ratio of the derived $\dot{M}$ using photometry in all bands over the derived $\dot{M}$ using photometry up to 8 $\mu$m for RSGs in the LMC. The green dashed line demonstrates the ratio of 1. The dotted and dot-dashed orange lines indicate the median (at 1.4) and mean for $\log{(L/L_\odot)}>4$, respectively.}
    \label{fig:ratio8mu}
\end{figure}

Finally, we examined the sample of RSGs presented in our previous study in the LMC \citep{Antoniadis_2024}, to investigate if it is secure to use photometry at bands up to 8 $\mu$m, as we did in our main result for NGC 6822. To achieve this, we calculated the ratio of the derived $\dot{M}$ using photometry in all bands in the LMC over the $\dot{M}$ using photometry up to \textit{Spitzer} [8.0]. We found that the two $\dot{M}$ agree within an order of magnitude for the majority of RSGs at $\log{(L/L_\odot)}>4$. The median value of the ratio is 1.4 and the mean is 7.4, driven by some outliers. We demonstrate this result in Fig.~\ref{fig:ratio8mu}. Thus, there is more confidence in our derived $\dot{M}$ of RSGs in the NGC 6822 than our estimated errors presented in Fig.~\ref{fig:Mdot_ngc}.

\clearpage

\section{M31 and M33}\label{app:m31}

\begin{figure}[h!]
    \centering
    \includegraphics[width=1\columnwidth]{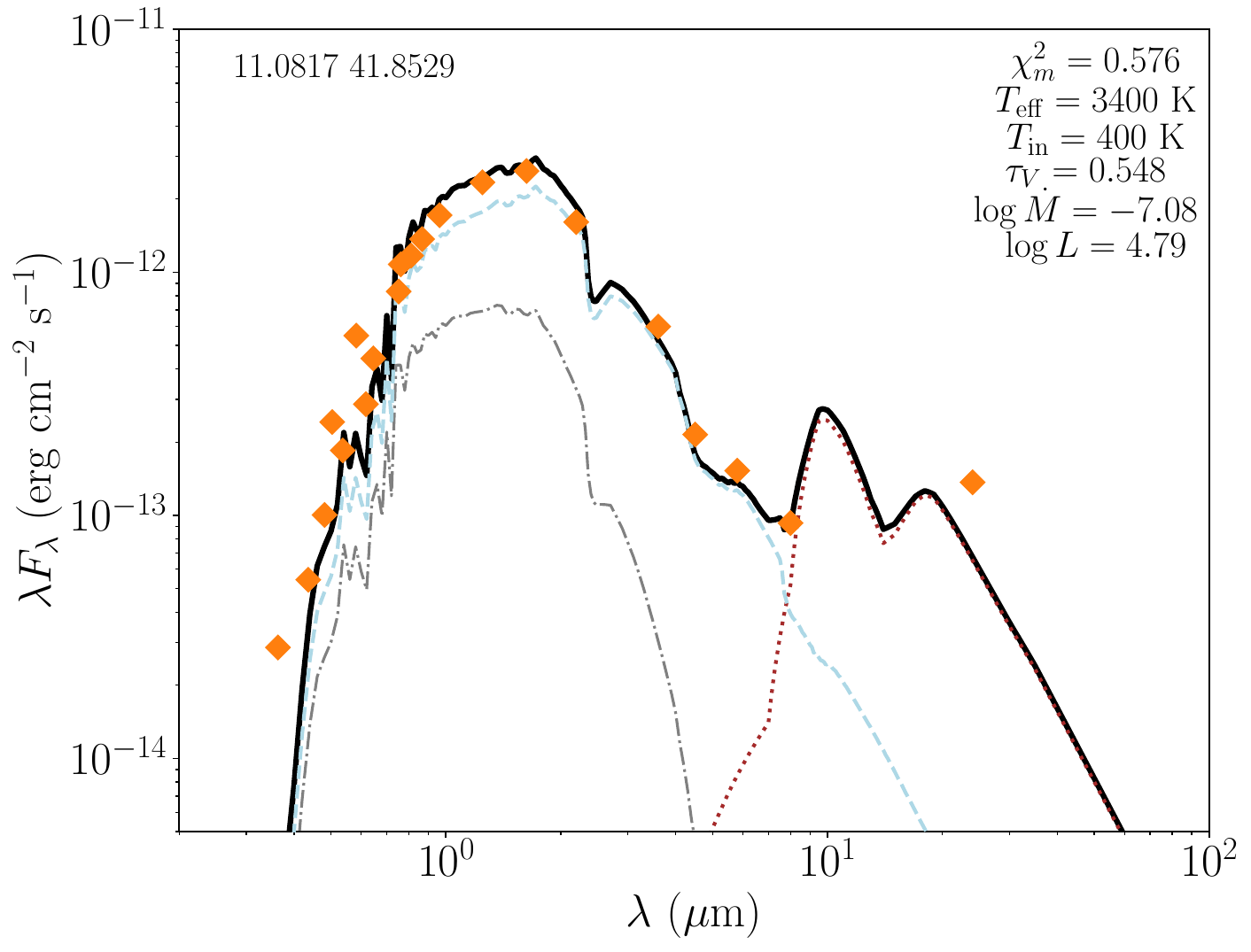}
    \caption{Example SEDs of two RSGs in M31. Symbols and lines are same as Fig.~\ref{fig:sed}.}
    \label{fig:sed_m31}
\end{figure}

\begin{figure}[h]
    \centering
    \includegraphics[width=1\columnwidth]{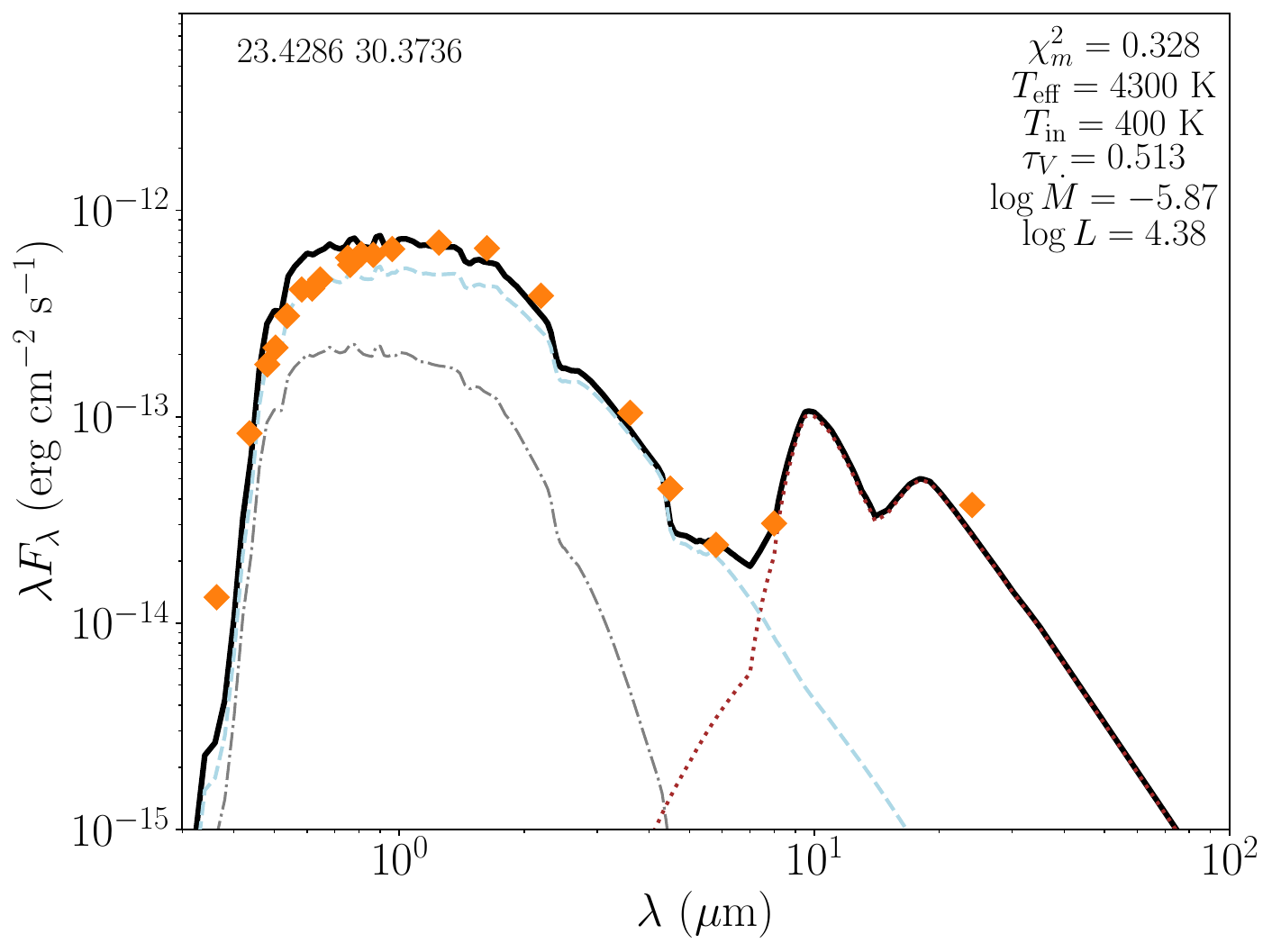}
    \caption{Example SEDs of two RSGs in M33. Symbols and lines are same as Fig.~\ref{fig:sed}.}
    \label{fig:sed_m33}
\end{figure}

\begin{figure}[h]
    \centering
    \includegraphics[width=1\columnwidth]{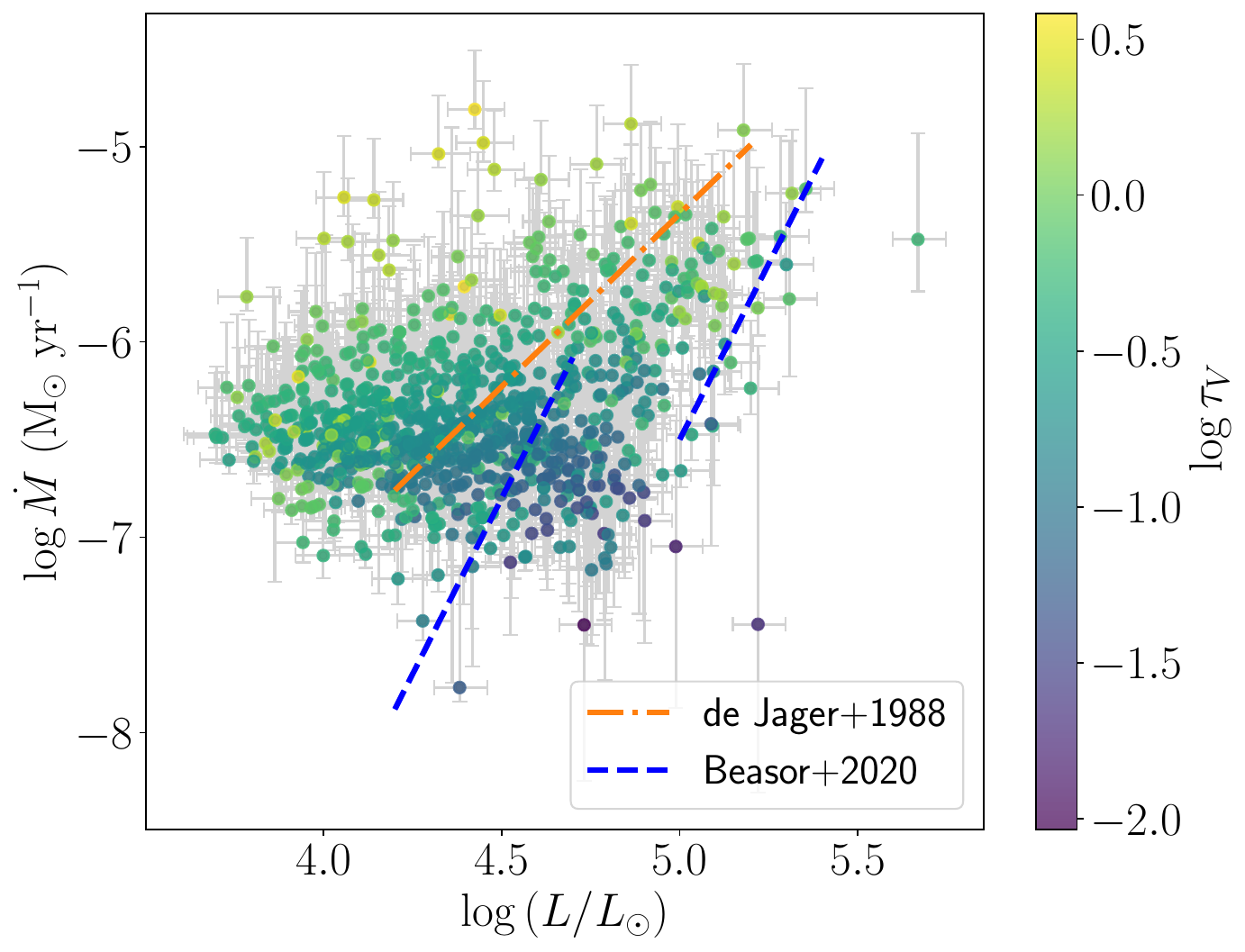}
    \caption{Mass-loss rate as a function of the luminosity of each RSG candidate in M31. The colour bar shows the best-fit $\tau_V$. The lines are the same as in Fig. \ref{fig:Mdot_ngc}.}
    \label{fig:mdot_M31}
\end{figure}

\begin{figure}[h]
    \centering
    \includegraphics[width=1\columnwidth]{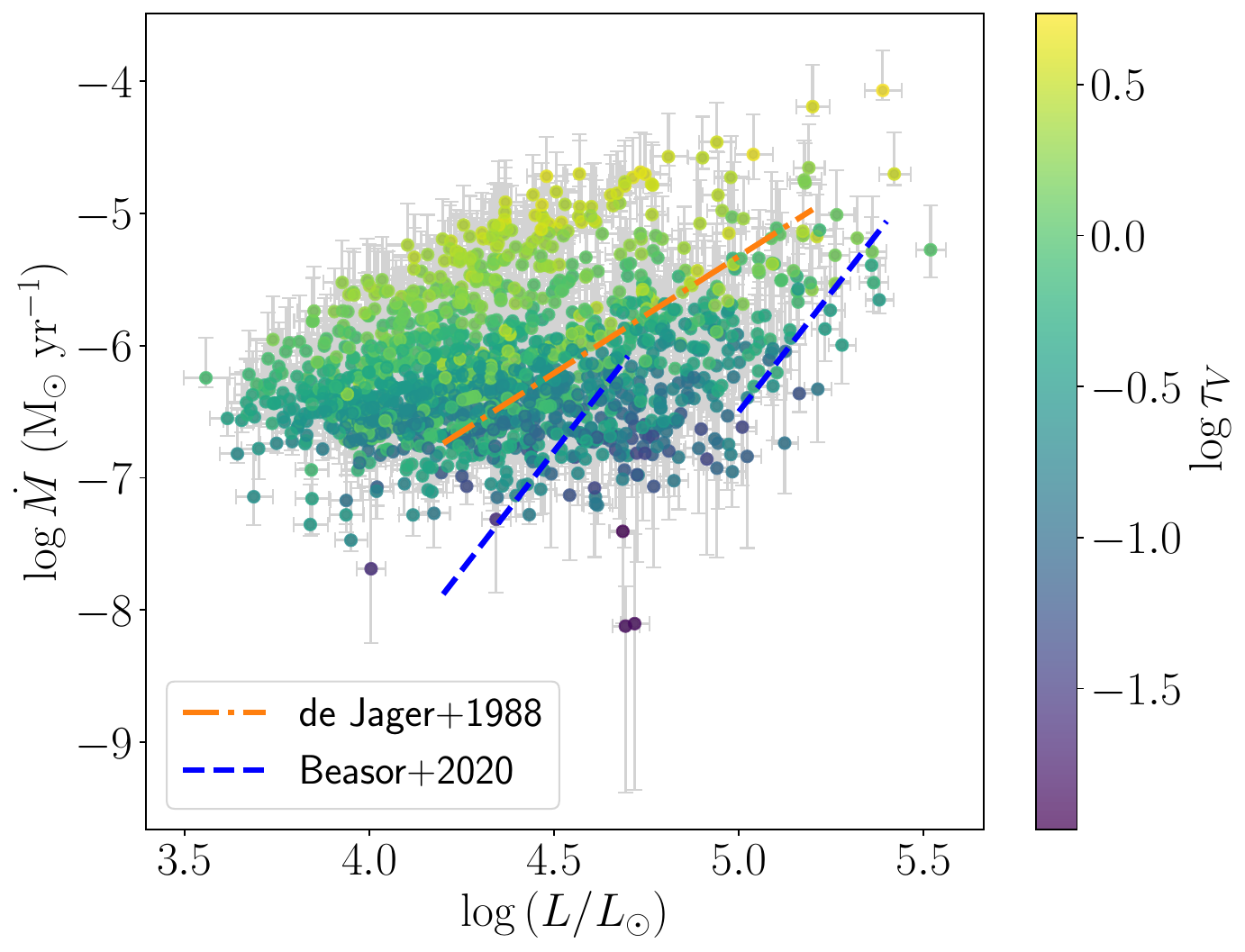}
    \caption{Same as Fig. \ref{fig:mdot_M31}, but for M33.}
    \label{fig:mdot_M33}
\end{figure}

We computed the mass-loss rates of the RSG candidates in M31 and M33 from the sample of \citet{Wang_2021}. We show example SEDs of RSGs in M31 and M33 in Fig.\ \ref{fig:sed_m31} and \ref{fig:sed_m33}, respectively. The \textit{Spitzer} photometry indicates abnormally high infrared excess compared to the other galaxies at closer distance, especially at [24], which is also seen in the example SEDs \citep[blending in \textit{Spitzer} photometry for M33 was also found by][]{Javadi_2015}. Correspondingly, Fig.\ \ref{fig:mdot_M31} and \ref{fig:mdot_M33} demonstrate the resulting $\dot{M}$ vs.\ $L$ in M31 and M33. It has to be mentioned that \citet{Wang_2021} found higher $\dot{M}$ than our work. The main reason is that they assumed an MRN grain size distribution in the range of [0.01, 1] which would result in higher $\dot{M}$ by a factor of 20-30 (see \citealt{Antoniadis_2024}) and used the relation for the outflow speed for the LMC from \citet{Goldman_2017}, which is steep and also results in higher $\dot{M}$.

\clearpage
\onecolumn
\section{RSG catalogues and results} \label{app:catalogues}

\begin{table}[htp]
\small
\centering
  \begin{threeparttable}
    \caption{Properties and derived parameters of the RSG candidates in the SMC.}
    \renewcommand{\arraystretch}{1.3}
        \begin{tabular}{lllllllllll}
        \hline\hline 
        ID &  ID$_\mathrm{Y19}$ &   R.A. (J2000) &    Dec. (J2000) &  $J_\mathrm{2MASS}$ (mag) &  ... &      $\log{(\dot{M}/M_\odot)}$          &  $T_\mathrm{eff}^{(J-K_s)_0}$ (K) &  $T_\mathrm{in}$ (K)  & $\tau_V$ &   \\
        \hline        
        1 &  3125   &  7.980026 & $-73.578530$ &  12.086 &  ... &       $-8.56_{-1.19}^{+ 1.76}$ &       4280 &        $ 1200^ {+0}  _{-800} $  &  $0.003_{-0.002}^{+0.008}$ \\
        2 &  3130   &  7.981623 & $-73.585580$ &  11.399 &  ... &       $-7.86_{-1.67}^{+ 1.55}$ &       4320 &        $ 1200^ {+0}  _{-800} $  &  $0.008_{-0.007}^{+0.013}$ \\
        3 &  3480   &  8.161747 & $-73.661354$ &  12.899 &  ... &       $-7.70_{-1.54}^{+ 0.90}$ &       4200 &        $ 1200^ {+0}  _{-600} $  &  $0.028_{-0.023}^{+0.005}$ \\
        4 &  3945   &  8.391373 & $-73.944916$ &  11.455 &  ... &       $-8.31_{-1.27}^{+ 1.82}$ &       4220 &        $ 1100^{+100} _{-700} $  &  $0.003_{-0.002}^{+0.012}$ \\
        5 &     - &  8.477611 & $-73.845380$ &  11.905 &  ... &         - &                          4220 &            - &    -  \\
        \vdots  &  \vdots & \vdots &  \vdots &    \vdots & \vdots  &  \vdots & \vdots & \vdots & \vdots \\
        \hline
        \end{tabular}
    \label{tab:results_smc} 
    \begin{tablenotes}
        \small
        \vspace{5pt}
        \item Notes: This table is available in its entirety in machine-readable format at the CDS. A portion of it is shown here for guidance regarding its form and content.
    \end{tablenotes}
  \end{threeparttable}
\end{table}

\begin{table}[htp]
\centering
\small
  \begin{threeparttable}
    \caption{Properties and derived parameters of the RSGs in the MW.}
    \renewcommand{\arraystretch}{1.3}
        \begin{tabular}{llllllll}
        \hline\hline 
        ID &     Alias &    SpType\_a &  Spectral Type &   ... &      $\log{(\dot{M}/M_\odot)}$     &  $T_\mathrm{in}$ (K)  &  $\tau_V$  \\
        \hline        
        1 &   V* V386 Cep &       M3 &          M4I                       , M3Ib, M3   Ib &   ... &  $-5.03_{-0.07}^{+0.30}$ &   1200$_{-  0}^{+  0}$ &   3.240$_{-0}^{+ 0}$    \\
        2 &   BD-16  2051 &       M3 &                                                M3I &   ... &  $-9.75_{-0.07}^{+1.10}$ &   1200$_{-500}^{+  0}$ &   0.001$_{-0}^{+ 0.001}$ \\
        3 &     HD 164264 &       K5 &                                           K5 (III) &   ... &  $-9.80_{-0.07}^{+1.49}$ &   1200$_{-700}^{+  0}$ &   0.001$_{-0}^{+ 0.002}$ \\
        4 &  [2018MZM] 77 &       M1 &                                               M1I: &   ... &  $-8.27_{-0.15}^{+0.37}$ &   1100$_{-  0}^{+100}$ &   0.062$_{-0}^{+ 0.011}$ \\
        5 &     HD  11092 &       K4 &  K5Iab-Ib   , K4+ Ib-IIa,   ...  &             ... &  $ -9.47_{-0.07}^{+0.37}$ &   1100$_{-  0}^{+  0}$ &   0.001$_{-0}^{+0}$    \\
        \vdots  &  \vdots & \vdots &  \vdots &    \vdots    & \vdots & \vdots & \vdots \\
        \hline
        \end{tabular}
    \label{tab:results_mw} 
    \begin{tablenotes}
        \small
        \vspace{5pt}
        \item Notes: This table is available in its entirety in machine-readable format at the CDS. A portion of it is shown here for guidance regarding its form and content.
    \end{tablenotes}
  \end{threeparttable}
\end{table}

\begin{table}[htp]
\small
\centering
  \begin{threeparttable}
    \caption{Properties and derived parameters of the RSG candidates in the NGC 6822.}
    \renewcommand{\arraystretch}{1.3}
        \begin{tabular}{lllllllllll}
        \hline\hline 
        ID &     R.A. (J2000) &    Dec. (J2000) &  $V_\mathrm{LGGS}$ (mag) & $\sigma_{V, \ \mathrm{LGGS}}$ (mag) &  ... &      $\log{(\dot{M}/M_\odot)}_{\mathrm{ML}}$    &  $T_\mathrm{eff}^{(J-K_s)_0} (K)$  &  $T_\mathrm{in}$ (K)  &  $\tau_V$ &   \\
        \hline        
        1 &      296.237200 & $-$14.875604 &    18.580 & 0.004 &  ... &    $-8.64^{+1.31}_{-0.48}$ &             4330 &        800$_{-400}^{+400}$ &      $0.001_{-0}^{+ 0.004}$ \\
        2 &      296.264900 & $-$14.727196 &    18.600 & 0.005 &  ... &    $-8.88^{+1.03}_{-0.21}$ &             4200 &       1100$_{-400}^{+100}$ &      $0.001_{-0}^{+ 0.003}$ \\
        3 &      296.318970 & $-$14.780240 &    18.394 & 0.005 &  ... &    $-8.54^{+1.38}_{-0.48}$ &             4180 &        800$_{-400}^{+400}$ &      $0.001_{-0}^{+ 0.005}$ \\
        4 &      296.218422 & $-$14.754016 &  -        &     - &  ... &    $-8.88^{+1.44}_{-0.21}$ &             4150 &       1100$_{-700}^{+100}$ &      $0.001_{-0}^{+ 0.003}$ \\
        5 &      296.299448 & $-$14.756322 &         - &     - &  ... &    $-8.88^{+1.58}_{-0.21}$ &             4210 &       1100$_{-700}^{+100}$ &      $0.001_{-0}^{+ 0.004}$ \\
        \vdots  &  \vdots & \vdots &  \vdots &    \vdots & \vdots  &  \vdots & \vdots & \vdots & \vdots \\
        \hline
        \end{tabular}
    \label{tab:results_6822} 
    \begin{tablenotes}
        \small
        \vspace{5pt}
        \item Notes: This table is available in its entirety in machine-readable format at the CDS. A portion of it is shown here for guidance regarding its form and content.
    \end{tablenotes}
  \end{threeparttable}
\end{table}

\clearpage
\twocolumn
\section{Spectroscopic RSGs in the LMC and SMC} \label{app:spec}

\begin{figure}[h]
    \centering
    \includegraphics[width=1\columnwidth]{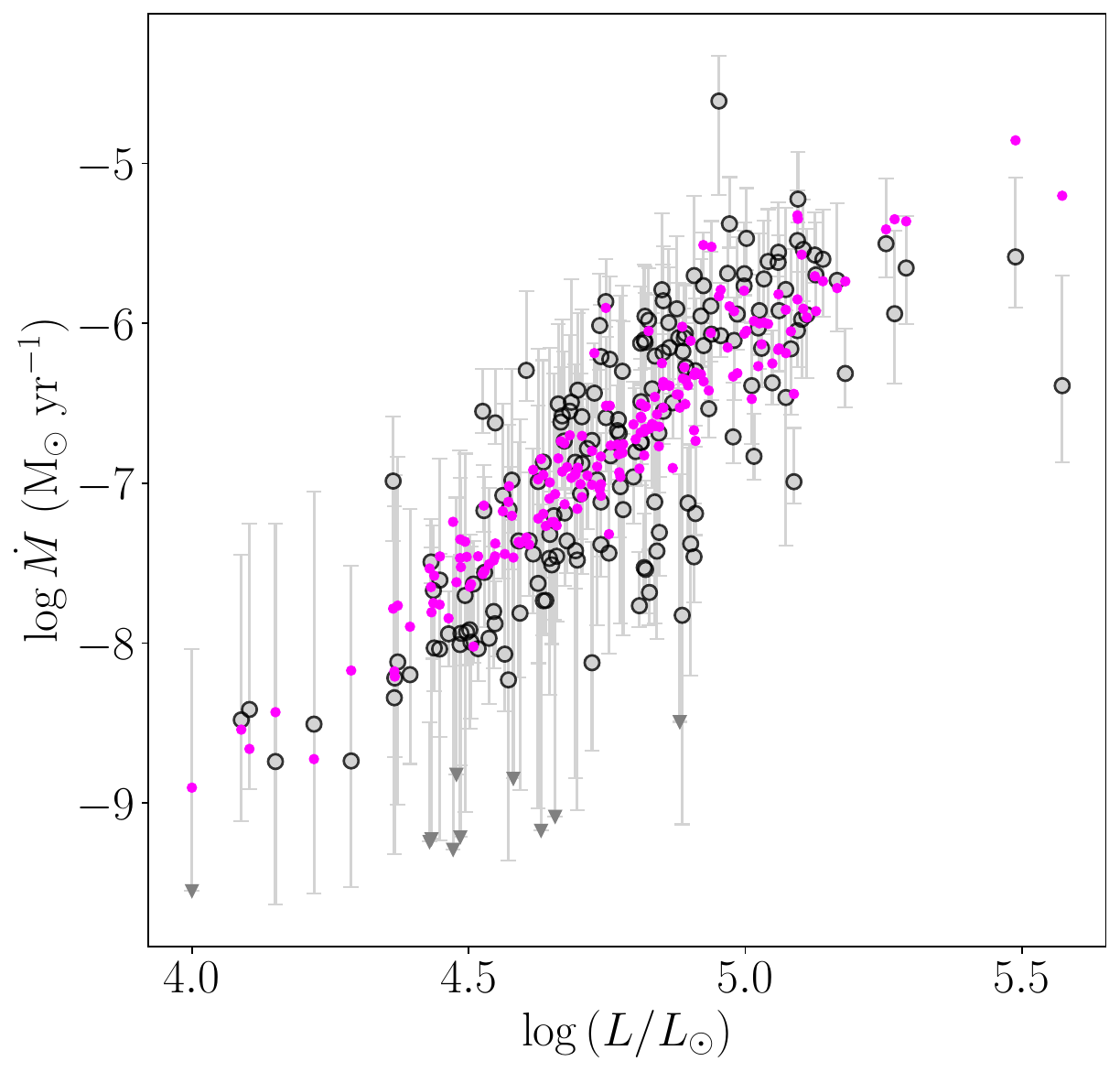}
    \caption{Mass-loss rates of RSGs in the LMC with spectroscopic classifications from the literature. Triangles indicate upper limits. The magenta points represent the prediction from fitted $\dot{M}(L,T_\mathrm{eff})$ relation.}
    \label{fig:mdot_specLMC_fit}
\end{figure}

We used the spectroscopic RSGs from the LMC and SMC to determine whether there is a turning point in the $\dot{M}$ vs. $L$ relation. However, there are insufficient spectroscopic RSGs in the literature for the LMC at low $L$ to determine a turning point. Figure~\ref{fig:mdot_specLMC_fit} shows the result from the LMC and our derived linear relation using Eq.~\ref{eq:Mdot}.

\begin{figure}[h]
    \centering
    \includegraphics[width=1\columnwidth]{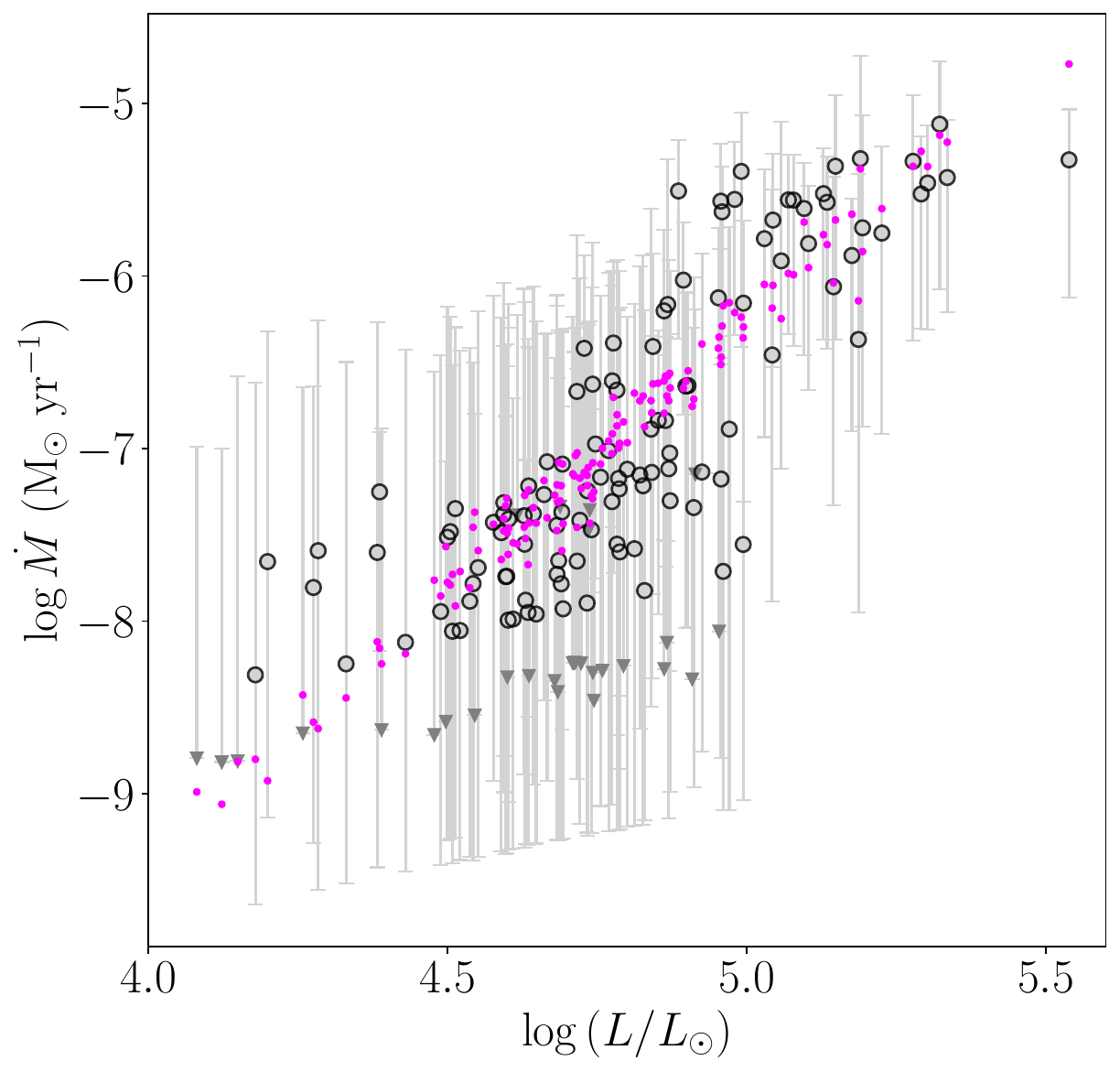}
    \includegraphics[width=1\columnwidth]{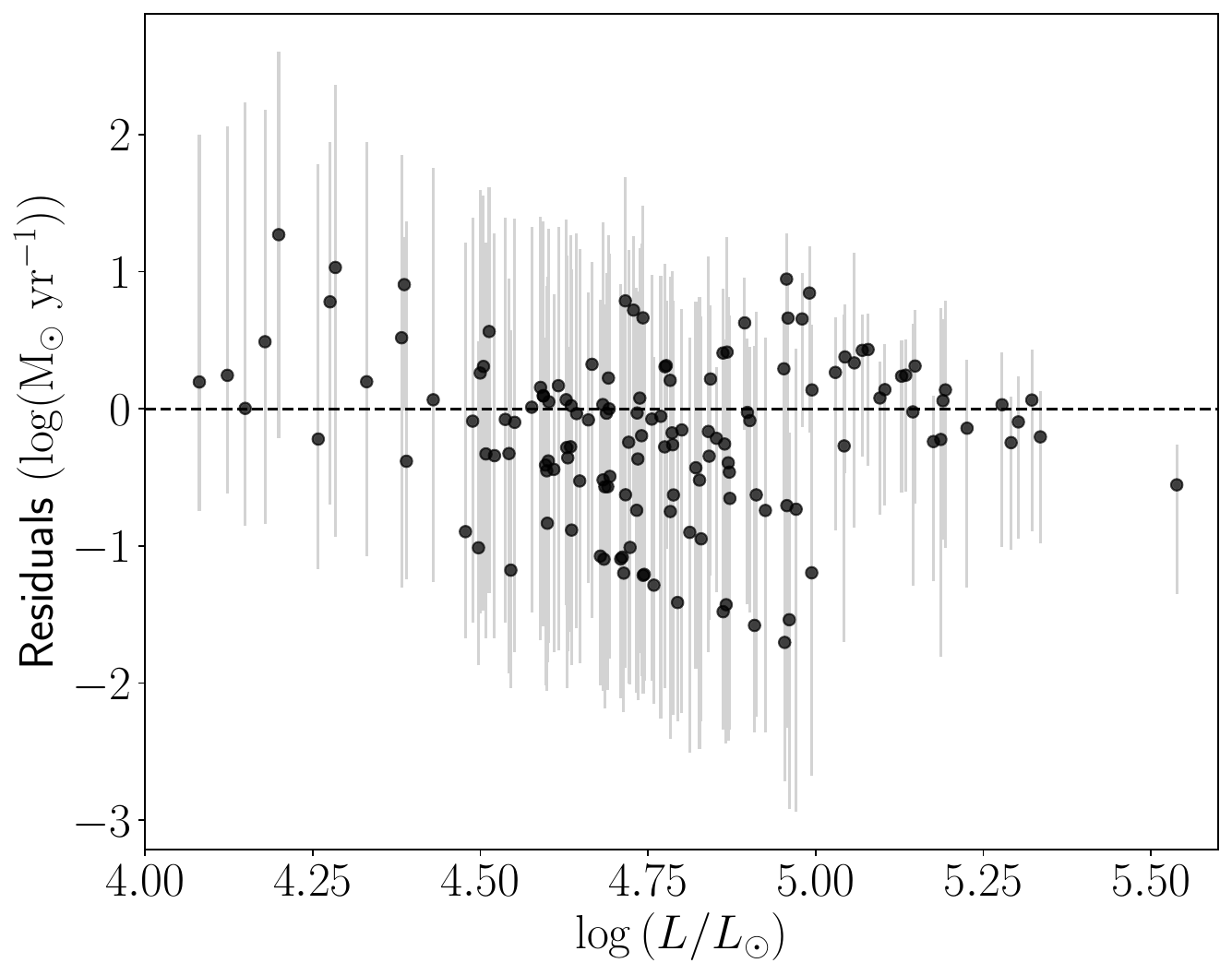}
    \caption{\textit{Top:} Mass-loss rates of RSGs in the SMC with spectroscopic classifications from the literature. Triangles indicate upper limits. \textit{Bottom:} Residual $\dot{M}$ values, defined as $\log\dot{M}_\mathrm{measured}-\log\dot{M}_\mathrm{prescription}$, fitting a linear relation in the sample of spectroscopic RSGs in the SMC.}
    \label{fig:mdot_specSMC_fitlin}
\end{figure}

\begin{figure}[h]
    \centering
    \includegraphics[width=1\columnwidth]{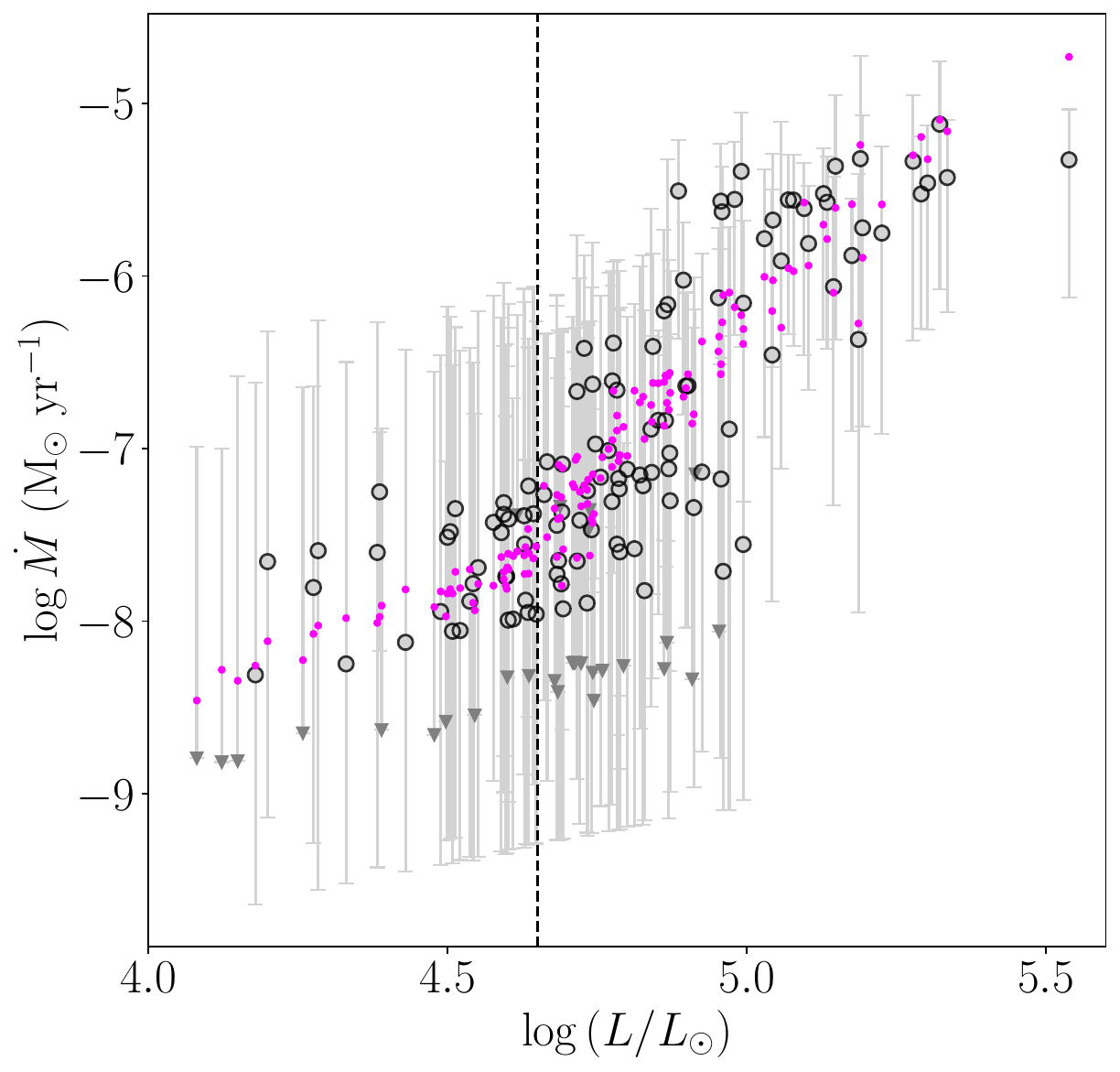}
    \includegraphics[width=1\columnwidth]{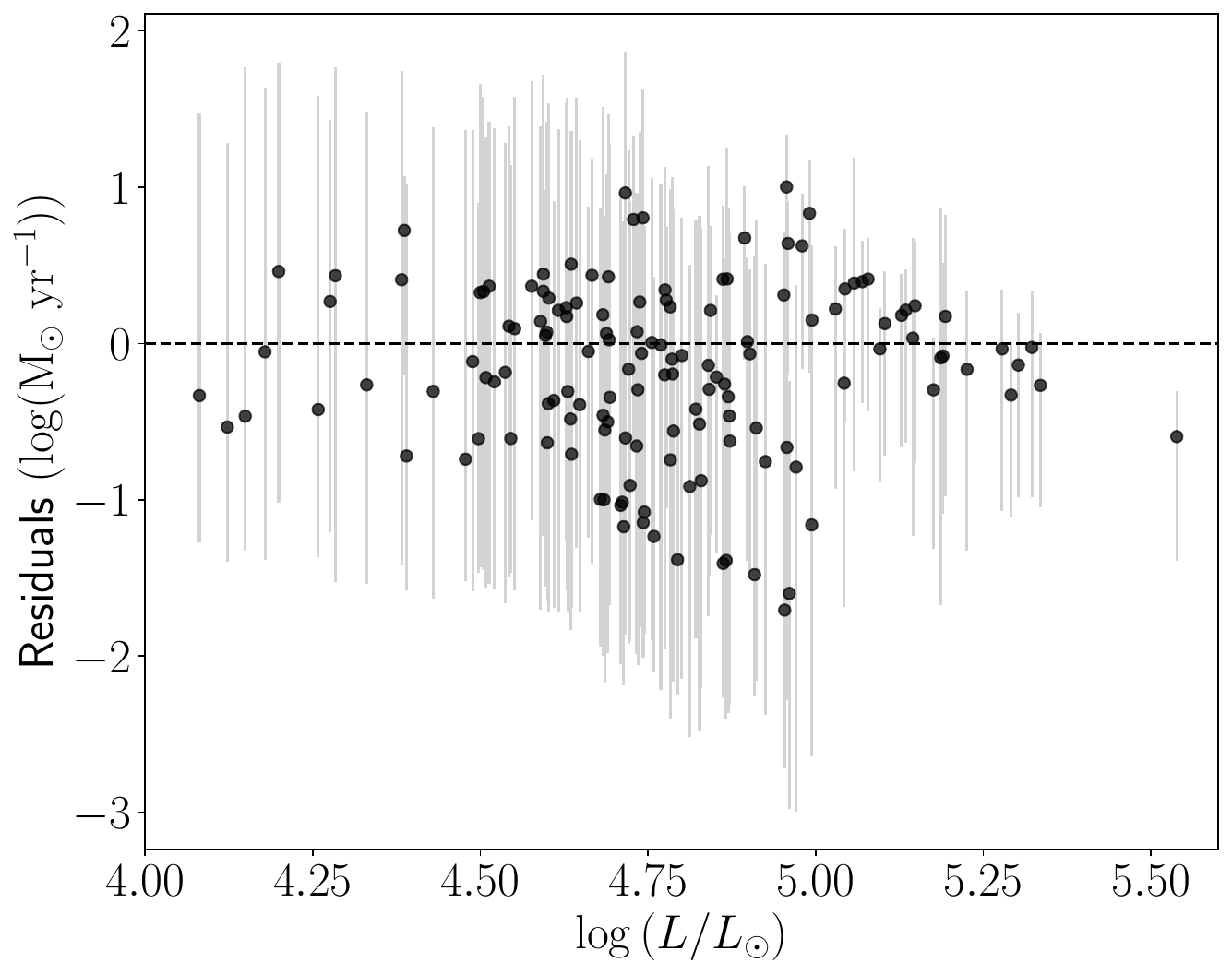}
    \caption{\textit{Top:} Mass-loss rates of RSGs in the SMC with spectroscopic classifications from the literature. Triangles indicate upper limits. \textit{Bottom:} Residual $\dot{M}$ values, defined as $\log\dot{M}_\mathrm{measured}-\log\dot{M}_\mathrm{prescription}$, fitting a broken relation in the sample of spectroscopic RSGs in the SMC.}
    \label{fig:mdot_specSMC_fitsplit}
\end{figure}

The SMC has more data at low $L$ than the LMC. Thus, we fitted both a linear relation and a broken law to examine which reproduces the measured $\dot{M}$ better. 
\autoref{tab:coef_spec} presents the best-fit coefficients of Eq.~(\ref{eq:Mdot}) for the spectroscopic RSGs in the LMC and SMC. Figure~\ref{fig:mdot_specSMC_fitlin} shows the derived linear relation and the residuals for the SMC, and Fig.~\ref{fig:mdot_specSMC_fitsplit} the corresponding relation and residuals using a broken-law at $\log{(L/L_\odot)}=4.65$. Finally, the upper limit for the fraction of dust-free spectroscopic RSGs in the SMC is 20\%. We should note that the value of $c_2$ for the SMC broken law for $\log{(L/L_\odot)}=4.65$ is positive, not showing an anti-correlation with $T_\mathrm{eff}$, as we would expect, and with substantial error. This is probably because of the low number of spectroscopic RSGs at that range and should not be considered. We suggest using the relation assuming an average $T_\mathrm{eff}$ or the derived $\dot{M}$ relation for the whole sample of RSGs in the SMC, presented in our main results.

\begin{table}[h]
    \centering
    \caption{Best-fit parameters of Eq. (\ref{eq:Mdot}) for the spectroscopic RSGs in the LMC and SMC.}
    \renewcommand{\arraystretch}{1.2}
    \begin{tabular}{c | c c c}
        \hline\hline
        $\log{(L/L_\odot)}$ & $c_1$   & $c_2$  & $c_3$ \\
         \hline
        \rowcolor{lightgray!50} \multicolumn{4}{c}{LMC} \\
        all & $2.25\pm0.13$  & $-31.48\pm3.18$  & $-17.82\pm0.61$ \\
       \hline
        \rowcolor{lightgray!50} \multicolumn{4}{c}{SMC linear} \\
        all & $2.94\pm0.32$  & $-15.28\pm10.25$  & $-20.84\pm1.61$ \\
       \hline
       \rowcolor{lightgray!50} \multicolumn{4}{c}{SMC broken-law} \\
        $<4.65$ & $1.45\pm1.37$  & $9.22\pm25.18$  & $-14.48\pm6.24$ \\
        $\geq4.65$ & $3.09\pm0.45$  & $-20.73\pm11.24$  & $-21.56\pm2.3$ \\
        \hline
    \end{tabular}
    \
    \label{tab:coef_spec}
\end{table}

\clearpage
\onecolumn
\section{Mass-loss rates as a function of mid-infrared colours} \label{app:colour}

\begin{figure}[h]
    \centering
    \includegraphics[width=0.45\columnwidth]{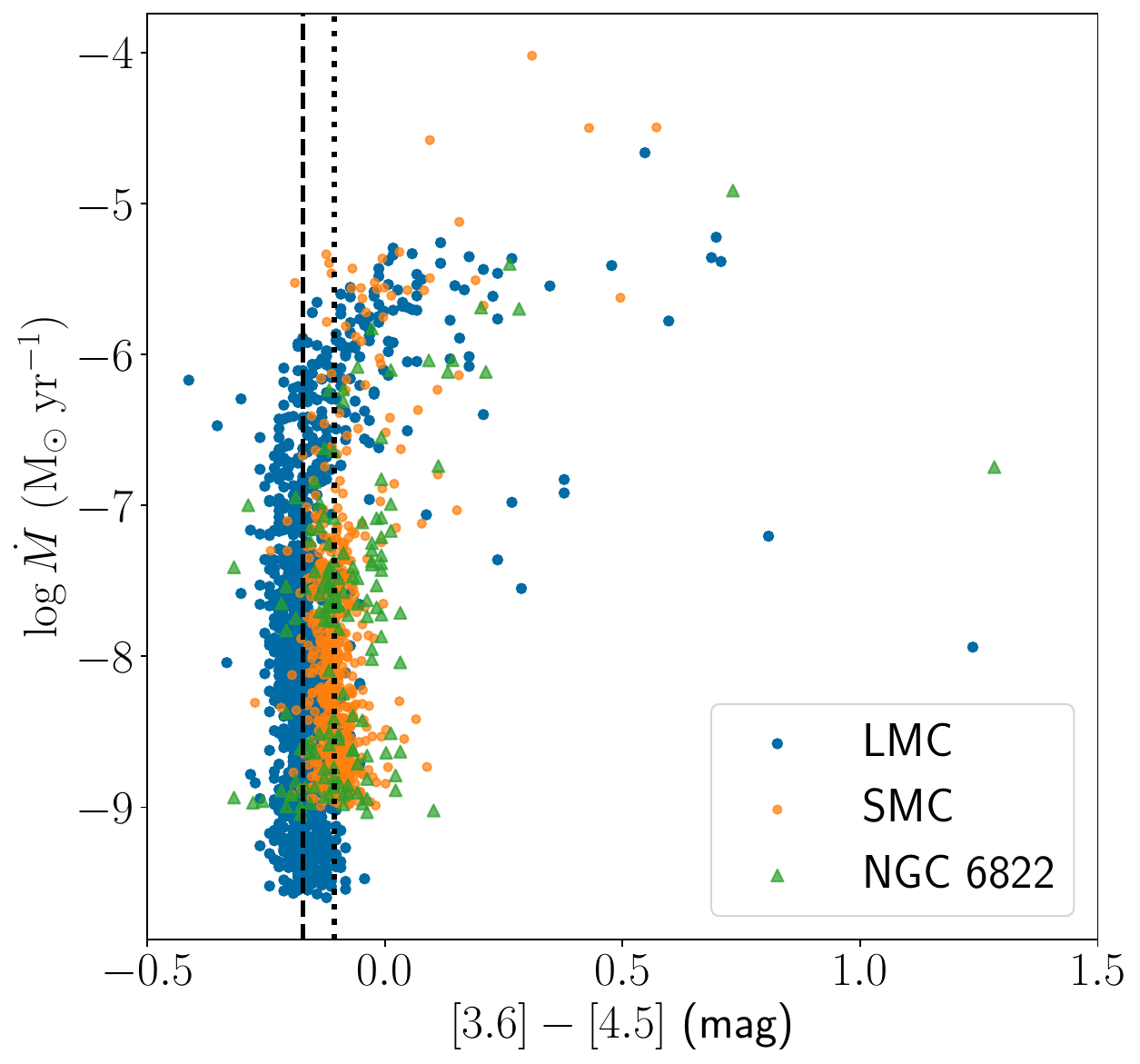}
    \includegraphics[width=0.45\columnwidth]{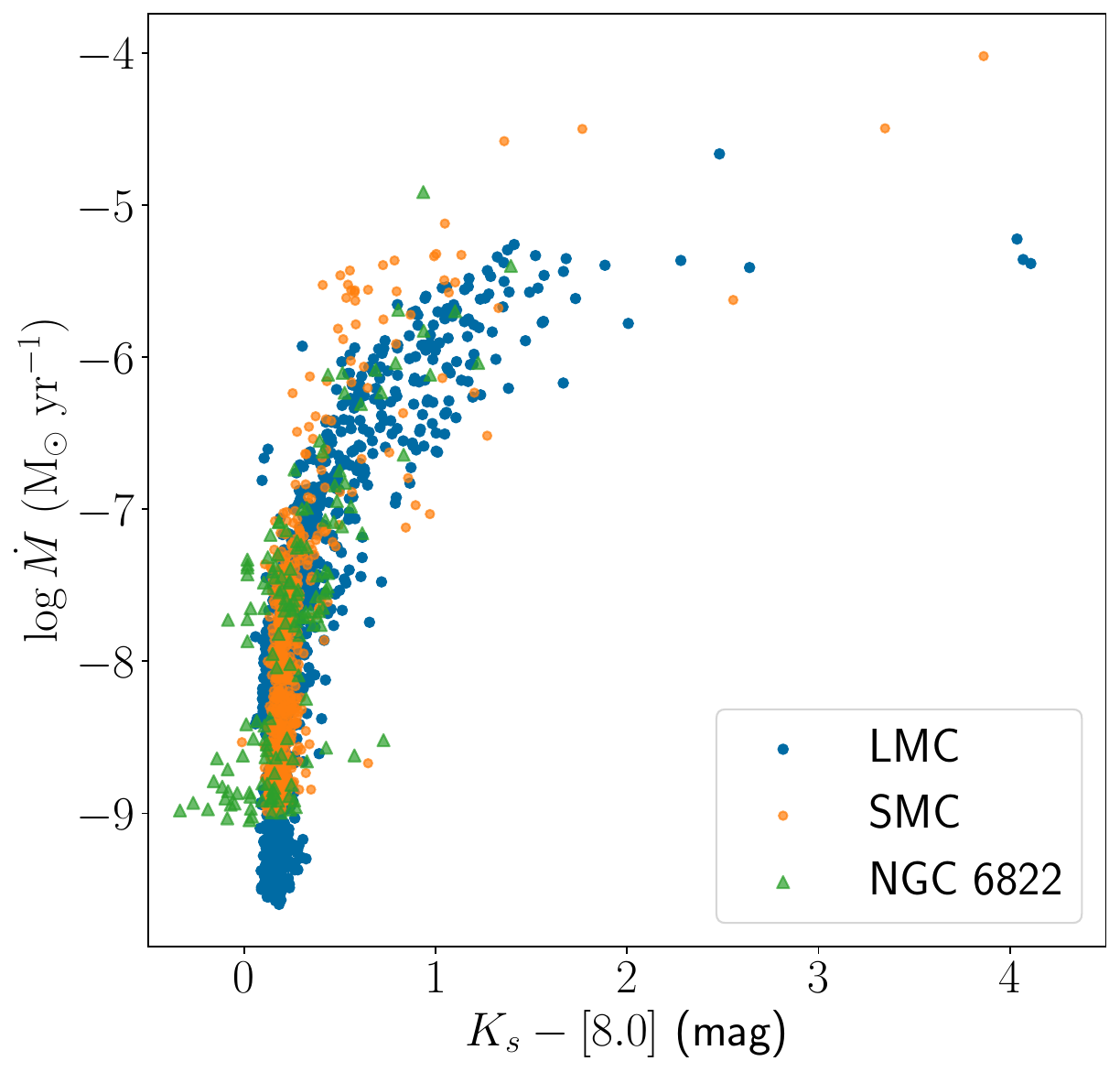}
    \includegraphics[width=0.45\columnwidth]{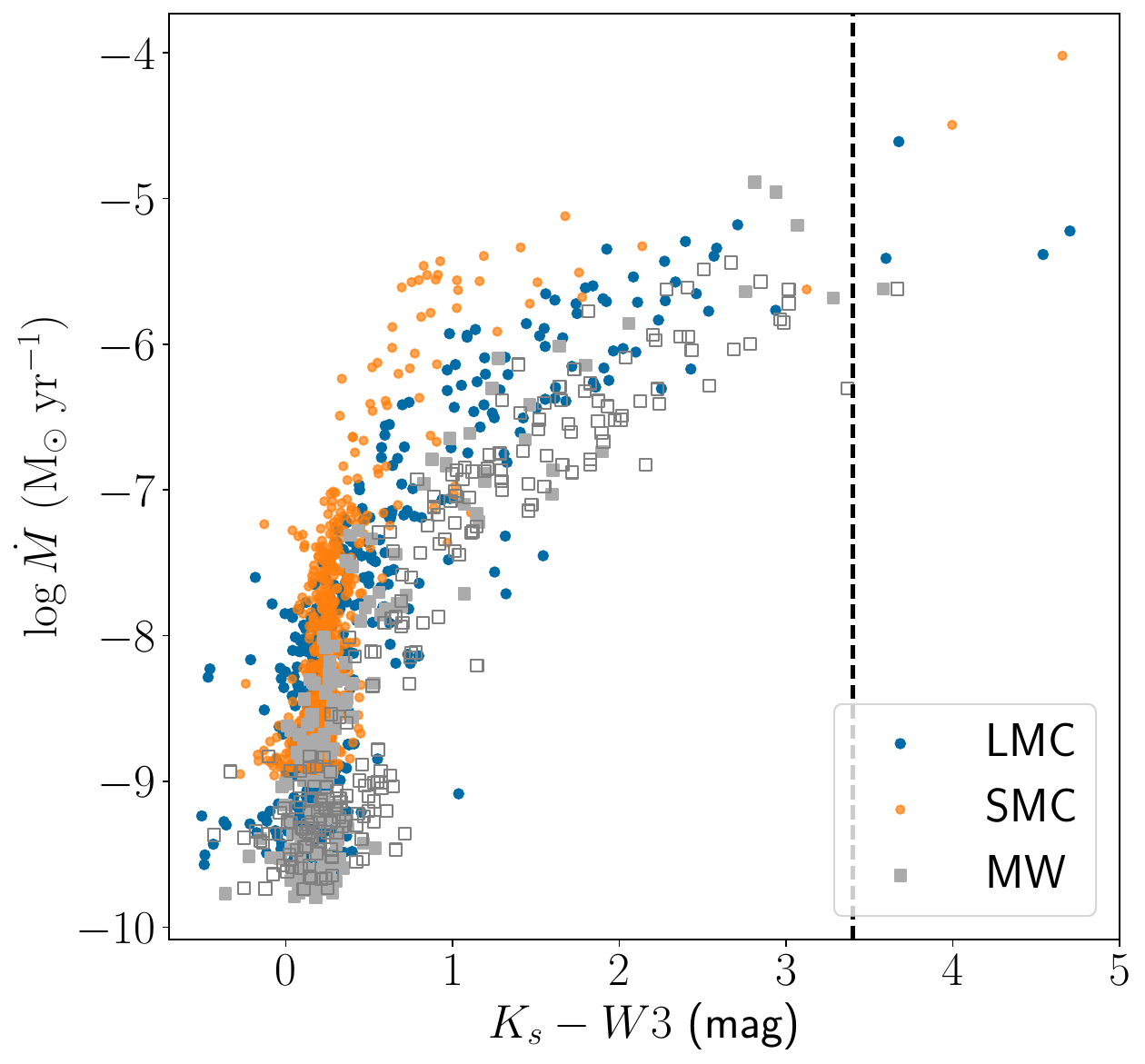}
    \includegraphics[width=0.45\columnwidth]{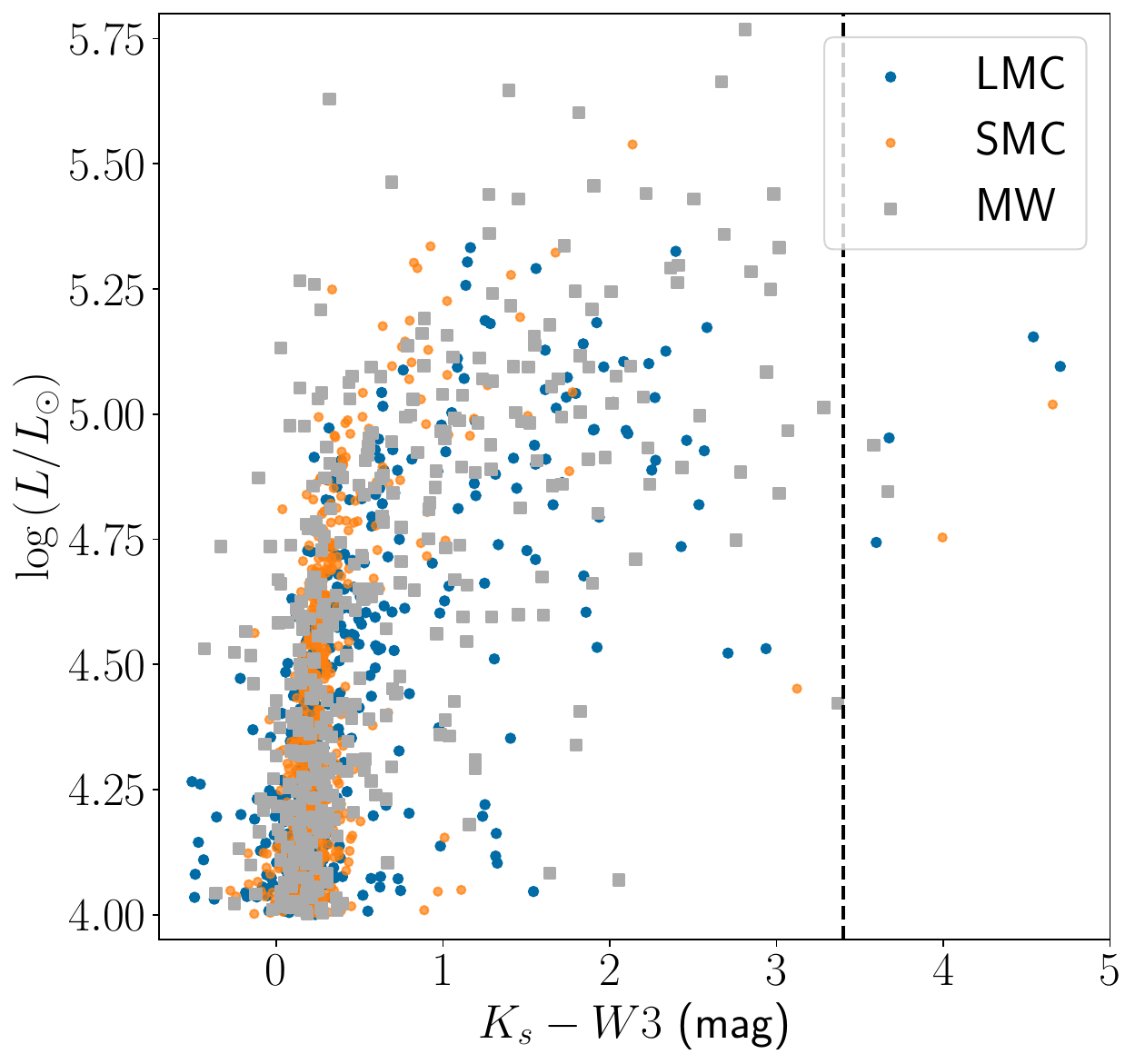}

    \caption{\textit{Top:} Mass-loss rate as a function of colour for the LMC (blue), SMC (orange), and NGC 6822 (green triangles). The dashed and dotted vertical lines on the left panel indicate the median values at $\log(\dot{M}/(\mathrm{M_\odot\ yr^{-1}}))<-7$ for the LMC and SMC, respectively. \textit{Bottom:} Colour mass-loss rate (left) and colour luminosity (right) diagrams for the LMC (blue), SMC (orange), and MW (grey squares). The open squares indicate the more uncertain $\dot{M}$ in the MW with not well-fit $JHK_s$ photometry. The dashed vertical line shows the limit above which there are candidate dust-enshrouded RSGs.}
    \label{fig:irac}
\end{figure}

Figure \ref{fig:irac} presents the mass-loss rates of RSGs with $\log(L/L_\odot)>4$ as a function of different mid-IR colours (corrected for foreground extinction). We have shown that dust production is correlated with metallicity, which affects the mid-IR excess, and it is also known that the intrinsic colour of RSGs is metallicity dependent \citep[e.g.][]{Li_2024}. The dashed and dotted vertical lines in the top left panel of Fig.\ \ref{fig:irac} indicate the median values at $\dot{M}<10^{-7}\ \mathrm{M_\odot\ yr^{-1}}$ for the LMC and SMC, respectively, which differ by 0.07 mag. The turn at $\dot{M}>10^{-7}\ \mathrm{M_\odot\ yr^{-1}}$ becomes sharper at higher metallicity. The results from NGC 6822 are more dispersed due to the worse quality of \textit{Spitzer} photometry at that distance but seem to agree with the SMC. This turn and the trend with metallicity is clearer in Fig.\ \ref{fig:irac} because the $W3$ band is connected with the silicate bump at around 10 $\mu$m. However, the open squares show the $\dot{M}$ of the Galactic RSGs with not well-fit near-IR photometry to the models, which could be underestimated, as we have mentioned. 

The bottom right panel of Fig.\ \ref{fig:irac} shows the luminosity as a function of the $K_s-W3$ colour. According to \citet{Beasor_2022}, $K_s-W3>3.4$~mag indicates candidate dust-enshrouded RSGs. These RSGs consist of roughly 0.6\% of our total sample or RSGs with $WISE3$ photometry. We should mention that the samples of RSG candidates in the LMC and SMC could be biased against the inclusion of dust-enshrouded stars due to the photometric selection criteria. Nevertheless, we found a similar fraction considering only the Galactic RSGs. The sample may still not be complete, but it supports the argument from \citet{Beasor_2022} of the extreme scarcity of such RSGs.

\end{appendix}

\end{document}